\documentclass[rmp,aps,nofootinbib,twocolumn]{revtex4-1}

\usepackage{graphicx}
\usepackage{multirow}
\usepackage{amsfonts,amsmath,amssymb,bm,textcomp}

\numberwithin{equation}{section}

\def\bmm{{\rm \bf m}}

\def\bY{{\rm \bf Y}}
\def\bU{{\rm \bf U}}
\def\bN{{\rm \bf N}}
\def\BR{{\rm BR}}
\def\Im{{\rm Im}}
\def\Re{{\rm Re}}
\def\Hc{{\rm H.c.}}
\def\dmsol{\Delta m^2_{21}}
\def\dmatm{\Delta m^2_{31}}
\def\CP{{\rm CP}}
\def\T{{\rm T}}
\def\diag{{\rm diag}}
\def\GeV{{\rm GeV}}
\def\eV{{\rm eV}}
\def\km{{\rm km}}
\def\ndbd{0\nu\beta\beta}

\begin{document}
\title{Leptonic CP violation}

\author{G.~C.~Branco}
\email{gbranco@ist.utl.pt}
\affiliation{\mbox{CERN, Theoretical Physics Division, CH-1211 Geneva 23, Switzerland}}
\affiliation{\mbox{Centro de F\'{\i}sica Te\'{o}rica de Part\'{\i}culas, Instituto Superior T\'{e}cnico,} \mbox{Universidade T\'{e}cnica de Lisboa, Avenida Rovisco Pais, 1049-001 Lisboa, Portugal}}

\author{R. Gonz\'{a}lez Felipe}
\email{ricardo.felipe@ist.utl.pt}
\affiliation{Instituto Superior de Engenharia de Lisboa, \mbox{Rua Conselheiro Em\'{\i}dio Navarro, 1959-007 Lisboa, Portugal}}
\affiliation{\mbox{Centro de F\'{\i}sica Te\'{o}rica de Part\'{\i}culas, Instituto Superior T\'{e}cnico,} \mbox{Universidade T\'{e}cnica de Lisboa, Avenida Rovisco Pais, 1049-001 Lisboa, Portugal}}

\author{F.~R.~Joaquim}
\email{filipe.joaquim@ist.utl.pt}
\affiliation{\mbox{Centro de F\'{\i}sica Te\'{o}rica de Part\'{\i}culas, Instituto Superior T\'{e}cnico,} \mbox{Universidade T\'{e}cnica de Lisboa, Avenida Rovisco Pais, 1049-001 Lisboa, Portugal}}

\begin{abstract}
Several topics on CP violation in the lepton sector are reviewed. A few theoretical aspects concerning neutrino masses, leptonic mixing, and CP violation will be covered, with special emphasis on seesaw models. A discussion is provided on observable effects which are manifest in the presence of CP violation, particularly, in neutrino oscillations and neutrinoless double beta decay processes, and their possible implications in collider experiments such as the LHC. The role that leptonic CP violation may have played in the generation of the baryon asymmetry of the Universe through the mechanism of leptogenesis is also discussed.
\end{abstract}


\maketitle \tableofcontents
\newpage

\section{INTRODUCTION}
\label{sec1}

The violation of the product of the charge conjugation (C) and parity (P) symmetries, i.e., CP violation (CPV), is well established in the quark sector of the standard model (SM). At present, there is clear evidence that the Cabibbo-Kobayashi-Maskawa (CKM) matrix is complex, even if one allows for the presence of new physics in the $B_d\,$-$\,\bar{B}_d$ and $B_s\,$-$\,\bar{B}_s$ mixings. From a theoretical point of view, the complex phase in the CKM matrix may arise from complex Yukawa couplings and/or from a relative CP-violating phase in the vacuum expectation values (VEV) of Higgs fields. In either case, one expects an entirely analogous mechanism to arise in the lepton sector, leading to leptonic CP violation (LCPV).

The discovery of neutrino oscillations provides evidence for nonvanishing neutrino masses and leptonic mixing. Therefore, it is imperative to look for possible manifestations of CP violation in leptonic interactions. The ideal playground for such a program relies on the phenomenon of neutrino oscillations. At present, several experiments are being planned to pursue such a task, including long-baseline facilities, superbeams, and neutrino factories. Hopefully, they will be able to measure the strength of CP violation and provide a knowledge of the leptonic mixing comparable to what is presently known about the quark sector. Yet, it is crucial to look for alternative manifestations of CP violation outside neutrino oscillations. In particular, the effects of Majorana-type phases may arise in neutrinoless double beta decay ($0\nu\beta\beta$) processes. The observation of such processes would establish the Majorana nature of neutrinos and, possibly, provide some information on the Majorana CP phases. In this review, we discuss the observable effects, which are manifest in the presence of leptonic CP violation. We present a short review of the neutrino oscillation formalism and summarize the prospects for the discovery of CP violation in the lepton sector. The possibility  of extracting information about Majorana phases from $0\nu\beta\beta$ decay processes is also discussed.

The fact that neutrino masses are so small constitutes one of the most puzzling problems of modern particle physics. From a theoretical point of view, the smallness of neutrino masses can be elegantly explained through the seesaw mechanism, which can be realized in several ways depending on the nature of the heavy states added to the SM particle content. One of the most popular variants is the one in which the tree-level exchange of heavy neutrino singlets mediates the process of neutrino mass generation. The mechanism can be equally implemented considering, for instance, heavy scalar or fermion triplets. We review some of the realizations of the seesaw mechanism and discuss different parametrizations, which are useful when establishing a bridge between low-energy and high-energy CP violation in the lepton sector. This analysis will be relevant for the discussion of the connection between low-energy neutrino physics and leptogenesis, one of the most appealing scenarios for the generation of the baryon asymmetry of the Universe.

After the discovery of neutrino oscillations, several models have been put forward to offer an explanation for the pattern of neutrino masses and leptonic mixing. Future data from several kinds of experiments, ranging from kinematical searches to cosmology, will probably shed some light on the ultimate structure of the neutrino mass and mixing. In this regard, there are still fundamental questions to be answered: Are neutrinos Dirac or Majorana particles? What is the absolute neutrino mass scale? How are neutrino masses ordered? How large is the 1-3 leptonic mixing angle?

The explanation of the cosmological matter-antimatter asymmetry observed in nature constitutes one of the greatest challenges for modern particle physics and cosmology. We have entered a new era marked by outstanding advances in experimental cosmology and an unprecedented precision in measuring several cosmological parameters. In particular, the seven-year data recently collected from the Wilkinson Microwave Anisotropy Probe (WMAP) satellite have placed the observed baryon asymmetry in a rather narrow window. These measurements have also made it clear that the current state of the Universe is very close to a critical density and that the primordial density perturbations that seeded large-scale structure formation in the Universe are nearly scale invariant and Gaussian, which is consistent with the inflationary paradigm. Since any primordial asymmetry would have been exponentially diluted during inflation, a dynamical mechanism must have been operative after this period, in order to generate the baryon asymmetry that we observe today. The present review is not aimed at covering all the theoretical ideas on baryogenesis extensively developed over the last few years. Instead, we focus our discussion on the simplest leptogenesis scenarios, putting the emphasis on the role that leptonic CP violation may have played in the origin of matter. After briefly reviewing the simplest seesaw leptogenesis mechanisms, we analyze the possibility of establishing a bridge between leptonic CP violation at high and low energies. As it turns out, there is no model-independent relation between CP violation in leptogenesis and the observable phases of the low-energy leptonic mixing matrix. Such a link can only be established by restricting the number of free parameters in the leptonic flavor sector. From the model-building viewpoint, these restrictions are also necessary to fully reconstruct the neutrino mass matrix from low-energy data measured in feasible experiments.

In the analysis of lepton flavor models, a useful approach when addressing the question of CP violation is the construction of the CP-odd weak basis (WB) invariants. Independent of the basis choice and phase convention, any of these quantities should vanish if CP is an exact symmetry of the theory. Thus, in CP-violating theories which contain several phases, invariants constitute a powerful tool to investigate whether a particular model leads to leptonic CP violation at high and/or low energies. In our review, we briefly present such an invariant approach, in an attempt at relating leptogenesis with low-energy leptonic mixing phases. Finally, other interesting issues that we address here include the connection of leptogenesis with flavor symmetries and its viability under the hypothesis of minimal lepton flavor violation.

The layout of the review is as follows. In Sec.~\ref{sec2}, we review several topics related with fundamental aspects of neutrino masses, mixing, and CP violation in the lepton sector. First, in Sec.~\ref{sec2.1}, we study leptonic mixing and CP violation in the case when neutrino masses are generated by new physics which breaks the difference between baryon (B) and lepton (L) numbers, i.e., $({\rm B}-{\rm L})$. Our analysis exclusively relies on the low-energy limit and not on any particular $({\rm B}-{\rm L})$-breaking mechanism to give masses to neutrinos. The construction of Dirac and Majorana unitarity triangles is presented in Sec.~\ref{sec2.2}, and the CP transformation properties in the lepton sector of the Lagrangian are discussed in Sec.~\ref{sec2.3}. The weak-basis invariants relevant for low-energy CP violation are then introduced in Sec~\ref{sec2.4}. In Sec.~\ref{sec2.5}, we recall the most popular versions of the seesaw mechanism for the neutrino mass generation, and in Sec.~\ref{sec2.6} we make a short digression on the origin of CP violation. We also briefly comment on the hypothesis of minimal lepton flavor violation in Sec.~\ref{sec2.7}. The present status of the neutrino mass and mixing parameters and the basic aspects of neutrino oscillations in vacuum and matter are briefly reviewed in Sec.~\ref{sec3.1} and~\ref{sec3.2}. In the latter section, we focus on aspects related to CPV in neutrino oscillations and on the prospects of establishing CPV in future experiments. The possibility of probing CPV in $0\nu\beta\beta$ decays is discussed in Sec.~\ref{sec3.3}. In the framework of the type II seesaw mechanism, the CP-violating phases play a crucial role in the predictions for lepton flavor-violating charged-lepton decays, and also in the scalar triplet decays at accelerators, as discussed in Sec.~\ref{sec3.4} and~\ref{sec3.5}, respectively. Nonunitarity effects in the lepton sector are discussed in Sec.~\ref{sec3.6}. Section~\ref{sec4} is devoted to the discussion of the possible role of leptonic CP violation in the origin of the matter-antimatter asymmetry in the context of leptogenesis. After reviewing the three main variants of this mechanism in Sec.~\ref{sec4.1}, we discuss in Sec.~\ref{sec4.2} how high-energy and low-energy CP violation can be related in some specific cases. We then briefly comment on the relevant CP-odd WB invariants for leptogenesis in Sec.~\ref{sec4.3}. Finally, our conclusions and outlook are drawn in Sec.~\ref{sec:conclusion}.

\section{NEUTRINO MASSES, MIXING, AND LEPTONIC CP VIOLATION}
\label{sec2}

Neutrinos are strictly massless in the SM. No Dirac mass can be written since the right-handed neutrino field $\nu_R$ is not introduced, and no Majorana mass term can be generated, either in higher orders of perturbation theory or by nonperturbative effects, due to an exact $({\rm B}-{\rm L})$ conservation. A Majorana mass term has the form $\nu^T_{L_i} C \nu_{L_j} m_{ij}$ and violates $({\rm B}-{\rm L})$ by two units, so it is forbidden by the exact $({\rm B}-{\rm L})$ symmetry. Because of the vanishing of neutrino masses, there is no leptonic mixing or leptonic CP violation in the SM. Any mixing generated in the diagonalization of the charged-lepton masses can be ``rotated away" through a redefinition of the neutrino fields. Therefore, the experimental discovery of neutrino oscillations, pointing to nonvanishing neutrino masses, is a clear indication of physics beyond the SM.

\subsection{The low-energy limit}
\label{sec2.1}

We start by studying leptonic mixing and CP violation in an extension of the SM with neutrino masses generated by new physics which breaks $({\rm B}-{\rm L})$. Our analysis follows an effective theory approach, without relying on any particular mechanism that breaks $({\rm B}-{\rm L})$ and gives masses to neutrinos. Later on we present several realizations in which the $({\rm B}-{\rm L})$-breaking occurs due to the decoupling of heavy states.

\subsubsection{Lepton mass terms}
We assume that the gauge symmetry breaking has taken place and charged-lepton masses have been generated through the Yukawa couplings with the Higgs doublet, while Majorana neutrino masses arise from some unspecified $({\rm B}-{\rm L})$-breaking new physics. The Lagrangian mass terms are
\begin{align}\label{Lmass}
    \mathcal{L}_\text{mass}=-\overline{l_{L}}\,\mathbf{m}_{l}\, l_{R} - \frac{1}{2} \nu^T_{L} C\, \mathbf{m}_{\nu}\, \nu_{L} + \text{H.c.},
\end{align}
where $l_{L,R} \equiv (e, \mu, \tau)_{L,R}^T$ stands for the SM charged-lepton fields, $\nu_{L} \equiv (\nu_e, \nu_\mu, \nu_\tau)_{L}^T$ are the left-handed neutrino fields, and $\mathbf{m}_{l,\nu}$ are arbitrary complex matrices, being $\mathbf{m}_{\nu}$ symmetric.

There is clear evidence in the quark sector that the CKM mixing matrix is complex, even if one allows for the presence of new physics~\cite{Botella:2005fc}. So, in analogy, we assume the existence of leptonic CP violation arising from complex lepton masses. The mass matrices of Eq.~\eqref{Lmass} are written in a weak basis, i.e., a basis for the lepton fields with real and flavor diagonal charged currents,
\begin{align}\label{Lcharged}
     \mathcal{L}_W=\frac{g}{\sqrt{2}} \overline{l_{L}} \gamma_\mu \nu_L W^\mu + \text{H.c.}.
\end{align}
The lepton mass matrices $\mathbf{m}_{l}$ and $\mathbf{m}_{\nu}$ encode all information about lepton masses and mixing. However, there is a redundancy of free parameters in these matrices so that not all of them are physical. This redundancy stems from the fact that one has the freedom to make a unitary WB transformation,
\begin{align}\label{WB1}
    \nu_L=\mathbf{W}_L\,\nu_L^\prime, \quad l_L=\mathbf{W}_L\, l_L^\prime, \quad l_R=\mathbf{W}_R \, l_R^\prime,
\end{align}
under which all gauge currents remain real and diagonal, but the matrices $\mathbf{m}_{l}$ and $\mathbf{m}_{\nu}$ transform in the following way:
\begin{align}\label{WB2}
    \mathbf{m}_{l}^{\prime} = \mathbf{W}_L^\dagger\, \mathbf{m}_{l}\, \mathbf{W}_R,\quad
    \mathbf{m}_{\nu}^{\prime} = \mathbf{W}_L^T\, \mathbf{m}_{\nu}\, \mathbf{W}_L.
\end{align}

One may also use the freedom to make WB transformations to go to a basis where $\mathbf{m}_{l} = \mathbf{d}_{l}$ is real and diagonal. In this basis, one can still make the rephasing
\begin{align}\label{rephasingK}
    l_{L,R}^{\prime\prime}=\mathbf{K}_L\, l_{L,R}^\prime, \quad   \nu_{L}^{\prime\prime}=\mathbf{K}_L\, \nu_{L}^\prime,
\end{align}
with $\mathbf{K}_L=\text{diag} (e^{i\varphi_1},e^{i\varphi_2},e^{i\varphi_3})$. Under this rephasing $\mathbf{d}_{l}$ remains invariant, but $\mathbf{m}_{\nu}$ transforms as
\begin{align}\label{mnu2prime}
    (\mathbf{m}_{\nu}^{\prime\prime})_{ij}=e^{i(\varphi_i+\varphi_j)}
    \, (\mathbf{m}_{\nu}^{\prime})_{ij}.
\end{align}
Since $\mathbf{m}_{\nu}^{\prime}$ is an arbitrary complex symmetric matrix it has $n(n+1)/2$ phases, where $n$ denotes the number of generations. One is still free to rephase Eq.~\eqref{mnu2prime} and further eliminate $n$ phases. One is then left with
\begin{align}\label{Nphi}
    N_\phi=\frac{1}{2} n(n-1)
\end{align}
physically meaningful phases\footnote{Alternatively, the parameter counting can be performed by analyzing the symmetries of the Lagrangian~\cite{Santamaria:1993ah}.}. It will be shown in the sequel that these phases in general violate CP. Note that the $N_\phi$ phases appear in a WB, prior to the diagonalization of both $\mathbf{m}_{l}$ and $\mathbf{m}_{\nu}$, and the generation of the leptonic mixing matrix. Note also that $N_\phi$ coincides with the number of physical phases appearing in the leptonic mixing.

For three generations, $N_\phi=3$, and one may use the rephasing of Eq.~\eqref{mnu2prime} in order to make, for example, all the diagonal elements of $\mathbf{m}_{\nu}$ real. For this choice, the three CP-violating phases can be identified with $\arg[(\mathbf{m}_{\nu})_{12}]$, $\arg[(\mathbf{m}_{\nu})_{13}]$ and $\arg[(\mathbf{m}_{\nu})_{23}]$. It is clear that the individual phases of $(\mathbf{m}_{\nu})_{ij}$ do not have any physical meaning, since they are not invariant under the rephasing given in Eq.~\eqref{mnu2prime}. One may, however, construct polynomials of $(\mathbf{m}_{\nu})_{ij}$ which are rephasing invariant~\cite{Farzan:2006vj}, such as
\begin{align} \label{rephasePi}
\begin{split}
    P_1 &=(\mathbf{m}_{\nu}^\ast)_{11}\, (\mathbf{m}_{\nu}^\ast)_{22}\, (\mathbf{m}_{\nu})_{12}^2,\\
    P_2 &=(\mathbf{m}_{\nu}^\ast)_{11}\, (\mathbf{m}_{\nu}^\ast)_{33}\, (\mathbf{m}_{\nu})_{13}^2,\\
    P_3 &=(\mathbf{m}_{\nu}^\ast)_{33}\, (\mathbf{m}_{\nu}^\ast)_{12}\, (\mathbf{m}_{\nu})_{13}\, (\mathbf{m}_{\nu})_{23}.
\end{split}
\end{align}

\subsubsection{Leptonic mixing}
The lepton mass matrices in Eq.~\eqref{Lmass} are diagonalized by the unitary transformations
\begin{align}\label{mlmnudiag}
       \mathbf{U}^{l\, \dagger}_L\, \mathbf{m}_{l}\, \mathbf{U}^l_R  = \mathbf{d}_{l}\,, \quad
        \mathbf{U}^{\nu\, T}\,\mathbf{m}_{\nu}\,\mathbf{U}^{\nu}   = \mathbf{d}_{m}\,,
\end{align}
where $\mathbf{U}^l_{L,R}$ and $\mathbf{U}^{\nu}$ are unitary matrices; $\mathbf{d}_{l}$ and $\mathbf{d}_{m}$ are diagonal matrices. In terms of the lepton mass eigenstates, the charged current becomes
\begin{align}\label{Lcharged1}
    \mathcal{L}_W=\frac{g}{\sqrt{2}} \overline{l_{L}} \gamma_\mu \mathbf{U}\, \nu_L W^\mu + \text{H.c.},
\end{align}
where $\mathbf{U}= \mathbf{U}^{l\, \dagger}_L\, \mathbf{U}^{\nu}$ is the Pontecorvo-Maki-Nakagawa-Sakata (PMNS) leptonic mixing matrix. The matrix $\mathbf{U}$ is unitary, so it has $n^2$ parameters; $n(n-1)/2$ of these parameters can be used to define the $O(n)$ rotation, while $n$ phases of $\mathbf{U}$ can be removed through the rephasing of $n$ charged-lepton fields. Thus, one is left with $n(n-1)/2$ phases characterizing CP violation in $\mathbf{U}$. As mentioned, this number of phases coincides with the number of physical phases $N_\phi$ in the neutrino mass matrix, counted in a WB in which the charged-lepton mass matrix is diagonal and real.

For three generations, the $3\times3$ matrix $\mathbf{U}$ is conveniently parametrized by~\cite{Nakamura:2010zzi}
\begin{align}\label{Uparam}
    \mathbf{U} = \mathbf{V}\,\mathbf{K}, \quad \mathbf{K} = \text{diag}(1, e^{i\alpha_1/2},e^{i\alpha_2/2}),
\end{align}
with $\alpha_{1,2}$ denoting the phases associated with the Majorana character of neutrinos~\cite{Bilenky:1980cx,Doi:1980yb,Schechter:1980gr,Bernabeu:1982vi}, and the unitary matrix $\mathbf{V}$ written, as in the case of the CKM quark mixing matrix, in terms of three mixing angles $(\theta_{12}, \theta_{23}, \theta_{13})$ and one phase $\delta$,
\begin{align}\label{VPDG}
&\mathbf{V}=\nonumber\\
&{\small
\begin{pmatrix}
c_{12} c_{13} & s_{12} c_{13} & s_{13}e^{-i\delta}\\
-s_{12} c_{23}-c_{12} s_{23} s_{13}e^{i\delta} & c_{12} c_{23}-s_{12} s_{23} s_{13}e^{i\delta} & s_{23} c_{13}\\
s_{12} s_{23}-c_{12} c_{23} s_{13}e^{i\delta} & -c_{12} s_{23}-s_{12} c_{23} s_{13}e^{i\delta} & c_{23} c_{13}
\end{pmatrix}}.
\end{align}
Hereafter, $s_{ij} = \sin \theta_{ij}$ and $c_{ij} = \cos \theta_{ij}$ with the mixing angles chosen to lie in the first quadrant, and $\delta$ is a Dirac-type CP-violating phase. An alternative parametrization of the mixing matrix $\bU$, which turns out to be more appropriate for the $\ndbd$ analysis, is given by
\begin{align}\label{Uparam2}
    \mathbf{U} = \mathbf{V}\,\mathbf{K}^\prime, \quad \mathbf{K}^\prime = \mathbf{K}\, \text{diag}\bigl(1, 1,e^{i\delta}\bigr).
\end{align}
In what follows, we also use the simplified notation
\begin{align}\label{Unotation}
    \mathbf{U}=\begin{pmatrix}
    \mathbf{U}_{e1} & \mathbf{U}_{e2} & \mathbf{U}_{e3}\\
    \mathbf{U}_{\mu1} & \mathbf{U}_{\mu2} & \mathbf{U}_{\mu3}\\
    \mathbf{U}_{\tau1} & \mathbf{U}_{\tau2} & \mathbf{U}_{\tau3}
    \end{pmatrix}
\end{align}
to denote the matrix elements of $\mathbf{U}$.

It is clear that the phase of a particular matrix element of $\mathbf{U}$ does not have any physical meaning. This reflects the fact that under a rephasing of the charged-lepton fields $l_{Lj} \rightarrow l_{Lj}^\prime = e^{i \phi_j} l_{Lj}$ the matrix $\mathbf{U}$ transforms as
\begin{align} \label{Urephasing}
\mathbf{U}_{jk} \rightarrow \mathbf{U}^\prime_{jk} =e^{i \phi_j} \mathbf{U}_{jk}.
\end{align}
This is entirely analogous to what one encounters in the quark sector. The novel feature in leptonic mixing with Majorana neutrinos is that one cannot rephase Majorana neutrino phases since this would not leave invariant the neutrino mass terms. Note that we consider real neutrino masses, which satisfy Majorana conditions that do not contain phase factors. It should also be emphasized that rephasing invariance is a requirement for any physical quantity. In the quark sector, the simplest rephasing invariant functions of the CKM matrix elements $\mathbf{V}_{ij}$, apart from the trivial example of moduli, are the rephasing invariant quartets $\mathbf{V}_{ij} \mathbf{V}_{kj}^\ast \mathbf{V}_{kl}
\mathbf{V}_{il}^\ast$. In the lepton sector with Majorana neutrinos, the simplest rephasing invariant functions of the PMNS matrix elements $\mathbf{U}_{ij}$ are the bilinears of the type $\mathbf{U}_{ij} \mathbf{U}_{ik}^\ast$~\cite{Nieves:1987pp,AguilarSaavedra:2000vr,Nieves:2001fc}, with $j \neq k$ and no summation over repeated indices. We then designate ``Majorana-type" phases the following quantities:
\begin{align}\label{Majphases}
    \gamma_{jk} \equiv \arg(\mathbf{U}_{ij} \mathbf{U}_{ik}^\ast).
\end{align}
From their definition, one can readily see that in the case of three generations there are six independent Majorana-type phases $\gamma_{jk}$. Using unitarity, one can then reconstruct the full matrix  $\mathbf{U}$ from these six Majorana-type phases~\cite{Branco:2008ai}.

\subsection{Dirac and Majorana unitarity triangles}
\label{sec2.2}

In a SM-like theory with an arbitrary number of generations, quark mixing is defined through the CKM matrix which is unitary by construction. For three standard generations, unitarity leads to various relations among the moduli of the CKM matrix and rephasing invariant angles. These relations provide a crucial test of the SM and its mechanism of mixing and CP violation. We assume, for the moment, that the PMNS matrix $\mathbf{U}$ is unitary. Then one can construct six unitarity triangles from the orthogonality of the rows and columns of $\mathbf{U}$~\cite{AguilarSaavedra:2000vr}. These triangles are analogous to the ones used in the quark sector to study various manifestations of CP violation. However, in the case of Majorana neutrinos there is an important difference. In the quark sector, the orientation of the unitarity triangles in the complex plane has no physical meaning, since under rephasing of the quark fields all triangles rotate. For example, one may choose in the quark sector, without loss of generality, any side of a given triangle to coincide with the real axis.

\begin{figure}[t]
\includegraphics[width=6cm]{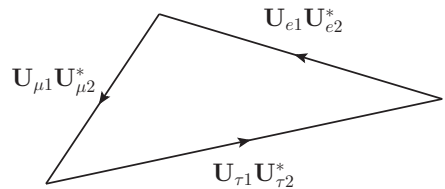}
\caption{\label{fig21} Majorana unitarity triangle $T_{12}$. The arrows indicate the orientation of the triangle, which is determined by the Majorana phases and cannot be rotated in the complex plane.}
\end{figure}

In the lepton sector with Majorana neutrinos there are two types of unitarity triangles: Dirac triangles that correspond to the orthogonality of rows,
\begin{align}\label{Diractriangles}
    \begin{split}
       T_{e\mu}& : \mathbf{U}_{e1} \mathbf{U}_{\mu 1}^\ast +  \mathbf{U}_{e2} \mathbf{U}_{\mu 2}^\ast + \mathbf{U}_{e3} \mathbf{U}_{\mu 3}^\ast = 0, \\
       T_{e\tau}& : \mathbf{U}_{e1} \mathbf{U}_{\tau 1}^\ast +  \mathbf{U}_{e2} \mathbf{U}_{\tau 2}^\ast + \mathbf{U}_{e3} \mathbf{U}_{\tau 3}^\ast = 0, \\
       T_{\mu\tau}& : \mathbf{U}_{\mu 1} \mathbf{U}_{\tau 1}^\ast +  \mathbf{U}_{\mu 2} \mathbf{U}_{\tau 2}^\ast + \mathbf{U}_{\mu 3} \mathbf{U}_{\tau 3}^\ast = 0,
     \end{split}
\end{align}
and Majorana triangles that are defined by the orthogonality of columns,
\begin{align}\label{Majotriangles}
    \begin{split}
       T_{12}& : \mathbf{U}_{e1} \mathbf{U}_{e2}^\ast +  \mathbf{U}_{\mu 1} \mathbf{U}_{\mu 2}^\ast + \mathbf{U}_{\tau 1} \mathbf{U}_{\tau 2}^\ast = 0, \\
       T_{13}& : \mathbf{U}_{e1} \mathbf{U}_{e3}^\ast +  \mathbf{U}_{\mu 1} \mathbf{U}_{\mu 3}^\ast + \mathbf{U}_{\tau 1} \mathbf{U}_{\tau 3}^\ast = 0, \\
       T_{23}& : \mathbf{U}_{e2} \mathbf{U}_{e3}^\ast +  \mathbf{U}_{\mu 2} \mathbf{U}_{\mu 3}^\ast + \mathbf{U}_{\tau 2} \mathbf{U}_{\tau 3}^\ast = 0.
     \end{split}
\end{align}
It is clear from Eq.~\eqref{Diractriangles} that the orientation of Dirac triangles has no physical meaning since under the rephasing of the charged-lepton fields these triangles rotate in the complex plane, $\mathbf{U}_{ik} \mathbf{U}_{jk}^\ast \rightarrow e^{i (\phi_i - \phi_j)} \mathbf{U}_{ik} \mathbf{U}_{jk}^\ast$, in accordance with Eq.~\eqref{Urephasing}. On the contrary, the orientation of Majorana triangles does have physical meaning since these triangles remain invariant under rephasing (cf. Fig.~\ref{fig21}).

Leptonic CP violation with Majorana neutrinos has some novel features, when compared to CP violation in the quark sector. In the latter case, there is CP violation if and only if the imaginary part of a rephasing invariant quartet of the CKM matrix elements does not vanish. It is an important consequence of the unitarity of the CKM matrix that the imaginary part of all invariant quartets have the same modulus. The only meaningful phases in the quark sector are the arguments of rephasing invariant quartets. In the lepton sector, one may have an entirely analogous CP violation from the nonvanishing of the imaginary part of an invariant quartet of $\mathbf{U}$. In the limit when $\mathbf{U}$ is unitary, again the imaginary part of all invariant quartets have the same modulus. Nevertheless, one may also  have Majorana-type CP violation associated with the Majorana-type phases, identified as arguments of the rephasing invariant bilinears defined in Eq.~\eqref{Majphases}.

In order to understand some of the special features of leptonic CP violation with Majorana neutrinos, it is instructive to study the limit of CP invariance. This case can be analyzed using the Majorana unitarity triangles of Eq.~\eqref{Majotriangles}, which provide the necessary and sufficient conditions for CP conservation:
\begin{itemize}
  \item[(i)] Vanishing of their common area $\mathcal{A} = 1/2\, |\text{Im}\, Q|$,
      with $Q = \mathbf{U}_{ij} \mathbf{U}_{kj}^\ast \mathbf{U}_{kl}
      \mathbf{U}_{il}^\ast $ standing for any invariant quartet of $\mathbf{U}$ (no
      sum over repeated indices and $i \neq k, j \neq l$);
  \item[(ii)] Orientation of all Majorana triangles along the direction of the real
      or imaginary axes.
\end{itemize}

The first requirement eliminates the possibility of Dirac-type CP violation while the second condition implies that Majorana phases do not violate CP. In order to understand requirement~(ii), we assume that condition~(i) is satisfied, i.e., all triangles collapse. If all Majorana triangles $T_{jk}$ collapse along the real axis then $\gamma_{jk}=0$ in Eq.~\eqref{Majphases}. It is obvious that CP is conserved in this case and the leptonic mixing matrix $\mathbf{U}$ is real. If one of the triangles $T_{jk}$ collapse along the imaginary axis, this means that the mass eigenstates $\nu_j$ and $\nu_k$ have opposite CP parities, but no CP violation is implied. One can make the triangle $T_{jk}$, which collapsed in the imaginary axis to collapse in the real axis instead, by multiplying the Majorana fields by $\pm i$ and rendering the corresponding mass eigenstate negative.

\subsection{Majorana neutrinos and CP violation}
\label{sec2.3}

In order to study CP violation in an extension of the SM with Majorana masses for left-handed neutrinos, it is convenient to consider the Lagrangian after the spontaneous gauge symmetry breaking. The relevant part of the Lagrangian reads
\begin{align}\label{LCP}
\mathcal{L}=-\overline{l_{L}}\, \mathbf{m}_{l}\, l_{R} - \frac{1}{2} \nu^T_{L} C\, \mathbf{m}_{\nu}\, \nu_{L} + \frac{g}{\sqrt{2}} \overline{l_{L}} \gamma_\mu \nu_L W^\mu + \text{H.c.}\,.
\end{align}

The CP transformation properties of the various fields are dictated by the part of the Lagrangian which conserves CP, namely, the gauge interactions. One should keep in mind that gauge interactions in a WB do not distinguish the different generations of fermions and, consequently, the Lagrangian of Eq.~\eqref{LCP} conserves CP if and only if there is a CP transformation defined by
\begin{align}\label{CPtransf}
    \begin{split}
    CP\, l_L\, (CP)^\dagger & = \mathbf{W}_L \gamma^0 C\, \overline{l_L}^{\,T},\\
    CP\, \nu_L\, (CP)^\dagger & = \mathbf{W}_L \gamma^0 C\, \overline{\nu_L}^T,\\
    CP\, l_R\, (CP)^\dagger & = \mathbf{W}_R \gamma^0 C\, \overline{l_R}^{\,T},
    \end{split}
\end{align}
where $\mathbf{W}_L$ and $\mathbf{W}_R$ are unitary matrices acting in generation space.

Often, in the literature, the transformations given in Eqs.~\eqref{CPtransf} are referred to as generalized CP transformation. This is a misnomer, since the inclusion of the unitary matrices $\mathbf{W}_L$ and $\mathbf{W}_R$ is mandatory for a correct definition the CP transformation, in view of the flavor symmetry of gauge interactions. The lepton fields $l_L$ and $\nu_L$ have to transform in the same way due to the presence of the left-handed charged-current interactions. Then, the Lagrangian of Eq.~\eqref{LCP} conserves CP if and only if the lepton mass matrices $\mathbf{m}_{\nu}$ and $\mathbf{m}_{l}$ satisfy the following relations:
\begin{align}\label{CPmass}
     \mathbf{W}_L^T \mathbf{m}_{\nu} \mathbf{W}_L = - \mathbf{m}_{\nu}^\ast,\quad
    \mathbf{W}_L^\dagger \mathbf{m}_{l} \mathbf{W}_R = \mathbf{m}_{l}^\ast.
\end{align}

The above CP conditions are WB independent in the sense that if there exist matrices $\mathbf{W}_L$ and $\mathbf{W}_R$ that satisfy Eq.~\eqref{CPmass} when $\mathbf{m}_{\nu}$ and $\mathbf{m}_{l}$ are written in a particular WB, they will also exist when the mass matrices are written in another WB. One can use this WB independence to study the CP restrictions in an appropriate WB.  We perform this analysis in two different basis. We first consider the basis of real and diagonal charged-lepton mass matrix. At this stage, $\mathbf{m}_{\nu}$ is an arbitrary complex symmetric matrix. While keeping $\mathbf{m}_{l}$ diagonal, real, and positive, one can still make a WB transformation which renders the diagonal elements of $\mathbf{m}_{\nu}$ real. In this basis, Eq.~\eqref{CPmass} constrains $\mathbf{W}_L$ to be of the form
\begin{align}\label{WLconstrained}
    \mathbf{W}_L = \text{diag}\,(\pm i, \pm i, \pm i).
\end{align}
Substituting Eq.~\eqref{WLconstrained} into Eq.~\eqref{CPmass}, one concludes that CP invariance constrains the elements of $\mathbf{m}_{\nu}$ to be either real or purely imaginary. Note, for instance, that the matrix \begin{align}\label{mnuCP}
\mathbf{m}_{\nu} = \begin{pmatrix}
  \,|m_{11}| & \,|m_{12}| & i |m_{13}|\\
   \,|m_{12}| & \,|m_{22}| & i |m_{23}|\\
   i |m_{13}| & i |m_{23}| & \,|m_{33}|
\end{pmatrix}
\end{align}
does not lead to CP violation, since Eqs.~\eqref{CPmass} can be satisfied with $\mathbf{W}_L = \text{diag}\,(i, i, -i)$. One could have also suspected that the matrix $\mathbf{m}_{\nu}$ defined in Eq.~\eqref{mnuCP} would correspond to CP invariance since $\text{Im}\,P_i=0$, where $P_i$ denotes the rephasing invariants given in Eqs.~\eqref{rephasePi}.

\subsection{Weak-basis invariants and low-energy CP violation}
\label{sec2.4}

We have seen that the existence of unitary matrices $\mathbf{W}_L$ and $\mathbf{W}_R$ satisfying Eqs.~\eqref{CPmass} is a necessary and sufficient condition for having CP invariance in the low-energy limit. We address now the question of finding CP-odd WB invariants which would detect CP violation in the lepton sector. Obviously, these WB invariants should be written in terms of $\mathbf{m}_\nu$ and $\mathbf{m}_l$. It is well known that, in the quark sector of the SM with three generations, there is only one CP-odd WB invariant which controls CP violation at low energies, namely,~\cite{Bernabeu:1986fc,Gronau:1986xb},
\begin{align}\label{ICPquark}
    \mathcal{J}^\text{CP}_\text{quark}=\text{Tr} \bigl[\mathbf{m}_u \mathbf{m}_u^\dagger,\mathbf{m}_d \mathbf{m}_d^\dagger\bigr]^3,
\end{align}
where $\mathbf{m}_u$ and $\mathbf{m}_d$ denote the up and down quark mass matrices, respectively\footnote{ This invariant can also be written in the equivalent form $\mathcal{J}^\text{CP}_\text{quark}=\text{Im Det} \bigl(\bigl[\mathbf{m}_u \mathbf{m}_u^\dagger,\mathbf{m}_d \mathbf{m}_d^\dagger\bigr]\bigr)$~\cite{Jarlskog:1985ht}.}.

In the case of three (Dirac or Majorana) neutrinos, one can write an entirely analogous CP-odd WB invariant which controls Dirac-type CP violation in the lepton sector:
\begin{align}\label{ICPlepton}
    \mathcal{J}^\text{CP}_\text{lepton}=\text{Tr} \left[(\mathbf{m}_\nu \mathbf{m}_\nu^\dagger)^\ast,\mathbf{m}_l \mathbf{m}_l^\dagger\right]^3.
\end{align}
This relation can be computed in any weak basis. The low-energy invariant \eqref{ICPlepton} is sensitive to the Dirac-type phase $\delta$ and vanishes for $\delta=0$. On the other hand, it does not depend on the Majorana phases $\alpha_1$ and $\alpha_2$ appearing in the leptonic mixing matrix $\mathbf{U}$. The quantity $\mathcal{J}^\text{CP}_\text{lepton}$ can be fully written in terms of physical observables,
\begin{align}\label{ICPleptonobs}
    \mathcal{J}^\text{CP}_\text{lepton}=&-6\,i \,({m_{\mu}}^2-{m_e}^2)\,({m_{\tau}}^2-{m_\mu}^2)\,({m_{\tau}}^2-{m_e}^2)\nonumber\\
    &\times \Delta m^2_{21}\,\Delta m^2_{31}\,\Delta m^2_{32}\,{\cal J}_{e\mu}^{21}\,,
\end{align}
where $\Delta m_{ji}^2=m_j^2-m_i^2$ are the light neutrino mass-squared differences.  As shown in Sec.~\ref{sec3.2}, the quantity ${\cal J}_{e\mu}^{21}$ is the imaginary part of an invariant quartet appearing in the difference of the CP-conjugated neutrino oscillation probabilities $P(\nu_e\rightarrow\nu_\mu)-P(\bar{\nu}_e\rightarrow \bar{\nu}_\mu)$. One can easily get
\begin{align}
{\cal J}_{e\mu}^{21} &\equiv {\rm Im}\left[\,\mathbf{U}_{11} \mathbf{U}_{22}
\mathbf{U}_{12}^\ast \mathbf{U}_{21}^\ast\,\right] \nonumber \\
&= \frac{1}{8} \sin(2\,\theta_{12})
\sin(2\,\theta_{13})\sin(2\,\theta_{23}) \sin \delta\,, \label{Jemu21}
\end{align}
where $\theta_{ij}$ are the mixing angles in the standard parametrization of Eq.~\eqref{VPDG}.

The requirement $\mathcal{J}^\text{CP}_\text{lepton} \neq 0$ is a necessary and sufficient condition for having Dirac-type leptonic CP violation, independent of whether neutrinos are Majorana or Dirac particles. However, in the case of Majorana neutrinos there is also the possibility of Majorana-type CP violation. It is therefore interesting to find CP-odd invariants which could directly detect this type of CP violation, even in the limit when there is no Dirac-type CP violation. An example of such an invariant is~\cite{Branco:1986gr}
\begin{align}\label{ICPMajorana}
    \mathcal{J}^\text{CP}_\text{Maj}=\text{Im Tr} \bigl(\mathbf{m}_l \mathbf{m}_l^\dagger \mathbf{m}_\nu^\ast \mathbf{m}_\nu \mathbf{m}_\nu^\ast \mathbf{m}_l^T \mathbf{m}_l^\ast \mathbf{m}_\nu \bigr).
\end{align}

The simplest way of verifying that $\mathcal{J}^\text{CP}_\text{Maj}$ is sensitive to Majorana phases is by evaluating it for the particular case of two Majorana neutrinos. In this situation, there is only one Majorana-type phase and no Dirac-type phase. The leptonic mixing matrix can be parametrized by
\begin{align}
    \mathbf{U}= \begin{pmatrix}
    \cos \theta & -\sin \theta\, e^{i \gamma}\\
    \sin \theta\, e^{-i \gamma} & \cos \theta
    \end{pmatrix},
\end{align}
where $\gamma$ denotes the Majorana phase. An explicit evaluation of $\mathcal{J}^\text{CP}_\text{Maj}$ gives
\begin{align}\label{ICPMajoranaex}
    \mathcal{J}^\text{CP}_\text{Maj}=\frac{1}{4}\,m_1 m_2 \Delta m^2_{21} (m_\mu^2-m_e^2)^2 \sin^2 2\theta \sin 2\gamma.
\end{align}
It is worth pointing out that $\mathcal{J}^\text{CP}_\text{Maj}$ shows explicitly some subtle points of Majorana-type CP violation. In particular, it shows that a phase $\gamma=\pi/2$ does not imply CP violation; it simply corresponds to CP invariance with the two neutrinos having opposite CP parities.

The invariants given in Eqs.~\eqref{ICPlepton} and \eqref{ICPMajorana} vanish if neutrinos are exactly degenerate in mass. In this limit, the parametrization of the mixing matrix $\bU$ requires, in general, two angles and one CP-violating phase. This is to be contrasted to the case of Dirac neutrinos, in which there is no mixing or CP violation in the exact degeneracy limit. Therefore, leptonic CP violation may arise even when the three Majorana neutrinos have identical mass~\cite{Branco:1986gr}. It is possible to construct a WB invariant which controls the strength of the CP violation in the latter case~\cite{Branco:1998bw}, namely,
\begin{align}
\label{ICPMajoranadeg}
\mathcal{J}^\text{CP}_\text{deg} = \text{Tr} \left[\mathbf{m}_\nu \mathbf{m}_l \mathbf{m}_l^\dagger \mathbf{m}_\nu^\ast,\mathbf{m}_l^\ast \mathbf{m}_l^T \right]^3.
\end{align}
A necessary and sufficient condition for CP invariance is $\mathcal{J}^\text{CP}_\text{deg} = 0$. The CP-odd invariant~\eqref{ICPMajoranadeg} can be expressed in terms of lepton masses and mixing parameters by choosing the WB in which $\mathbf{m}_l \mathbf{m}_l^\dagger=\text{diag}\, (m_e^2, m_\mu ^2, m_\tau ^2)$. Parametrizing the mixing matrix $\bU$ in the standard form of Eqs.~\eqref{Uparam} and ~\eqref{VPDG}, with vanishing $\theta_{13}$ and $\delta$, and $\alpha_1=2\pi$, so that in the limit of CP invariance one of the Majorana neutrinos has relative CP parity opposite to the other two, one obtains
\begin{align}
\label{ICPMajoranadeg1}
\mathcal{J}^\text{CP}_\text{deg} =&-\frac{3i}{2} \, m^6 (m_\tau ^2-m_\mu ^2)^2(m_\tau ^2-m_e^2)^2(m_\mu ^2-m_e^2)^2 \nonumber\\
&\times \cos (2 \theta_{12}) \sin^2(2\theta_{12}) \sin^2(2\theta_{23}) \sin \alpha_2,
\end{align}
where $m$ denotes the common neutrino mass. The special feature of the WB invariant of Eq.~\eqref{ICPMajoranadeg1} is the fact that, in general, it does not vanish, even in the limit of exact degeneracy of the three Majorana neutrino masses.

\subsection{Seesaw mechanisms for neutrino mass generation}
\label{sec2.5}

In the SM, quarks and charged fermions get masses through renormalizable Yukawa couplings with the Higgs doublet $\phi=(\phi^+,\phi^0)^T$, and their corresponding mass terms break the $SU(2)_L$ gauge symmetry as doublets. In contrast, a Majorana neutrino mass term, as the one given in Eq.~\eqref{LCP}, breaks $SU(2)_L$ as a triplet, and therefore it cannot be generated in the same way. This term is most likely to arise from higher dimensional operators, such as the lepton number violating ($\Delta {\rm L}=2$) dimension-five operator $\mathcal{O}=(\ell_\alpha \phi) (\ell_\beta \phi)/M$~\cite{Weinberg:1980bf}, where $\ell=(l_L,\nu_L)$ is the SM lepton doublet. Once the Higgs field acquires a nonzero vacuum expectation value, $\langle \phi^0 \rangle = v$, Majorana neutrino masses proportional to $v^2/M$ are induced, in contrast to the quark and charged-lepton masses which are linear in $v$. Thus, if the mass scale $M$ is much heavier than the electroweak breaking scale $v$, neutrinos could naturally get masses much smaller than all the other SM fermions.

\begin{figure}[t]
\includegraphics[width=9cm]{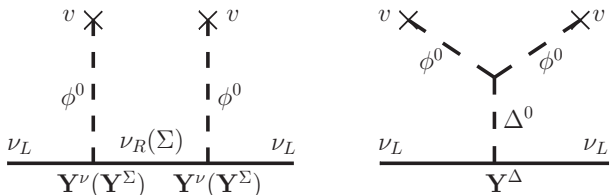}
\caption{\label{fig22} Canonical seesaw mechanisms for neutrino mass generation. The left diagram corresponds to type I and type III seesaw masses, mediated by the tree-level exchange of singlet ($\nu_R$) and triplet ($\Sigma$) fermions, respectively. The right diagram leads to type II seesaw neutrino masses via the exchange of a triplet scalar $\Delta$.}
\end{figure}

The simplest and perhaps most attractive realization of the operator $\mathcal{O}$ in gauge theories is through the so-called seesaw mechanism. In this approach, the effective operator $\mathcal{O}$ is induced by the exchange of heavy particles with a mass scale $M$. Such heavy states are commonly present in grand unified theories (GUT). Several seesaw realizations are conceivable for neutrino mass generation~\cite{Nunokawa:2007qh,Mohapatra:2005wg}. The following three types, schematically depicted in Fig.~\ref{fig22}, are among the most popular ones:
\begin{itemize}
\item \emph{Type
    I}~\cite{Minkowski:1977sc,Yanagida:1979as,GellMann:1980vs,Glashow:1980,Mohapatra:1979ia},
    mediated by heavy fermions, singlets under the $SU(3)\times SU(2)\times U(1)$
    gauge symmetry;
\item \emph{Type
    II}~\cite{Konetschny:1977bn,Mohapatra:1980yp,Cheng:1980qt,Lazarides:1980nt,Schechter:1980gr},
    mediated by the exchange of $SU(2)$-triplet scalars;
\item \emph{Type III}~\cite{Foot:1988aq}, mediated by the exchange of
    $SU(2)$-triplet fermions.
\end{itemize}
Below we briefly describe each of these realizations.

Adding two or three singlet fermions $\nu_R$ to the SM particle content is one of the simplest and rather natural possibilities to generate neutrino masses. Since the $\nu_R$ states are electroweak singlets, their masses are not protected by the electroweak symmetry and therefore can be very large. In the basis of diagonal charged-lepton Yukawa couplings, the relevant terms in the neutrino sector of the Lagrangian are
\begin{align}\label{LtypeIseesaw}
    -\mathcal{L}^\text{I}=\mathbf{Y}_{\alpha i}^{\nu \ast}\, \overline{\ell}_{\alpha} \tilde{\phi}\, \nu_{Ri} + \frac{1}{2}\, \overline{\nu_{Ri}}\, (\mathbf{m}_R)_{ij}\, \nu_{Rj}^c + \text{H.c.},
\end{align}
where $\tilde{\phi} = i \sigma_2 \phi^\ast$, $\mathbf{Y}^\nu$ is the Dirac-neutrino Yukawa coupling matrix, and $\mathbf{m}_R$ is the right-handed neutrino mass matrix. Notice that we have not included a Majorana mass term for left-handed neutrinos since this would require an enlargement of the scalar sector. For 3 generations and $n_R$ heavy Majorana states, the type I seesaw Lagrangian of Eq.~\eqref{LtypeIseesaw} contains altogether $7 n_R-3$ free parameters. The counting can be done as follows. In the mass basis of the singlet fermions, $N_i= \mathbf{U}_R^T\, \nu_R$, or, more precisely, in the basis where the $n_R \times n_R$ symmetric matrix $\mathbf{m}_R$ is diagonal, with positive and real eigenvalues $M_i$, i.e.,
\begin{align}\label{mRdiag}
    \mathbf{U}_R^T\, \mathbf{m}_R\, \mathbf{U}_R = \mathbf{d}_M = \text{diag}(M_1,M_2,\cdots, M_{n_R}),
\end{align}
the Majorana mass term in Eq.~\eqref{LtypeIseesaw} contains only $n_R$ free parameters. In this basis, the Yukawa coupling matrix $\mathbf{Y}^\nu$ is an arbitrary $3 \times n_R$ complex matrix with $6 n_R$ parameters. Of those, 3 phases can be removed by phase redefinitions of the charged-lepton fields $l_L$, thus remaining $3 (2n_R-1)$ physical parameters, to wit $3n_R$ moduli and $3(n_R-1)$ phases.

After integrating out the heavy Majorana fields in the Lagrangian of Eq.~\eqref{LtypeIseesaw}, the effective mass matrix of the light neutrinos is given by the standard seesaw formula
\begin{align}\label{mnutypeIseesaw}
    \mathbf{m}_\nu = - v^2\, \mathbf{Y}^\nu\, \mathbf{m}_R^{-1}\, \mathbf{Y}^{\nu T},
\end{align}
with the matrix $\mathbf{m}_\nu$ being diagonalized by the PMNS leptonic mixing matrix $\mathbf{U}$,
\begin{align}\label{mnutypeIdiag}
    \mathbf{U}^T\, \mathbf{m}_\nu\, \mathbf{U} = \mathbf{d}_m = \text{diag}(m_1,m_2,m_3),
\end{align}
where $m_i$ are the light neutrino masses.

The general type I seesaw framework introduces many more parameters than those required at low energies. Indeed, the effective neutrino mass matrix $\mathbf{m}_\nu$ can be written in terms of only nine physical parameters: the three light neutrino masses, and the three mixing angles and three phases that parametrize the mixing matrix $\mathbf{U}$.

We now consider the type II seesaw framework. In this case, the SM scalar sector is extended by introducing a scalar triplet $\Delta$ with hypercharge $+1$ (in the normalization of hypercharge $-1/2$ for the lepton doublets) and mass $M_{\Delta}$. In the $SU(2)$ representation,
\begin{align}
\Delta=\begin{pmatrix}
         \Delta^0 & -\Delta^+/\sqrt{2} \\
         -\Delta^+/\sqrt{2} & \Delta^{++} \\
       \end{pmatrix}.
\end{align}
The relevant Lagrangian terms are, in this case,
\begin{align}\label{LtypeIIseesaw}
-\mathcal{L}^\text{II} = & \bigl(\mathbf{Y}_{\alpha\beta}^{\Delta}\, \ell_\alpha^T C \Delta\, \ell_\beta - \mu M_{\Delta} \tilde{\phi}^T\Delta\, \tilde{\phi} +\text{H.c.}\bigr)\nonumber\\
&+ M^2_{\Delta}\,\text{Tr}(\Delta^\dagger \Delta),
\end{align}
where $\mathbf{Y}^{\Delta}$ is a $3\times3$ symmetric complex coupling matrix, and $\mu$ is a dimensionless coupling, which can be taken real without loss of generality. When compared to the type I seesaw Lagrangian of Eq.~\eqref{LtypeIseesaw}, the Lagrangian terms in Eq.~\eqref{LtypeIIseesaw} contain less free parameters. Indeed, only eleven parameters are required to fully determine the type II seesaw Lagrangian. Besides the two ``unflavored" parameters $\mu$ and $M_\Delta$, there are nine ``flavored" parameters contained in the Yukawa matrix $\mathbf{Y}^\Delta$. In this sense, the type II seesaw is more economical, since the flavor structure of the neutrino mass matrix $\mathbf{m}_\nu$ is uniquely determined by the flavor structure of $\mathbf{Y}^\Delta$. The exchange of the heavy triplet leads to the effective neutrino mass matrix
\begin{align}\label{mnutypeIIseesaw}
    \mathbf{m}_\nu = \frac{\mu v^2}{M_\Delta}\, \mathbf{Y}^\Delta.
\end{align}
Leptonic CP violation is thus encoded in the phases of the matrix $\mathbf{Y}^\Delta$.

Neutrino masses can also be generated by the tree-level exchange of two or three $SU(2)$-triplet fermions $\Sigma_i$ with zero hypercharge,
\begin{align}
\Sigma_i =\begin{pmatrix}
         \Sigma_i^0 & \sqrt{2}\, \Sigma^+_i \\
         \sqrt{2}\, \Sigma^-_i & -\Sigma_i^0 \\
       \end{pmatrix}.
\end{align}

The Lagrangian that leads to the effective matrix $\mathbf{m}_\nu$ is similar to the type I seesaw Lagrangian of Eq.~\eqref{LtypeIseesaw}, but with different contractions of the $SU(2)$ indices
\begin{align}\label{LtypeIIIseesaw}
    -\mathcal{L}^\text{III}=(\mathbf{Y}^{\Sigma})^\ast_{\alpha i}\, \bar{\ell}_{\alpha} \tilde{\phi}\, \Sigma_i + \frac{1}{2}\,(\mathbf{m}_{\Sigma})_{ij} \text{Tr}(\overline{\Sigma}_i \Sigma_j^c) + \text{H.c.}.
\end{align}
The parameter counting is analogous to the type I case. In particular, eighteen (eleven) parameters are required to fully determine the high-energy neutrino sector in a model with three (two) triplet fermions. The effective light neutrino mass matrix exhibits the same seesaw structure of Eq.~\eqref{mnutypeIseesaw}, with the obvious substitutions $\mathbf{Y}^\nu \rightarrow \mathbf{Y}^{\Sigma}$ and $\mathbf{m}_R \rightarrow \mathbf{m}_\Sigma$.

It is worth noting that, besides the three seesaw realizations discussed above, there are other types of unconventional seesaw schemes~\cite{Nunokawa:2007qh}. For instance, in the so-called double-seesaw models~\cite{Mohapatra:1986aw,Mohapatra:1986bd}, in addition to the conventional singlet fermions $\nu_{R}$, one or more singlet fields $S_i$ with lepton number ${\rm L}=1$ are added to the SM particle content. The relevant double-seesaw Lagrangian terms are
\begin{align}\label{Ltypeinvseesaw}
    -\mathcal{L}^\text{IS}=&\mathbf{Y}_{\alpha i}^{\nu \ast}\, \overline{\ell}_{\alpha} \tilde{\phi}\, \nu_{Ri} + \overline{S_{i}} (\mathbf{m}_{RS})_{ij} \nu_{Rj}\nonumber\\
     & + \frac{1}{2}\, \overline{S_{i}^c}\, (\mathbf{m}_S)_{ij}\, S_{j} + \text{H.c.},
\end{align}
where $\mathbf{m}_{RS}$ is an arbitrary complex matrix and $\mathbf{m}_S$ is a complex symmetric matrix. In this case, the effective mass matrix of the light neutrinos is given by
\begin{align}\label{mnutypeinvseesaw}
    \mathbf{m}_\nu = - v^2\, \mathbf{Y}^\nu\, (\mathbf{m}_{RS}^T)^{-1}\mathbf{m}_S\, \mathbf{m}_{RS}^{-1}\, \mathbf{Y}^{\nu T}.
\end{align}
The inverse seesaw is a variant of the double seesaw with a Majorana mass matrix $\mathbf{m}_S \ll v \mathbf{Y}^\nu \ll \mathbf{m}_{RS}$. Since in the limit $\mathbf{m}_S \rightarrow 0$ lepton number is conserved, this is a natural scenario in the 't Hooft sense~\cite{tHooft:1980xb}.

Finally, there is a variety of models of neutrino masses with the operator $\mathcal{O}$ being induced from physics at TeV or even lower energy scales~\cite{Chen:2011de}. In such scenarios, loop and Yukawa coupling suppression factors typically guarantee the smallness of neutrino masses. Furthermore, $\Delta {\rm L}=2$ effective operators with dimension higher than five can give a dominant contribution to neutrino Majorana masses, if the leading effective operator $\mathcal{O}$ is forbidden due to a new symmetry or selection rule~\cite{Babu:2001ex}.

\subsection{On the origin of CP violation}
\label{sec2.6}

CP violation plays a central role in particle physics and has profound implications for cosmology. Yet the origin of CP violation is an entirely open question~\cite{Branco:1999fs}. It is well known that, if one allows for complex Yukawa couplings, CP violation arises in the SM with three or more fermion generations.

An alternative possibility is having CP as a good symmetry of the Lagrangian, only broken spontaneously by the vacuum. This is an attractive scenario which may be the only choice at a fundamental level, if one keeps in mind that pure gauge theories necessarily conserve CP~\cite{Grimus:1995zi}. The first model with spontaneous CP violation was suggested by T.D.~Lee~\cite{Lee:1973iz} at a time when only two incomplete generations were known. Obviously, in the original Lee model with two generations, CP violation arises exclusively through the Higgs exchange. The Lee model has two Higgs doublets and no extra symmetry is introduced. As a result, fermions of a given charge receive contributions to their mass from the two Higgs fields. It can be readily verified that a nontrivial CKM mixing matrix is then generated by the relative phase between the two neutral Higgs VEV. However, since natural flavor conservation (NFC) is not implemented in the Higgs sector, there are dangerous Higgs-mediated flavor-changing neutral currents (FCNC) at tree level. One can implement NFC in the Higgs sector~\cite{Glashow:1976nt,Paschos:1976ay}, but then three Higgs doublets are required in order to achieve spontaneous CP violation~\cite{Branco:1980sz}. The CKM matrix is, however, real in this model, which is in disagreement with the experimental evidence for a complex mixing matrix, even if one allows for the presence of new physics~\cite{Botella:2005fc}.

One can envisage a simple model of spontaneous CP violation, which avoids the above difficulties while providing a possible common source for the various manifestations of CP violation~\cite{Branco:2003rt} in the quark and lepton sectors, as well as a solution to the strong CP problem. We outline below the main features of such a model~\cite{Branco:2003rt} in which all CP-breaking effects share the same origin, namely, the VEV of a complex singlet scalar field. This minimal model consists of an extension of the SM with the following additional fields: three right-handed neutrinos $\nu_R$, a neutral scalar singlet $S$, and a singlet vectorial quark $D$ with charge $-1/3$. Furthermore, one imposes on the Lagrangian a $Z_4$ symmetry, under which the fields transform in the following manner:
\begin{align}\label{Z4sym}
   \ell \rightarrow i\, \ell,\,\, l_R \rightarrow i\, l_R,\,\, \nu_R \rightarrow i\, \nu_R,\,\,D \rightarrow -D,\,\,  S \rightarrow -S.
\end{align}
Under the above $Z_4$ symmetry, all other fields remain invariant. Furthermore, we impose CP invariance at the Lagrangian level. In the quark sector, the most general $SU(3)\times SU(2)\times U(1) \times Z_4$ invariant Yukawa couplings can be written as
\begin{align}\label{Lquark}
    \mathcal{L}_\text{quark}= & \overline{Q}_i \mathbf{Y}^u_{ij} \phi\, u_{Rj} + \overline{Q}_i \mathbf{Y}^d_{ij} \tilde{\phi}\, d_{Rj} + \tilde{M}\, \overline{D}_L D_R \nonumber \\
    & + \overline{D}_L (\mathbf{f}_{qi} S + \mathbf{f}_{qi}^\prime S^\ast)\, d_{Ri} +\text{H.c.},
\end{align}
while for the lepton sector they are
\begin{align}\label{Llepton}
    \mathcal{L}_\text{lepton}&=\overline{\ell_{i}}\,  \mathbf{Y}^\ell_{ij} \phi \, l_{Rj} + \overline{\ell_i}\, \mathbf{Y}^\nu_{ij} \tilde{\phi}\, \nu_{Rj} \nonumber\\
    &+ \frac{1}{2} \nu_{Ri}^T C\, [\,(\mathbf{f}_\nu)_{ij} S + (\mathbf{f}_\nu^\prime)_{ij} S^\ast\,]\, \nu_{Rj} +\text{H.c.}.
\end{align}
Here $Q$, $u_R$, and $d_R$ are the SM quark fields; $\mathbf{Y}^{u,d}$, $\mathbf{Y}^{\ell}$, $\mathbf{f}_{q,\nu}$ and $\mathbf{f}_{q,\nu}^\prime$ are Yukawa coupling matrices.
All couplings are assumed to be real so that the full Lagrangian is CP invariant. However, CP is spontaneously broken by the vacuum. Indeed, the Higgs potential contains terms of the form
\begin{align}\label{Vhiggs}
    V \propto (\mu^2 + \lambda_1 S^\ast S + \lambda_2\, \phi^\dagger \phi) (S^2 + S^{\ast 2}) + \lambda_3 (S^4 + S^{\ast 4}),
\end{align}
and, for an appropriate region of the parameter space, the scalar fields acquire VEV of the form $\langle \phi \rangle = v$ and $\langle S \rangle = V e^{i\alpha}$.

It is possible to show that the phase $\alpha$ generates all CP violations, namely, nontrivial complex CKM and PMNS matrices, as well as the leptonic CP violation at high energies needed for leptogenesis. In order to verify that a nontrivial phase is generated in the CKM matrix $\mathbf{V}_\text{CKM}$, one has to recall that the mixing matrix connecting standard quarks is determined by the relation
\begin{align}\label{Vckmrel}
    \mathbf{V}_\text{CKM}^{-1} \mathbf{h}_d \mathbf{V}_\text{CKM} = \mathbf{d}_d^2,
\end{align}
where
\begin{align}\label{eq:hd}
    \mathbf{h}_d = \mathbf{m}_d \mathbf{m}_d^\dagger - \frac{\mathbf{m}_d\, \mathbf{M}_D^\dagger \mathbf{M}_D\, \mathbf{m}_d^\dagger}{\overline{M}^{\,2}},
\end{align}
$\mathbf{d}_d^2=\text{diag}\,(m_d^2, m_s^2, m_b^2)$, $\mathbf{m}_d=v \mathbf{Y}^d$, $\overline{M}^{\,2}= \mathbf{M}_D \mathbf{M}_D^\dagger + \tilde{M}^2$, and $\mathbf{M}_D = V (\mathbf{f}_q^{+} \cos\alpha + i\, \mathbf{f}_q^{-} \sin\alpha)$ with $\mathbf{f}_q^{\pm}\equiv \mathbf{f}_q  \pm \mathbf{f}_q^\prime$. Note that without loss of generality, we have chosen a weak basis with a diagonal and real up-quark mass matrix. The crucial point is then the following: the first term contributing to $\mathbf{h}_d$ in Eq.~\eqref{eq:hd} is real since the matrix $\mathbf{m}_d$ is real due to the CP invariance of the Lagrangian; the second term in $\mathbf{h}_d$ is however complex, and of the same order of magnitude as the first one. As a result, $\mathbf{h}_d$ is a generic complex $3\times3$ Hermitian matrix, leading to a complex $\mathbf{V}_\text{CKM}$ matrix. For any specific model, one can explicitly check that CP violation \emph{\`{a} la} Kobayashi-Maskawa is generated by computing the CP-odd WB invariant given in Eq.~\eqref{ICPquark}. Having $\mathcal{J}^\text{CP}_\text{quark} \neq 0$ is a necessary and sufficient condition to have CP violation through the Kobayashi-Maskawa mechanism.

In the lepton sector, the neutrino mass matrix $\mathbf{m}_\nu$ is generated after the spontaneous symmetry breaking through the standard type I seesaw mechanism given in Eq.~\eqref{mnutypeIseesaw}, with $\mathbf{m}_R= V (\mathbf{f}_\nu^{+} \cos\alpha + i\, \mathbf{f}_\nu^{-} \sin\alpha)$ and $\mathbf{f}_\nu^{\pm}\equiv \mathbf{f}_\nu  \pm \mathbf{f}_\nu^\prime$. Although the Dirac-neutrino Yukawa coupling matrix $\mathbf{Y}^\nu$ is real, the matrix $\mathbf{m}_R$ is a generic complex symmetric matrix. As a result, the effective neutrino mass matrix $\mathbf{m}_\nu$ is a generic complex symmetric matrix, and the PMNS leptonic mixing matrix has, in general, three CP-violating phases. One can also check that the model has the CP violation necessary for leptogenesis to work.

An important constraint on models with spontaneous CP violation is related with the so-called domain-wall problem~\cite{Vilenkin:1984ib}. As pointed out in the seminal papers~\cite{Zeldovich:1974uw,Kibble:1976sj}, the spontaneous breaking of a discrete global symmetry in the early Universe leads to the formation of domain walls with an energy density proportional to the inverse of the cosmological scale factor. Therefore, those objects could dominate over matter and radiation, overclosing the Universe. Although this represents a serious problem, several solutions have been put forward in order to solve it. One possible way to avoid the crippling effects of domain walls is to invoke an inflationary period that dilutes them away~\cite{Langacker:1987ft}. Note that this does not prevent the complex phase of $\langle S \rangle$ from generating a complex CKM matrix [see Eq.~\eqref{eq:hd}]. An alternative way out relies on considering the existence of a (small) bare $\theta_{\rm QCD}$ term~\cite{Krauss:1992gf}. In this case, it can be shown that the vacuum degeneracy connecting the two sides of the CP domain wall is lifted, resulting in a wall annihilation driven by the decay of a false vacuum. More interestingly, assuming that gravity breaks global discrete symmetries explicitly, then there is probably no domain-wall problem at all~\cite{Dvali:1995cc,Rai:1992xw}. These few examples show that although this problem arises whenever CP is spontaneously broken, it is possible to overcome it independently of the dynamics behind the symmetry breaking. In particular, simple scenarios as the one outlined above could in principle generate complex CKM and PMNS matrices at low energies regardless of the solution chosen to the domain-wall problem.

\subsection{The hypothesis of minimal lepton flavor violation}
\label{sec2.7}

One of the proposals for the description of flavor-changing processes in the quark sector is the so-called hypothesis of minimal flavor violation (MFV)~\cite{Chivukula:1987py,Buras:2000dm,D'Ambrosio:2002ex}. It consists of assuming that even if there is new physics beyond the SM, Yukawa couplings are the only source flavor-changing processes. More precisely, the MFV hypothesis assumes that Yukawa couplings are the only source of the breaking of the large $U(3)^5$ global flavor symmetry present in the gauge sector of the SM with three generations.

If one assumes the presence of two Higgs doublets, the MFV principle can be implemented under the assumption of NFC in the Higgs sector~\cite{Glashow:1976nt,Paschos:1976ay} or through the introduction of a discrete symmetry which leads to naturally suppressed FCNC in the Higgs sector~\cite{Branco:1996bq,Botella:2009pq}. One of the interesting features of MFV in the quark sector is the prediction of the ratio of branching ratios of low-energy processes, which do not depend on the specific MFV model.

The MFV hypothesis has also been extended to the lepton sector~\cite{Cirigliano:2005ck} but, in contrast to the quark sector, this extension is not unique and requires additional input from physics at high energies. The reason is that the total lepton number may not be a symmetry of the theory since neutrinos can be Majorana particles. In order to extend MFV to the lepton sector, one has to make a choice between two possibilities\\

\indent (i) Minimal field content: No new fields are introduced beyond the SM content and it is just assumed that some new physics at a high-energy scale generates an effective Majorana mass for the left-handed neutrinos.\\

\indent (ii) Extended field content: Two or more right-handed neutrinos are introduced with gauge-invariant lepton number violating mass terms, which generate an effective seesaw neutrino mass matrix for light neutrinos.\\

CP violation was not considered at either low or high energies in Ref.~\cite{Cirigliano:2005ck}. The inclusion of CP violation in a minimal lepton flavor violation (MLFV) scenario is crucial, for instance, in order to have a consistent framework to generate the baryon asymmetry through leptogenesis~\cite{Branco:2006hz}. Subsequent suggestions~\cite{Cirigliano:2006nu,Cirigliano:2007hb,Davidson:2006bd,Gavela:2009cd,Alonso:2011jd} for MLFV did include CP violation in the lepton sector.

For definiteness, we analyze the MLFV hypothesis in the context of a minimal extension of the SM with three right-handed neutrinos $\nu_R$. In this case, the relevant leptonic Yukawa coupling and right-handed Majorana mass terms are those given by Eq.~\eqref{LtypeIseesaw} plus the usual charged-lepton Yukawa term $\bar{\ell}_i\,\phi \,\mathbf{Y}^\ell_{ij} l_{Rj}$. In the limit when these terms vanish, the Lagrangian of this extension of the SM has a large global flavor symmetry $SU(3)_{\ell} \times SU(3)_{l_R} \times SU(3)_{\nu_R} \times U(1)_{\ell} \times U(1)_{l_R} \times U(1)_{\nu_R}$. An interesting proposal for MLFV assumes that the physics leading to lepton-number violation through the generation of the mass matrix $\mathbf{m}_R$ is lepton blind, thus leading to an exact degenerate spectrum for the right-handed neutrinos at a high-energy scale. In this MLFV framework, the Majorana mass terms break $SU(3)_{\nu_R}$ into $O(3)_{\nu_R}$. Note that, even in the limit of exact degeneracy, $\mathbf{m}_R$ is not a WB invariant. Indeed, for a WB transformation under which $\nu_R \rightarrow \mathbf{V}_R\,\nu_R$, it transforms as $\mathbf{m}_R \rightarrow \mathbf{m}_R^\prime = \mathbf{V}_R\,\mathbf{m}_R\,\mathbf{V}_R^T$. This transformation does not leave $\mathbf{m}_R$ invariant, even in the limit of exact degeneracy, since in general $\mathbf{V}_R \mathbf{V}_R^T \neq \openone$.

It is worth emphasizing that MLFV in a framework with right-handed neutrinos is not as predictive as MFV in the quark sector~\cite{Branco:2006hz}. A rich spectrum of possibilities is allowed for lepton flavor-violating (LFV) processes and their correlation with low-energy neutrino physics and LHC physics [see, for example,~\cite{Deppisch:2010fr}].

\section{OBSERVABLE EFFECTS FROM LEPTONIC CP VIOLATION}
\label{sec3}

Establishing the existence of LCPV is one of the main goals of the future neutrino physics program. The most promising way to search for CPV effects in the lepton sector is through the study of neutrino oscillations, which are sensitive to the Dirac CP phase $\delta$ entering the neutrino mixing matrix $\bU$ of Eq.~\eqref{Uparam}. The experimental sensitivity to LCPV depends strongly on the value of the reactor neutrino mixing angle $\theta_{13}$, and also on the type of neutrino mass spectrum. In particular, if  $\theta_{13}$ is not too small, then future experiments will be able to establish soon the existence (or not) of LCPV.

There are however other phenomena which, although being CP conserving, are also sensitive to the presence of CP phases in the lepton mixing matrix. For instance, predictions regarding neutrinoless beta decay rates change depending on the values of the Majorana phases $\alpha_{1,2}$. Other phenomena, which are triggered by the presence of new physics directly connected with neutrino masses and mixing may also be impacted from the fact that CP is violated in the lepton sector. A typical example is rare lepton flavor-violating decays like $l_j\rightarrow l_i\gamma$ ($j\neq i$), $l_j\rightarrow l_il_kl_k$ ($i,k\neq j$) and $\mu-e$ conversion in nuclei~\cite{Raidal:2008jk}. Ultimately, if the physics responsible for neutrino mass generation is close to the electroweak scale, then LCPV may also affect phenomena which could be observed at high-energy colliders like the LHC or a linear collider.

In this section, we present a general discussion about the possible direct and indirect effects of LCPV, with special emphasis on neutrino oscillations.

\subsection{Neutrino oscillation parameters: Present status}
\label{sec3.1}

\begin{figure}[b]
\includegraphics[width=6.95cm]{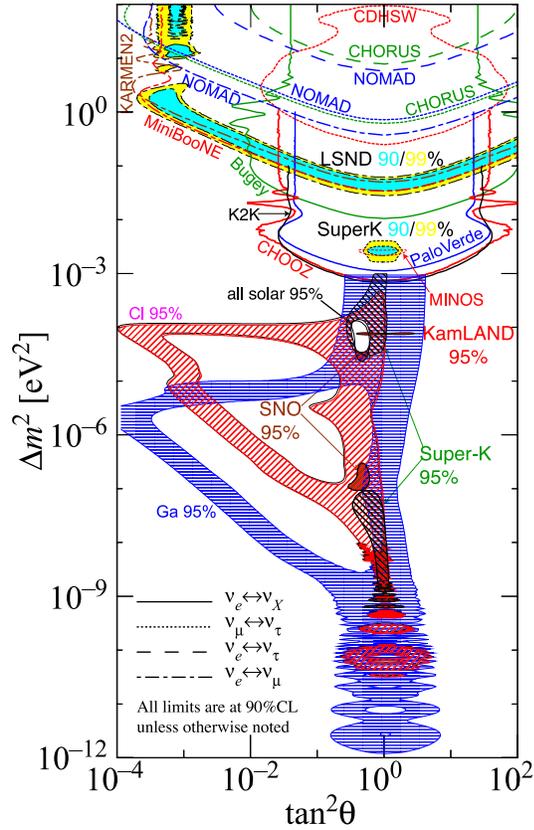}
\caption{Favored and excluded regions of neutrino mass-squared differences and mixing
angles taken into account the data of several neutrino experiments. From~\cite{Nakamura:2010zzi}.}
\label{nuexcl}
\end{figure}

\begin{table*}
\caption{\label{Tabnudata}Best-fit values with $1\sigma$ and $3\sigma$ errors for the three-flavor neutrino oscillation parameters, obtained by Gonzalez-Garcia, Maltoni and Salvado (GMS)~\cite{GonzalezGarcia:2010er}, Schwetz, T\'{o}rtola and Valle (STV)~\cite{Schwetz:2011zk} and the Bari group~\cite{Fogli:2011qn}.}
\begin{ruledtabular}
\begin{tabular}{lccc}
&GMS~\cite{GonzalezGarcia:2010er} &STV~\cite{Schwetz:2011zk}  &Bari~\cite{Fogli:2011qn}
\\ \hline
$\dmsol\,[10^{-5}]$ eV$^{2}$ &$7.59\pm 0.20\,(^{+0.61}_{-0.69})$
&$7.59^{+0.20}_{-0.18}(^{+0.60}_{-0.50})$ & $7.58^{+0.22}_{-0.26}(^{+0.60}_{-0.59})$
\\
$\dmatm\,[10^{-3}]$ eV$^{2}$ \;\;\;\;\; (NO)&$2.46\pm 0.12\,(\pm 0.37)$
&$2.50^{+0.09}_{-0.16}(^{+0.26}_{-0.36})$ & $2.35^{+0.12}_{-0.09}(^{+0.32}_{-0.29})$\\
\hspace{2.95cm} (IO)& $-2.36\pm 0.11 (\pm 0.37)$ & $-\left[2.40^{+0.08}_{-0.09}(\pm 0.27)\right]$
& $-[2.35^{+0.12}_{-0.09}(^{+0.32}_{-0.29})]$\\
$\sin^2\theta_{12}$ &$0.319\pm 0.016\,(^{+0.053}_{-0.046})$
&$0.312 ^{+0.017}_{-0.015} (^{+0.048}_{-0.042})$ &$0.312^{+0.017}_{-0.016}(^{+0.052}_{-0.047})$\\
$\sin^2\theta_{23}$ \hspace{1.8cm} (NO)
&$0.46^{+0.08}_{-0.05}(^{+0.18}_{-0.12})$
&$0.52^{+0.06}_{-0.07}\, (^{+0.12}_{-0.13})$ & $0.42^{+0.08}_{-0.03}(^{+0.22}_{-0.08})$\\
\hspace{2.95cm} (IO) &$0.46^{+0.08}_{-0.05}(^{+0.18}_{-0.12})$ &$0.52\pm 0.06\, (^{+0.12}_{-0.13})$
&$0.42^{+0.08}_{-0.03}(^{+0.22}_{-0.08})$\\
$\sin^2\theta_{13}$ \hspace{1.8cm} (NO)&  $0.0095^{+0.013}_{-0.007}(\le 0.047)$
& $0.013^{+0.007}_{-0.005}(^{+0.022}_{-0.012})$ &$0.025\pm 0.007(^{+0.025}_{-0.02})$\\
 \hspace{2.95cm} (IO) & $0.0095^{+0.013}_{-0.007}(\le 0.047)$
& $0.016^{+0.008}_{-0.006}(^{+0.023}_{-0.015})$ &$0.025\pm 0.007(^{+0.025}_{-0.02})$\\
\end{tabular}
\end{ruledtabular}
\end{table*}

The observation of a solar-neutrino deficit with respect to standard solar model predictions at the Homestake experiment~\cite{Cleveland:1998nv} provided the first hint in favor of neutrino oscillations. This observation has been confirmed by several other solar-neutrino experiments like SAGE~\cite{Abdurashitov:2002nt}, Gallex~\cite{Hampel:1998xg}, GNO~\cite{Altmann:2005ix}, Kamiokande~\cite{Fukuda:2002pe}, Super-Kamiokande~\cite{Smy:2003jf}, and the Sudbury Neutrino Observatory (SNO)~\cite{Ahmad:2001an}. The data collected from these experiments led to the large mixing angle solution to the solar neutrino problem, which was confirmed in 2002 by the KamLAND reactor neutrino experiment~\cite{Eguchi:2002dm}.

A similar anomaly has also been observed in the atmospheric neutrino sector by the IMB~\cite{BeckerSzendy:1992hq}, Kamiokande~\cite{Hirata:1992ku}, MACRO~\cite{Ambrosio:2003yz}, Soudan-2~\cite{Sanchez:2003rb} and Super-Kamiokande~\cite{Fukuda:1998mi} experiments, which detected a $\nu_\mu$ to $\nu_e$-induced event ratio smaller than the expected. Atmospheric neutrino parameters are also constrained by the K2K (KEK to Kamioka)~\cite{Aliu:2004sq} and MINOS (Fermilab to Soudan mine)~\cite{Michael:2006rx} accelerator long-baseline experiments. Both experiments observed that a fraction of the $\nu_\mu$ neutrinos in the original beam disappear consistently with the hypothesis of neutrino oscillations.

Other experiments have provided useful data in constraining the neutrino parameter space. An illustrative way to present these data is given in Fig.~\ref{nuexcl}, where the favored and excluded regions of neutrino mass-squared differences and mixing angles are shown, taking into account the results of several experiments. In Table~\ref{Tabnudata}, we summarize the results of three global analyses performed by Gonzalez-Garcia, Maltoni, and Salvado (GMS)~\cite{GonzalezGarcia:2010er}, Schwetz, T\'{o}rtola, and Valle (STV)~\cite{Schwetz:2011zk}, and the Bari group~\cite{Fogli:2011qn}.

In contrast with the quark sector, there are two large mixing angles in the lepton sector: $\theta_{12}$ and $\theta_{23}$, sometimes referred as the {\em solar} and {\em atmospheric} neutrino mixing angles (see Table~\ref{Tabnudata}). The current data indicate that, at their best-fit values, $\theta_{12}\simeq 34$\textdegree and $\theta_{23}\simeq 45$\textdegree (maximal atmospheric mixing), while the value of the remaining mixing angle, $\theta_{13}$, is mainly constrained by reactor and accelerator neutrino experiments to be small. Recent data from the T2K~\cite{Abe:2011sj} and MINOS~\cite{Adamson:2011qu} experiments also indicate a relatively large value for $\theta_{13}$. At 90\% C.L., the T2K data are consistent with $0.03\, (0.04)< \sin^2 2\theta_{13} < 0.28\, (0.34)$ for normal (inverted) hierarchy in the absence of Dirac CP violation. The MINOS Collaboration reports the best-fit values $2\sin^2(\theta_{23})\sin^2(2\theta_{13})\,\mathord{=}\,0.041^{+0.047}_{-0.031}\ \bigl(0.079^{+0.071}_{-0.053}\bigr)$. These results have been taken into account in the global analyses performed by STV and the Bari group. As it is apparent from Table~\ref{Tabnudata}, there is now an evidence for $\theta_{13}>0$ at more than $3\sigma$.

Neutrino oscillation experiments are not sensitive to the absolute neutrino mass scale since the oscillation frequency is controlled by the neutrino mass-squared differences $\Delta m_{ji}^2$ and the neutrino energy. The current data are consistent with a three-neutrino scenario with $\dmsol\sim 7.6\times 10^{-5}\,\eV^2$ and $|\dmatm|\sim 2.5\times 10^{-3}\,\eV^2$, which implies a hierarchy among these two quantities such that
\begin{align}\label{rdef}
r \equiv \frac{\dmsol}{\dmatm} \simeq \pm 0.03.
\end{align}
The sign of $\dmatm$ is not yet determined and therefore two types of neutrino mass spectrum are possible, namely,
\begin{align}
{\rm Normally-ordered\,(NO):}&\,\,m_1<m_2 <m_3\,,\nonumber\\
{\rm Invertedly-ordered\,(IO):}&\,\,m_3<m_1 <m_2\,.
\label{NOIOdef}
\end{align}
For each case, the neutrino masses can be expressed in terms of the lightest mass ($m_1$ and $m_3$ for the NO and IO cases, respectively), and the mass-squared differences $\Delta m_{ji}^2$:
\begin{eqnarray}
 {\rm NO:}\,\,m_2&=&\sqrt{m_1^2+\dmsol}\,,\nonumber\\
m_3&=&\sqrt{m_1^2+|\dmatm|}\,,\nonumber\\
 {\rm IO:}\,\,m_1&=&\sqrt{m_3^2+|\dmatm|}\,,\nonumber\\
m_2&=&\sqrt{m_3^2+\dmsol+|\dmatm|}\,.
\label{NOIOms}
\end{eqnarray}

Depending on the value of the lightest neutrino mass, one can further classify the neutrino mass spectrum as being hierarchical (HI): $m_1 \ll m_2 < m_3$, inverted-hierarchical (IH): $m_3 \ll m_1 < m_2$, or quasidegenerate (QD): $m_1\simeq m_2 \simeq m_3 \simeq m_0 \gg |\dmatm|, m_0\gtrsim 0.1\,\eV$. In the HI and IH limits, the neutrino masses are
\begin{eqnarray}
m_2^{\rm HI}&\simeq& \sqrt{\dmsol}\simeq 0.009\,\eV\,,\nonumber\\
m_3^{\rm HI}&\simeq& m_{1,2}^{\rm IH}\simeq \sqrt{|\dmatm|}\simeq 0.05\,\eV\,.
\label{HIIHms}
\end{eqnarray}

A direct kinematical bound is available for the effective electron-neutrino mass in $\beta$-decay, $m_\beta=\sqrt{\sum_i |\mathbf{U}_{ei}|^2\,m_i^2}$. From the Mainz~\cite{Bonn:2002jw} and Troitzk~\cite{Lobashev:2001uu} experiments, $m_{\beta}<2.3\,\eV$ at $95\%$  C.L., which implies $m_i<2.3$~eV. In the future, the KATRIN experiment~\cite{Osipowicz:2001sq} is expected to reach a sensitivity of $m_{\beta}\simeq 0.2\,\eV$. The current 7-year WMAP data constrain the sum of neutrino masses to be less than 1.3 eV at $95\%$ C.L.~\cite{Komatsu:2010fb} (within the standard cosmological model). Less conservative bounds can be obtained combining the data of several cosmological and astrophysical experiments~\cite{Abazajian:2011dt}. The future Planck satellite data alone will allow one to set an upper bound on $\sum_i m_i$ of 0.6 eV at $95\%$ C.L.~\cite{Hannestad:2010kz}. Concerning CPV in the lepton sector, the presently available neutrino data do not provide any information on the CP phases $\delta$ (Dirac) and $\alpha_{1,2}$ (Majorana). In the following we discuss how LCPV can be probed in future experiments.

\subsection{LCPV in neutrino oscillations}
\label{sec3.2}

The existence of more than two neutrino flavors opens the possibility for the existence of CP-violating effects in the lepton sector, characterized by the CP phases of the neutrino mixing matrix $\bU$. Since neutrino oscillations depend directly on the way neutrinos mix among themselves and, consequently, on the existence of CP phases, it is not surprising that they represent the golden path for the search of LCPV . Yet, establishing CPV in the neutrino sector turns out to be a rather hard task. In the last years, several ideas have been brought together with the aim of overcoming these difficulties and finding the best strategy to detect CPV effects in neutrino oscillations. In particular, new experimental setups have been proposed in order to improve our knowledge of the neutrino parameters.

In this section, we review some basic aspects related to the formalism of LCPV and neutrino oscillations and discuss possible ways to search for CPV, pointing out the main difficulties inherent to this investigation. Moreover, we intend to draw a general picture of the prospects for the discovery of LCPV in future neutrino oscillation experiments. For more complete discussions about theoretical aspects of neutrino oscillations, we address the reader to other dedicated reviews~\cite{Mohapatra:2006gs, Bilenky:1987ty,Bilenky:1998dt,Akhmedov:1999uz,Akhmedov:2010ua,Strumia:2006db,GonzalezGarcia:2007ib} and textbooks~\cite{Fukugita:2003en,Giunti:2007ry}.

\subsubsection{CPV in vacuum oscillations}
\label{sec3.2.1}

If neutrinos are massive and mix, then a neutrino state produced via weak interactions (like nuclear beta and pion decays) is not a mass eigenstate. In this case, the weak eigenstates $\nu_\alpha$ are a unitary linear combination of the mass eigenstates $\nu_k$, in such a way that
\begin{equation}
\label{WMass}%
|\nu_\alpha\rangle=\sum_{k=1}^n \bU_{\alpha k}^\ast |\nu_k\rangle\,,
\end{equation}
where $\bU$ is the lepton mixing matrix defined in Eq.~\eqref{Unotation}. As first pointed out by Pontecorvo, the fact that mass and flavor eigenstates are different leads to the possibility of neutrino oscillations~\cite{Pontecorvo:1967fh}. The time evolution of a neutrino produced with a specific flavor is governed by
\begin{equation}
\label{evol}%
|\nu_\alpha(t)\rangle=\sum_{k=1}^n \bU_{\alpha k}^\ast e^{-iE_k t} |\nu_k\rangle\,,
\end{equation}
where $E_k$ is the energy of the neutrino mass eigenstate $\nu_k$. For relativistic neutrinos, $E_k=\sqrt{p_k^2+m_k^2}\simeq p_k+m_k^2/(2E_k)$. The $\nu_\alpha\rightarrow \nu_\beta$ transition amplitude is then given by
\begin{equation}
\label{amp}%
\mathcal{A}_{\alpha\beta}(t)=
\sum_{k=1}^n \bU_{\beta k}e^{-iE_k t}\bU_{\alpha k}^\ast\,,
\end{equation}
and the corresponding transition probability by $P_{\alpha\beta}=|\mathcal{A}_{\alpha\beta}|^2$. For $t=0$ and $\alpha \neq \beta$, the above equation is equivalent to the definition of the Dirac unitarity triangles $T_{\alpha\beta}$ given in Eqs.~\eqref{Diractriangles}. The time evolution of $\mathcal{A}_{\alpha\beta}$ can then be interpreted as a time-dependent rotation of the sides of these triangles.

Considering that for ultrarelativistic neutrinos $t\simeq L$ (where $L$ is the distance traveled by neutrinos) and assuming equal momenta for all of the neutrino mass eigenstates ($p_k\equiv p\simeq E$ for any $k$), the $\nu_\alpha\rightarrow \nu_\beta$ oscillation probabilities can be further expressed as
\begin{eqnarray}
\label{tramp2}
P_{\alpha\beta}(L,E)&=&\delta_{\alpha\beta}-4\sum_{k>j}
\mathcal{R}_{\alpha\beta}^{kj}
\sin^2\frac{\Delta_{kj}}{2}\nonumber\\
&+&2\sum_{k>j}\mathcal{J}_{\alpha\beta}^{kj}
\sin\Delta_{kj}\,,
\end{eqnarray}
where
\begin{align}\label{Deltakj}
\Delta_{kj}=\frac{\Delta m^2_{kj}L}{2E}.
\end{align}
The quantities $\mathcal{R}_{\alpha\beta}^{kj}$ and $\mathcal{J}_{\alpha\beta}^{kj}$ are invariant combinations of the elements of $\bU$ given by
\begin{eqnarray}
\mathcal{R}_{\alpha\beta}^{kj}=
\Re\left[\bU_{\alpha k}^\ast\bU_{\beta k}\, \bU_{\alpha j}\bU_{\beta j}^\ast\right]\,,\nonumber \\
\label{RI}
\mathcal{J}_{\alpha\beta}^{kj}= \Im\left[\bU_{\alpha k}^\ast\bU_{\beta k}\, \bU_{\alpha
j}\bU_{\beta j}^\ast\right]\,.
\end{eqnarray}

The above formulas show that the transition probabilities $\nu_\alpha\rightarrow \nu_\beta$ depend on the elements of the mixing matrix $\bU$, $n-1$ independent mass-squared differences, and  the ratio $L/E$, which depends on the specific experimental setup. Within the simplest framework of two neutrinos, the oscillation probability is given by
\begin{equation}
\label{tramp3}
P_{\alpha\beta}=\sin^2(2\theta)\sin^2\left(\frac{\Delta m^2 L}{4E}\right)\,,
\; \alpha\neq\beta\,,
\end{equation}
being the survival probability $P_{\alpha\alpha}= 1-P_{\alpha\beta}$. Therefore, to be sensitive to neutrino oscillations, experiments must be designed in such a way that $L\sim L_{\rm osc}$, with
\begin{equation}
\label{losci}
L_{\rm osc}=\frac{4\pi E}{\Delta m^2}=2.47\frac{E\,[\GeV]}{\Delta m^2\,[\eV^2]}\,\km\,.
\end{equation}

\begin{figure}[b]
\includegraphics[width=6.00cm]{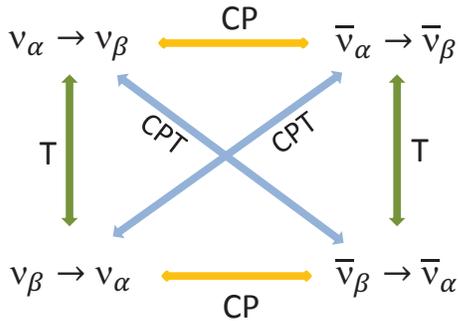}
\caption{Transformations of the different flavor-transition channels under CP, T, and CPT.}
\label{diagCPT}
\end{figure}

The fact that CP violation in the lepton sector can be tested in neutrino oscillation experiments was first noted by Cabibbo~\cite{Cabibbo:1977nk} and Barger {\em et. al.}~\cite{Barger:1980jm}. Such tests require the comparison of transitions $\nu_\alpha\rightarrow \nu_\beta$ with the corresponding CP-conjugate channel $\bar{\nu}_\alpha\rightarrow \bar{\nu}_\beta$, or with $\nu_\beta\rightarrow\nu_\alpha$ if CPT invariance holds. For antineutrinos, the equivalent of Eqs.~\eqref{WMass} and \eqref{evol} reads
\begin{equation}
\label{WMassbar}
|\bar{\nu}_\alpha\rangle=\sum_{k=1}^n \bU_{\alpha k} |\bar{\nu}_k\rangle\,,\quad
|\bar{\nu}_\alpha(t)\rangle=\sum_{k=1}^n \bU_{\alpha k} e^{-iE_k t} |\bar{\nu}_k\rangle\,,
\end{equation}
which lead to the following $\bar{\nu}_\alpha \rightarrow \bar{\nu}_\beta$ transition amplitudes and probabilities in vacuum:
\begin{equation}
\label{ampbar}
\bar{\mathcal{A}}_{\alpha\beta}(t)=
\sum_{k=1}^n \bU_{\beta k}^\ast e^{-iE_k t}\bU_{\alpha k}\,,\quad
\bar{P}_{\alpha\beta}=|\bar{\mathcal{A}}_{\alpha\beta}|^2\,,
\end{equation}
respectively. It is straightforward to see that, due to CPT conservation, $\bar{P}_{\alpha\beta}=P_{\beta\alpha}$~\cite{Cabibbo:1977nk}. The transformation properties of the neutrino flavor transitions under CP, T and CPT are shown in Fig.~\ref{diagCPT}.

Under CP, neutrinos transform into their antiparticles $(\nu_\alpha \leftrightarrow \bar{\nu}_\alpha)$. Depending on whether we consider the case of Dirac or Majorana neutrinos, CP invariance in the lepton sector implies~\cite{Bilenky:1984fg}
\begin{eqnarray}
\label{conditions}
\bU_{\alpha k}&=&\bU^\ast_{\alpha k}\hspace*{1.6cm}{\rm (Dirac)}\,,\\
\bU_{\alpha k}&=&-i\rho_k\,\eta_k^{\CP}\,\bU^\ast_{\alpha k}\;{\rm (Majorana)}\,,
\end{eqnarray}
where $\eta_k^{\CP}=\pm i$ is the CP parity of the neutrino mass eigenstate with mass $m_k$, and $\rho_k$ is an arbitrary phase factor present in the Majorana condition $C \, \overline{\nu_k}^T = \rho_k \nu_k$. Therefore, CP invariance automatically leads to $\bar{P}_{\alpha\beta}=P_{\alpha\beta}$. Obviously, due to CPT conservation, CP invariance is equivalent to T invariance.

The most obvious way to measure CP violation in the neutrino sector is by looking at the differences $\Delta P_{\alpha\beta}^\CP=P_{\alpha\beta}-\bar{P}_{\alpha\beta}$. Taking into account that $\bar{P}_{\alpha\beta}$ is obtained replacing $\bU$ by $\bU^\ast$ in Eq.~(\ref{tramp2}), one has~\cite{Barger:1980jm,Pakvasa:1980bz}
\begin{eqnarray}\label{deltaP2}
\Delta P_{\alpha\beta}^\CP=4\sum_{k>j}\mathcal{J}_{\alpha\beta}^{kj}\sin\Delta_{kj}\,,
\end{eqnarray}
which coincides with the T-violating probability differences $\Delta P_{\alpha\beta}^\T=P_{\alpha\beta}-P_{\beta\alpha}$. The above equation can be rewritten as
\begin{equation}
\label{deltaP3}
\Delta P_{\alpha\beta}^\CP=-16\,\mathcal{J}_{\alpha\beta}^{21}\sin\frac{\Delta_{21}}{2}\sin
\frac{\Delta_{13}}{2} \sin\frac{\Delta_{32}}{2}\,,
\end{equation}
with $\Delta P_{e\mu}^\CP=\Delta P_{\mu\tau}^\CP=\Delta P_{\tau e}^\CP=\Delta P^\CP$, and
\begin{equation}
\label{deltaP}
\Delta P^\CP=4\,\mathcal{J}_{e\mu}^{21}\left(\sin\Delta_{21}+\sin\Delta_{32}+\sin\Delta_{13}\right)\,.
\end{equation}
The invariant quantity $\mathcal{J}_{e\mu}^{21}$ has been defined in Eq.~\eqref{Jemu21}. From these results, it is clear that CP violation is absent in neutrino oscillations, if two (or more) neutrinos are degenerate in mass, or if one of the mixing angles is zero. Therefore, CPV in vacuum oscillations occurs as a pure three-flavor effect, and thus is suppressed by small mixing angles. Moreover, since $\bar{\nu}_\alpha \rightarrow \bar{\nu}_\alpha$ is the CPT transformed of $\nu_\alpha \rightarrow \nu_\alpha$, CPV cannot be observed in disappearance channels. Experimentally, the measurement of LCPV in neutrino oscillations requires sensitivity to the oscillatory behavior of the neutrino and antineutrino transition probabilities. In other words, $L$ and $E$ have to be such that at least one of the phases $\Delta_{kj}$ is of order 1. Indeed, if $\Delta_{kj}\ll 1$ for all $k$ and $j$, then the transition probabilities are too small to be observed. On the other hand, in the limit $\Delta_{kj}\gg 1$, the averaged $\Delta P_{\alpha\beta}$ goes to zero. It is also important to note that, if the order-one phase corresponds to the largest $\Delta m^2_{kj}$, then $\Delta P_{\alpha\beta}^\CP \simeq 0$~\cite{Bilenky:1981hf,Barger:1980hs}. This can be readily understood considering the case $\Delta m^2_{32} \simeq  \dmatm \gg \dmsol$. If $\Delta_{31}\simeq \Delta_{32} \simeq 1$ (short baseline) then $\Delta_{21}\ll 1$ and $\Delta P_{\alpha\beta}^\CP \simeq 4 (\mathcal{J}_{\alpha\beta}^{31}+\mathcal{J}_{\alpha\beta}^{32})\sin \Delta_{31}=0$, due to the fact that $\mathcal{J}_{\alpha\beta}^{31}=-\mathcal{J}_{\alpha\beta}^{32}$ [see Eq.~\eqref{RI}]. Therefore, a measurement of the CP-odd asymmetry in neutrino oscillations can be performed only in long-baseline experiments~\cite{Minakata:1997td,Arafune:1996bt,Tanimoto:1996ky,Bilenky:1997dd}, as long as $|\mathcal{J}_{e \mu}^{21}|$ is not too small.

\subsubsection{Matter-induced CP violation}
\label{sec.3.2.2}

The discussion presented in the previous section raises the question on whether a measurement of a nonzero $\Delta P^\CP$ automatically implies that CP is violated in the lepton sector. Although this would be true in vacuum oscillations, matter effects in neutrino propagation~\cite{Mikheev:1986gs,Wolfenstein:1977ue,Barger:1980tf} can fake CP violation~\cite{Kuo:1987km,Krastev:1988yu}. Indeed, the presence of matter violates C, CP, and CPT due to the unequal number of particle and antiparticles (electrons and positrons) in the medium. In matter, the relevant effective Hamiltonian for neutrinos can be written as
\begin{equation}
\label{Hmatter}
{\rm \bf H}^\prime=\frac{1}{2E}\left[\bU\, {\rm \bf M}^2\bU^\dag+{\rm \bf A}\right]\,,
\end{equation}
where ${\rm \bf M}^2=\diag(0,\dmsol,\dmatm)$ and ${\rm \bf A}=\diag(A(L),0,0)$ with
\begin{eqnarray}
A(L)&\equiv& 2\sqrt{2}E G_F N_e(L)\nonumber \\
\label{ACC}
&\simeq& 2.3\times 10^{-4}\,{\rm eV}^2\frac{\rho(L)}{3\,{\rm g\,cm}^{-3}}\frac{E}{{\rm GeV}} \,.
\end{eqnarray}
Here, $N_e(L)$ and $\rho(L)$ are the electron number and matter densities of the medium, respectively, as a function of the distance $L$. In the above estimate, the electron fraction number in matter has been considered to be 1/2. Note that for an average density of $3\,{\rm g\,cm}^{-3}$ (which corresponds approximately to the Earth's lithosphere density), $AL/(2E)\simeq 0.6\times 10^{-3}(L/{\rm km})$, meaning that matter effects are expected to be large for baselines $L\gtrsim 1000$~km.

For antineutrinos, the corresponding Hamiltonian ${\rm \bf \bar{H}^\prime}$ is obtained replacing $\bU$ by $\bU^\ast$ and $A$ by $-A$ on the right-hand side of Eq.~\eqref{Hmatter}. Taking into account the neutrino evolution equation, one can show that the oscillation probabilities in matter do not depend on the Majorana phases $\alpha_{1,2}$~\cite{Langacker:1986jv}, just as in the vacuum oscillation regime.

The effective masses and mixing matrix for neutrinos and antineutrinos are obtained by diagonalizing ${\rm \bf H^\prime}$ and ${\rm \bf \bar{H}^\prime}$, respectively. The neutrino (antineutrino) transition probability in matter is then obtained replacing $\bU$ by $\bU^\prime$ ($\bar{\bU}^\prime$) and $\Delta m^2_{kj}$ by $\Delta m^{\prime 2}_{kj}$ ($\Delta \bar{m}^{\prime 2}_{kj}$) in Eq.~(\ref{tramp2}), where the primes refer to quantities in matter. As a result, one obtains for a constant matter-density profile
\begin{eqnarray}
\label{trampsmat}
P^\prime_{\alpha\beta}(L,E)&=&\delta_{\alpha\beta}-4\sum_{k>j}
\mathcal{R}_{\alpha\beta}^{\prime kj}
\sin^2\frac{\Delta^\prime_{kj}}{2}\nonumber\\
&+&2\sum_{k>j}\mathcal{J}_{\alpha\beta}^{\prime kj}
\sin\Delta^\prime_{kj}\,,\\
\label{trampsmatbar}
\bar{P}^\prime_{\alpha\beta}(L,E)&=&\delta_{\alpha\beta}-4\sum_{k>j}
\bar{\mathcal{R}}_{\alpha\beta}^{\prime kj}
\sin^2\frac{\bar{\Delta}^\prime_{kj}}{2}\nonumber\\
&+&2\sum_{k>j}\bar{\mathcal{J}}_{\alpha\beta}^{\prime kj}
\sin\bar{\Delta}_{kj}^\prime\,,
\end{eqnarray}
where $\mathcal{R}_{\alpha\beta}^{\prime kj}$ and $\mathcal{J}_{\alpha\beta}^{\prime kj}$
are now the invariants analogous to those defined in the vacuum regime [cf. Eqs.~\eqref{RI}],
\begin{eqnarray}
\mathcal{R}_{\alpha\beta}^{\prime kj}&=& \Re\left[\bU_{\alpha k}^{\prime\ast}\bU_{\beta
k}^\prime\, \bU_{\alpha j}^\prime\bU_{\beta j}^{\prime\ast}\right]\,, \nonumber\\
\label{RJmat}
\mathcal{J}_{\alpha\beta}^{\prime kj}&=& \Im\left[\bU_{\alpha k}^{\prime\ast}\bU_{\beta
k}^\prime\, \bU_{\alpha j}^\prime\bU_{\beta j}^{\prime\ast}\right]\,,
\end{eqnarray}
and $\Delta^\prime_{kj}=\Delta m^{\prime 2}_{kj}L/(2E)$. The corresponding quantities $\bar{\Delta}^\prime_{kj}$, $\bar{\mathcal{R}}_{\alpha\beta}^{\prime kj}$ and $\bar{\mathcal{J}}_{\alpha\beta}^{\prime kj}$ are obtained replacing $\Delta m^{\prime 2}_{kj}$ and $\bU$ by  $\Delta \bar{m}^{\prime 2}_{kj}$ and $\bar{\bU}$, respectively, in the previous expressions. It can be shown that the quantities $\mathcal{J}_{\alpha\beta}^{\prime kj}$ and $\bar{\mathcal{J}}_{\alpha\beta}^{\prime kj}$ are as good as $\mathcal{J}_{\alpha\beta}^{kj}$ for the proof of CP violation~\cite{Bilenky:1997dd,Harrison:1999df}. However, the measurement of a CP-odd asymmetry in matter does not necessarily imply the existence of intrinsic CPV. From Eqs.~(\ref{trampsmat}) and (\ref{trampsmatbar}), it is straightforward to show that $\Delta P^{\prime {\rm CP}}_{\alpha\beta}=P^\prime_{\alpha\beta}- \bar{P}^\prime_{\alpha\beta}\neq 0$ even if $\mathcal{J}_{\alpha\beta}^{\prime kj}=\bar{\mathcal{J}}_{\alpha\beta}^{\prime kj}=0$, since the transition probabilities for neutrinos and antineutrinos are different in the CP-conserving limit~\cite{Langacker:1986jv}. CP-odd effects can also be observed in two-flavor neutrino oscillations due to the fact that the presence of matter may enhance, for instance, $\nu_e \leftrightarrow\nu_\mu$ oscillations and suppress the $\bar{\nu}_e \leftrightarrow\bar{\nu}_\mu$, giving rise to a nonzero $\Delta P^{\prime {\rm CP}}_{e\mu}$. As for the survival probabilities, in general one has $P^\prime_{\alpha\alpha}\neq \bar{P}^\prime_{\alpha\alpha}$, contrarily to what happens in a vacuum. In conclusion, these fake CPV effects complicate the study of fundamental CPV in neutrino oscillations since CP-odd asymmetries can be observed even if $\delta=0,\pi$.

Because of the CPT-violating character of the medium, CP and T-violation effects in matter are not directly connected\footnote{Some interesting relations between CP and T-odd asymmetries can still be obtained for the matter-oscillation case~\cite{Koike:1999hf,Akhmedov:2001kd,Minakata:1997td}.}. Therefore, T-odd effects in matter can be analyzed independently of the CP-odd ones. The first simple observation is that there is no T violation in the two-flavor case. Taking the two flavors to be $e$ and $\mu$, unitarity implies $P^\prime_{ee}+P^\prime_{e\mu}=P^\prime_{ee}+P^\prime_{\mu e}=1$, which in turn leads to the equality $P^\prime_{e\mu}=P^\prime_{\mu e}$. Thus, T-odd effects are present only for a number of neutrino flavors larger than 2. Moreover, in the presence of a symmetric matter-density profile, one can show that there are no matter-induced T-violating effects~\cite{Kuo:1987km}, since interchanging the final and initial neutrino flavors is equivalent to reversing the matter-density profile. In long-baseline neutrino oscillation experiments, matter effects due to the passage of neutrinos through the Earth are important. Since the Earth's matter density is not perfectly symmetric, the matter-induced T violation affects the T-odd asymmetries and therefore contaminates the determination of the fundamental T and CP asymmetries. Nevertheless, the asymmetries present in the Earth's density profile do not affect much the determination of the fundamental CP-violating phase $\delta$~\cite{Akhmedov:2001kd}.

It has been known for quite a long time that the most prominent oscillation channel for the study of three-flavor and matter effects in long-baseline experiments like neutrino factories is the so-called golden channel $\nu_e\rightarrow \nu_\mu$~\cite{Cervera:2000kp,Barger:1980hs,Freund:1999gy,Dick:1999ed,Donini:1999jc, Tanimoto:1996ky,Minakata:1997td,DeRujula:1998hd}. The exact formulas for the oscillation probabilities in matter are quite cumbersome due to the large number of parameters involved~\cite{Zaglauer:1988gz,Ohlsson:1999xb}. It is therefore convenient to consider expansions of $P_{\alpha\beta}$ and $\bar{P}_{\alpha\beta}$ in parameters which are known to be small. In the case of three-flavor neutrino oscillations, there are two rather obvious expansion parameters, namely, the mixing angle $\theta_{13}$ and the ratio $r$ defined in Eq.~\eqref{rdef}. Approximate expressions for the oscillation probabilities in matter of constant density have been obtained for $\dmsol \ll A,\dmatm$~\cite{Freund:2001pn,Cervera:2000kp,Asano:2011nj}. Treating $\theta_{13}$ and $r$ as small parameters, one has for the golden channel $\nu_e\rightarrow \nu_\mu$~\cite{Cervera:2000kp}
\begin{equation}
\label{Pmueap}
P_{e\mu }\simeq T_1\sin^22\theta_{13}+r\,(T_2 +T_3 )\sin 2\theta_{13}+r^2T_4\,,
\end{equation}
at second order in $\sin2\theta_{13}$ and $r$. The terms $T_i$ in the above equation
are~\cite{Huber:2006wb}
\begin{align}
T_1&\equiv s_{23}^2\, f_\Delta^2 (1-\hat{A})\,,\nonumber\\
T_2&\equiv  \sin \delta \sin \Delta \sin(2\theta_{12})\sin(2\theta_{23})f_\Delta(\hat{A}) f_\Delta(1-\hat{A})\,,\nonumber\\
T_3&\equiv  \cos\delta \cos\Delta\sin(2\theta_{12})\sin(2\theta_{23})  f_\Delta(\hat{A}) f_\Delta(1-\hat{A})\,,\nonumber\\
T_4&\equiv \cos^2(2\theta_{23}) \sin^2(2\theta_{12})f_\Delta(1-\hat{A})\,,
\end{align}
where $f_\Delta(x) \equiv \sin(x \Delta)/x$ and
\begin{eqnarray}
\Delta &\equiv& \frac{\dmatm L}{4E} \simeq 1.27\,\frac{\dmatm}{\eV^2}
\frac{L}{\km}\frac{\GeV}{E}\,,\nonumber\\
\label{Ts}
\hat{A}&\equiv& \frac{A}{\dmatm}\,,
\end{eqnarray}
with $A$ defined in Eq.~(\ref{ACC}). The corresponding antineutrino oscillation probability $\bar{P}_{e\mu}$ is obtained from $P_{e\mu}$, performing the replacements $(\delta\rightarrow -\delta,\hat{A} \rightarrow -\hat{A})$ in the coefficients $T_i$ defined above. The sign of $\hat{A}$ is determined by the sign of $\dmatm$, and by whether one considers neutrino or antineutrino oscillations. The above approximate expressions are accurate as long as $\theta_{13}$ is not too large and $E\gtrsim 0.5\,\GeV$~\cite{Barger:2001yr}. They are commonly used to illustrate some of the general features of the matter effects in the neutrino oscillation probabilities. In general, complete analyses are performed by integrating the evolution equations in matter and taking into account the Earth's matter-density profile provided by the preliminary reference Earth model~\cite{Dziewonski:1981xy}.

\subsubsection{Degeneracy problems}
\label{sec.3.2.3}

In the previous sections, we reviewed the basics of the neutrino oscillation formalism and how leptonic CP violation enters into the oscillation probabilities. The determination of the yet unknown neutrino parameters $\delta$, $\theta_{13}$ and the sign of $\dmatm$, sgn($\dmatm$), from the knowledge of $P_{e\mu}$ and $\bar{P}_{e\mu}$ is usually plagued by degeneracies and correlations among the different parameters in the oscillation probabilities. Consequently, one cannot determine unambiguously the values of $\delta$ and $\theta_{13}$~\cite{BurguetCastell:2001ez,Minakata:2001qm} from a given measurement of the probabilities $P$ and $\bar{P}$. The three twofold degeneracies related with the determination of the oscillation parameters in long-baseline neutrino experiments can be briefly summarized as follows.

\paragraph{CP degeneracy: $(\delta,\theta_{13})$ ambiguity}
\noindent\\

The CP degeneracy occurs as a consequence of the fact that two different sets $(\delta,\theta_{13})$ can lead to the same oscillation probabilities for fixed values of the remaining parameters~\cite{BurguetCastell:2001ez,Koike:2000jf}. For instance, there might be CP-conserving solutions which are degenerate with a CP-violating one. In the $(P,\bar{P})$ bi-probability space, the CP trajectories (for $\delta \neq n\pi/2$, with $n$ integer) are ellipses~\cite{Minakata:2001qm} and therefore the degeneracy can be geometrically understood as the intersection of two ellipses with distinct values of $\theta_{13}$. As a result, neutrino oscillation analysis relying on a monoenergetic beam at a fixed baseline $L$ will necessarily lead to parameter ambiguities. If $\delta = n\pi$ or $(n-1/2)\pi$, then the ellipses collapse to a line and, in principle, $\theta_{13}$ can be determined. Nevertheless, a $(\delta,\pi-\delta)$ or $(\delta,2\pi-\delta)$ ambiguity still remains~\cite{Barger:2001yr}. Instead, if $\delta \simeq n\pi/2$, the ambiguous values of $\theta_{13}$ are very close to each other, being this case qualitatively similar to the previous ones.

\paragraph{Mass-hierarchy degeneracy: sgn($\dmatm$) ambiguity}
\noindent\\

In certain cases, the same values of $P$ and $\bar{P}$ can be obtained for different pairs $(\theta_{13},\delta)$ and $(\theta_{13}^\prime,\delta^\prime)$ when considering $\dmatm>0$ or $\dmatm<0$~\cite{Minakata:2001qm}. This is commonly known as the sign or mass-hierarchy degeneracy. As in the previous case, CP-conserving solutions with $\dmatm > 0$ may be degenerate with CP-violating ones with $\dmatm < 0$. The sgn($\dmatm$) ambiguity is only present for some values of $\delta$ and tends to disappear when matter effects become large, i.e., when $L$ and $\theta_{13}$ are sufficiently large~\cite{Barger:2000cp,Lipari:1999wy}. Unlike the $(\delta,\theta_{13})$ ambiguity discussed above, where $\theta_{13}$ is resolved in the case $\delta=n\pi/2$, the sgn($\dmatm$) ambiguity can lead to different values of $\delta$ and $\theta_{13}$, even if the condition $\delta=n\pi/2$ is verified. In total, this ambiguity can lead to a fourfold degeneracy since there may be four sets of $(\theta_{13},\delta)$ (two for $\dmatm>0$ and two for $\dmatm<0$) which give the same values of $P$ and $\bar{P}$.

\paragraph{$\theta_{23}$ degeneracy: $(\theta_{23},\pi/2-\theta_{23})$ ambiguity}
\noindent\\

The extraction of $\delta$ and $\theta_{13}$ is affected by another ambiguity which is related with the atmospheric neutrino mixing angle $\theta_{23}$~\cite{Fogli:1996pv,Barger:2001yr}. Since only $\sin^2 2\theta_{23}$ enters in the $\nu_\mu$ survival probabilities, it is straightforward to conclude that $\theta_{23}$ cannot be distinguished from $\pi/2-\theta_{23}$. Obviously, for $\theta_{23}\simeq \pi/4$, which corresponds to the present best-fit value of this angle, the ambiguity is not present. Once again, CP-conserving and CP-violating solutions cannot be disentangled due to the $\theta_{23}$ ambiguity. Moreover, different values of $\theta_{13}$ can give the same $P$ and $\bar{P}$, even if $\delta=n\pi/2$.\\

\begin{figure}[t]
\begin{tabular}{c}
\includegraphics[width=7.8cm]{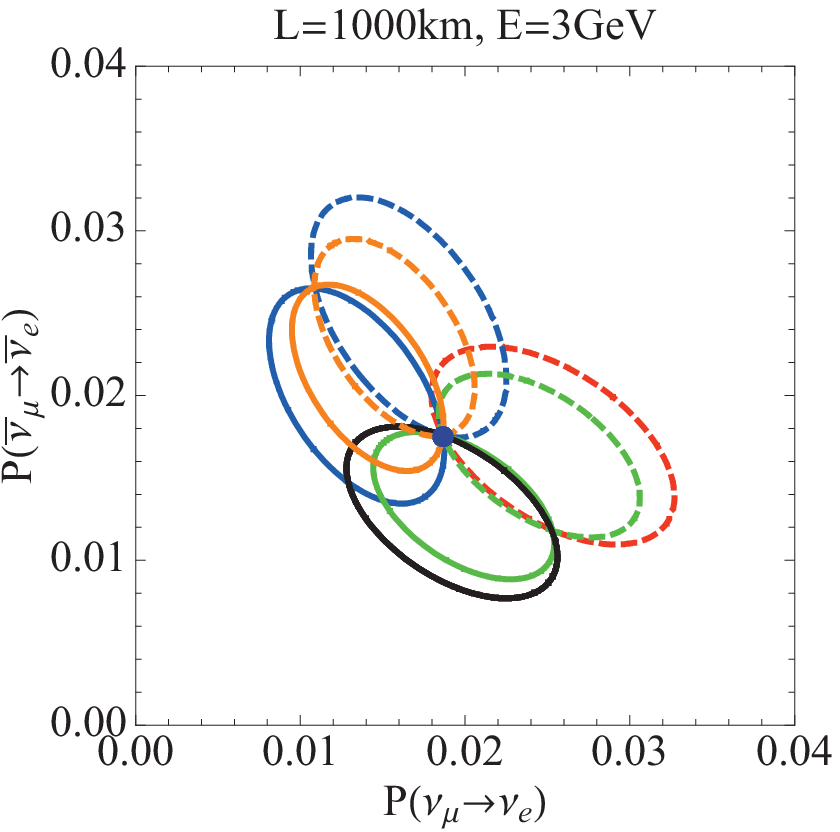} \\
\hspace*{0.5cm}\includegraphics[width=7.8cm]{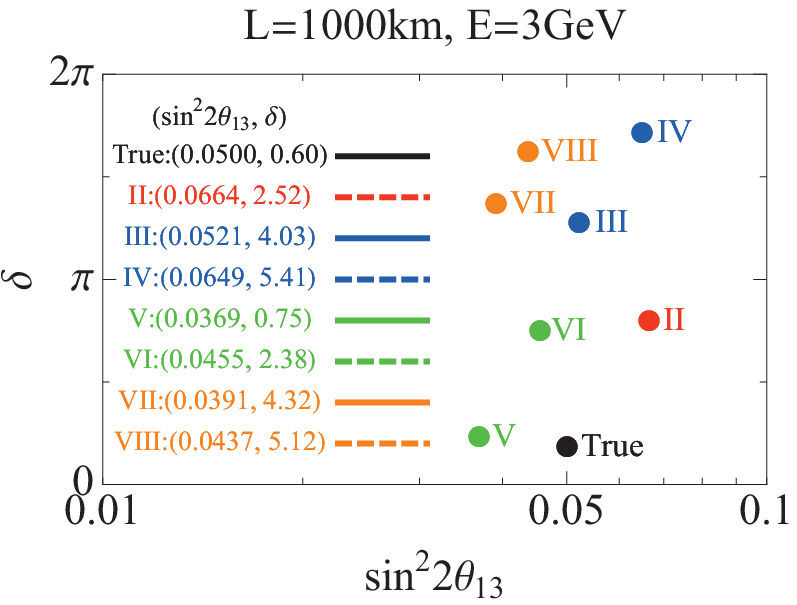}
\end{tabular}
\caption{Top panel: An illustrative example of the eightfold degeneracy  in terms of the bi-probability plot in $P_{\mu e}$-$\bar{P}_{\mu e}$ space~\cite{Minakata:2001qm}. Bottom panel: Values of ($\sin^2 2\theta_{13}$, $\delta$) for the true solutions and the clone solutions II-VIII in $\sin^2 2\theta_{13}$-$\delta$ space. The correspondence between the ellipses (top panel) and the solution labels are made manifest by using the same color lines and/or symbols in both panels. From~\cite{Minakata:2010zn}.}
\label{eightfold}
\end{figure}

From the above discussion, one concludes that, in the worst case, there can be an eightfold degeneracy~\cite{Barger:2001yr} when determining $\delta$ and $\theta_{13}$ from the measurement of the probabilities $P$ and $\bar{P}$, at a fixed baseline $L$ and neutrino energy $E$. Moreover, for all the ambiguities, one may not be able to distinguish a CP-conserving solution from a CP-violating one. An example of the eightfold degeneracy is pictorially represented in Fig.~\ref{eightfold}, where the point corresponding to the true solution is degenerate with the clone ones (points II to VIII shown in the bottom panel) at the intersection of the corresponding ellipses in the bi-probability space. A complete analysis of the parameter degeneracy in neutrino oscillations can be found in~\cite{Donini:2003vz,Minakata:2010zn}, where the degeneracies are interpreted as being a result of the invariance of the oscillation probabilities under discrete mapping of the mixing parameters. Moreover, the analytical solution of all the clone solutions has been obtained as a function of the true one.

The existence of parameter ambiguities represents a major difficulty in the extraction of the neutrino parameters from the experimental measurements of oscillation probabilities. To overcome this limitation, a set of complementary measurements have to be performed for distinct oscillation channels, baselines, and energies~\cite{BurguetCastell:2001ez,Kajita:2006bt,Ishitsuka:2005qi}. It has also been shown that a good energy resolution is also important to resolve the degeneracies~\cite{Freund:2001ui,Bueno:2001jd,Kajita:2001sb}. A powerful method to reduce the impact of ambiguities is to perform measurements at the so-called ``magic baseline''~\cite{Huber:2003ak} which satisfies the condition $\sin(\hat{A}\,\Delta)=0$. This choice leads to a simplified form of the oscillation probabilities since all terms in Eq.~(\ref{Pmueap}) will vanish, except the first one. This allows for a determination of $\sin^22\theta_{13}$ and sgn$(\dmatm)$, which is free of correlations with the CP phase $\delta$~\cite{Barger:2001yr,Lipari:1999wy}. It is straightforward to see that the first solution to the magic condition corresponds to $\sqrt{2}G_F n_e L=2\pi$ which, for a constant matter density profile, leads to
\begin{equation}
\label{Lmagic}
L_{\rm magic} \simeq 32726 \frac{1}{\rho\, [{\rm g/cm}^3]}\,\km\,.
\end{equation}
The magic baseline only depends on the matter density and taking an average $\rho \simeq 4.3\, {\rm g/cm}^3$ one has $L_{\rm magic} \simeq 7630$ km. The above baseline has the disadvantage that does not allow for the study of CP violation since the oscillation probabilities are independent from $\delta$ for $L \simeq L_{\rm magic}$. For this reason, the combination of the magic baseline with a shorter one (with better statistics) opens the possibility for the measurement of $\theta_{13}$, sgn($\dmatm$), and $\delta$ without much correlations.  In particular, a detailed optimization study reveals that the combination of two baselines $L_1=4000\,\km$ and $L_2=7500\,\km$ is optimal for these studies~\cite{Kopp:2008ds}.

The study of additional oscillation channels may also reduce the uncertainty in the determination of the neutrino oscillation parameters. For instance, it has been shown that the analysis of the ``silver'' channel $\nu_e \rightarrow \nu_\tau$~\cite{Donini:2002rm} can be used to reduce the number of clone solutions and better determine $\theta_{13}$ and $\delta$. In this case, the different behavior of the probability curves of different channels should reduce (or ideally eliminate) the impact of the degeneracies on the simultaneous fitting of the two sets of data. The combination of two superbeam facilities, one of them with a sufficiently long baseline and the other with a good $\theta_{13}$ sensitivity, could help to resolve the sgn($\dmatm$) degeneracy~\cite{Minakata:2003ca}. One of these superbeam experiments could be combined  with a reactor detector to determine the $\theta_{23}$ octant~\cite{Huber:2003pm,Minakata:2002jv}. An upgraded version of the NO$\nu$A experiment~\cite{Ayres:2004js} with a second detector off axis at a shorter baseline would also allow the determination of the neutrino mass hierarchy free of degeneracies~\cite{MenaRequejo:2005hn,Mena:2005ri}. Another possibility relies on combining long-baseline and atmospheric neutrino data to solve the $\theta_{23}$ and sgn($\dmatm$) degeneracies~\cite{Huber:2005ep}. These examples reveal the importance of working in the direction of establishing the optimum experimental facilities which reduce or even eliminate the impact of the ambiguities on the determination of the neutrino parameters in future neutrino oscillation experiments.

\subsubsection{Future prospects for leptonic CPV in neutrino oscillation experiments}
\label{sec3.2.4}

Even though neutrino physics has witnessed a series of successes in the last decade, there are still fundamental open questions about neutrinos. Among the ones for which neutrino oscillation experiments will seek an answer are

\begin{itemize}
\item{How large is the $\theta_{13}$ mixing angle?}
\item{Is there CPV in the lepton sector and, if so, what is the value of $\delta$?}
\item{How are neutrino masses ordered: is $\dmatm>0$ (NO) or $\dmatm<0$ (IO)?}
\item{Is the atmospheric neutrino mixing angle $\theta_{23}$ exactly equal to $\pi/4$?}
\item{Are there subdominant nonstandard interactions in the lepton sector?}
\end{itemize}

From the theoretical perspective, a better knowledge of the oscillation parameters could give some hints about the origin of flavor in the lepton sector and, perhaps, on the neutrino mass generation mechanism. With this goal in mind, the major challenge for the upcoming neutrino oscillation experiments will be to probe for subleading effects in neutrino oscillations. In the last years, there has been an intense activity towards finding the optimal experimental conditions and configurations that will allow one to answer the above questions.

It is beyond the scope of this review to give an exhaustive discussion of the physics reach of all future experiments. Instead, we aim at presenting a brief overview of the sensitivities and prospects in the measurement of $\theta_{13}$, $\delta$, and sgn$(\dmatm)$, in future neutrino oscillation facilities. For further details, we refer the reader to other works exclusively dedicated to the subject~\cite{Apollonio:2002en,Bandyopadhyay:2007kx,Bernabeu:2010rz,Nunokawa:2007qh,Mezzetto:2010zi}.

\paragraph{Upcoming reactor neutrino and superbeam experiments}
\noindent\\

Reactor neutrino experiments observe the disappearance of $\bar{\nu}_e$ antineutrinos produced in nuclear fission reactions in the core of a nuclear reactor. The neutrinos are detected through  the inverse beta decay reaction $\bar{\nu}_e+p\rightarrow e^+ + n$ with an energy threshold of approximately 1.8~MeV. Low-baseline reactor neutrino experiments like G\"{o}sgen~\cite{Zacek:1986cu}, Bugey~\cite{Declais:1994su}, Palo Verde~\cite{Boehm:2001ik}, and CHOOZ~\cite{Apollonio:2002gd} searched for $\bar{\nu}_e$ disappearance without success\footnote{Recently, the improved predictions of the reactor antineutrino fluxes show that these experiments may have observed less neutrinos than expected~\cite{Mueller:2011nm,Mention:2011rk}.}. In the case that the detector is placed at a distance $L \sim 100$~km, the experiment becomes sensitive to the solar-neutrino oscillation parameters $\dmsol$ and $\theta_{12}$. The ongoing KamLAND experiment in Japan uses a 1~kton liquid-scintillator detector to measure the flux of $\bar{\nu}_e$ coming from a complex of 53 surrounding nuclear plants located at an average distance $L \sim 180$~km. The KamLAND data indicated a $\bar{\nu}_e$ disappearance, in agreement with the large mixing angle solution of the solar-neutrino data~\cite{Eguchi:2002dm}.

Upcoming reactor neutrino experiments like Double CHOOZ in France~\cite{Ardellier:2006mn}, Daya Bay in Japan~\cite{Guo:2007ug}, and RENO in Korea~\cite{Ahn:2010vy} will have a typical baseline $L \sim 1$~km and therefore they will be looking for $\bar{\nu}_e$ disappearance driven by $\dmatm$ and the small mixing angle $\theta_{13}$. Consequently, the observation of a neutrino deficit in these experiments could be an indication for a nonzero $\theta_{13}$.  To increase the $\theta_{13}$ sensitivity, all these experiments will operate as multidetector setups. Double CHOOZ will be able to measure $\sin^2 2\theta_{13}$ down to 0.03, while Daya Bay and RENO aim at a sensitivity of $\sin^2 2\theta_{13}\sim 0.01$. Double CHOOZ has started to take data with one detector at the end of 2010 and it is expected to start operating with its two detectors by the middle of 2012. Daya Bay is currently under construction and full data taking is planned to start in 2012, while RENO has recently started its physics program.

In superbeam experiments, an intense proton beam is directed to a target, producing pions and kaons which subsequently decay into neutrinos. The resulting neutrino beam consists mainly of $\nu_\mu$ with a small $\nu_e$ component. Because of the increased statistics, the precision of the leading atmospheric neutrino parameters is improved and the sensitivity to $\theta_{13}$ may become comparable (or slightly better) to that of reactor neutrino experiments after a long running period. Moreover, under some circumstances, superbeam facilities may be able to provide some information regarding CP violation and the type of neutrino mass spectrum. The presence of $\nu_e$ in the original beam, which cannot be distinguished from the ones coming from the appearance process $\nu_\mu \rightarrow \nu_e$, is the main limitation of this kind of experiment. There are presently two superbeam experiments, namely, the ``NuMI'' (neutrinos at the main injector) off axis $\nu_e$ appearance experiment (NO$\nu$A) in the United States~\cite{Ayres:2004js} which is still under construction, and the ``Tokai to Kamioka'' (T2K) experiment in Japan~\cite{Itow:2001ee}.  In NO$\nu$A, the neutrino beam is provided by the NuMI Fermilab facility and its far detector is planned to be located at a distance of 812~km. For T2K, the neutrino beam is produced at the Japan Research Complex (J-PARC), and the far detector (the Super-Kamiokande one) is located at a distance of 295~km. In order to reduce the systematic uncertainties, both experiments will have near detectors dedicated to study the unoscillated neutrinos.

\begin{figure}[!t]
\begin{tabular}{c}
 \includegraphics[width=7.2cm]{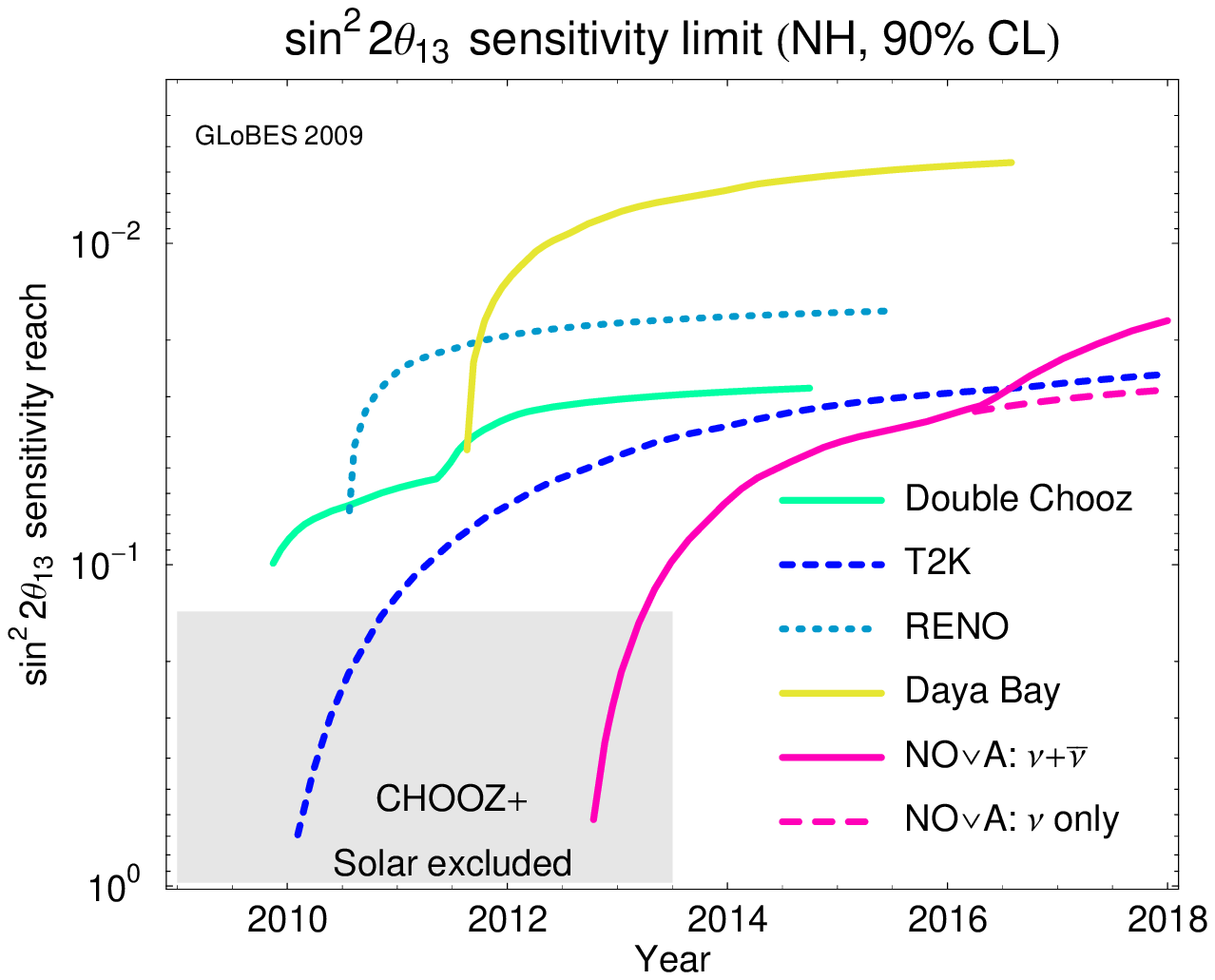} \\
 \includegraphics[width=7.2cm]{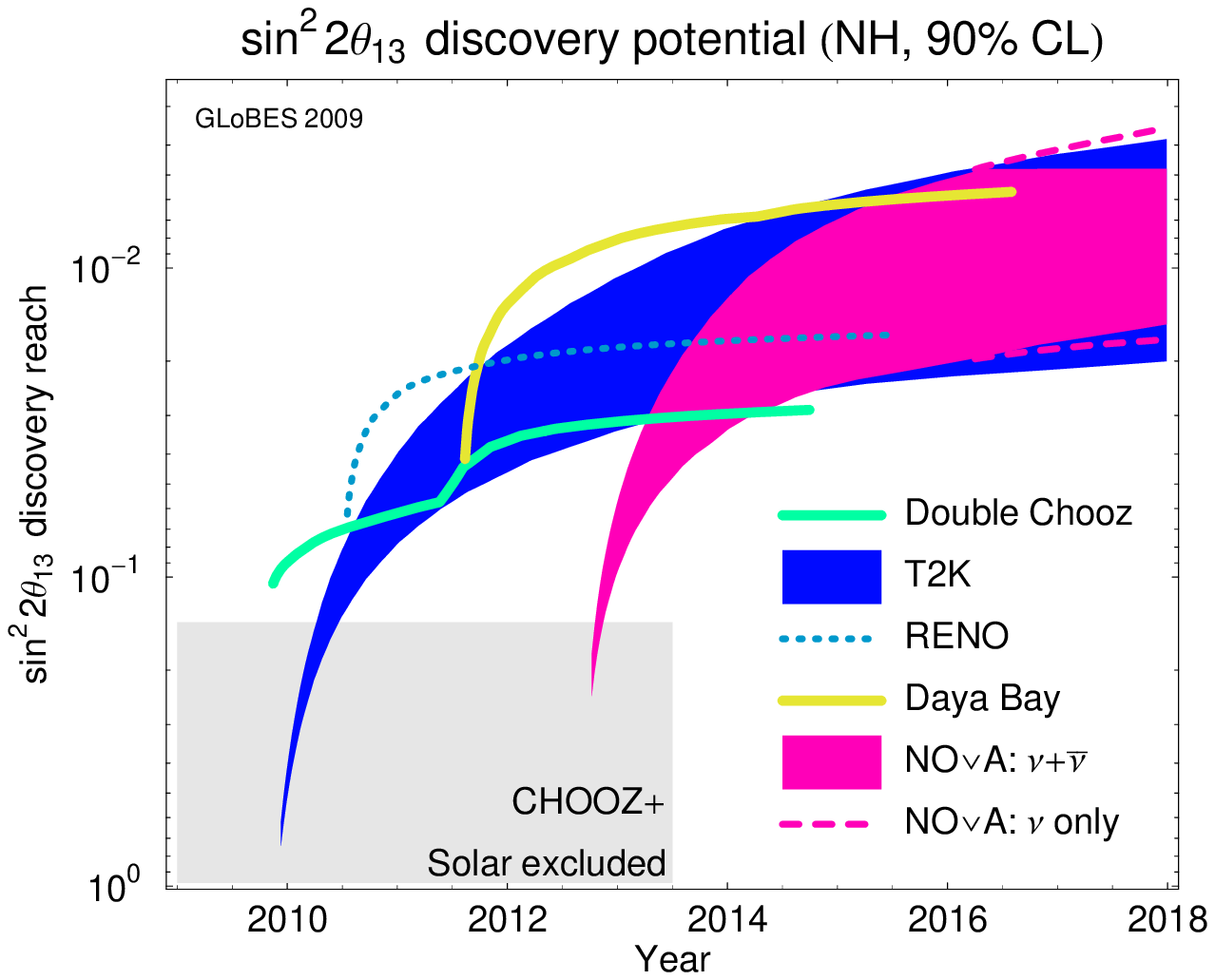} \\
 \includegraphics[width=7.2cm]{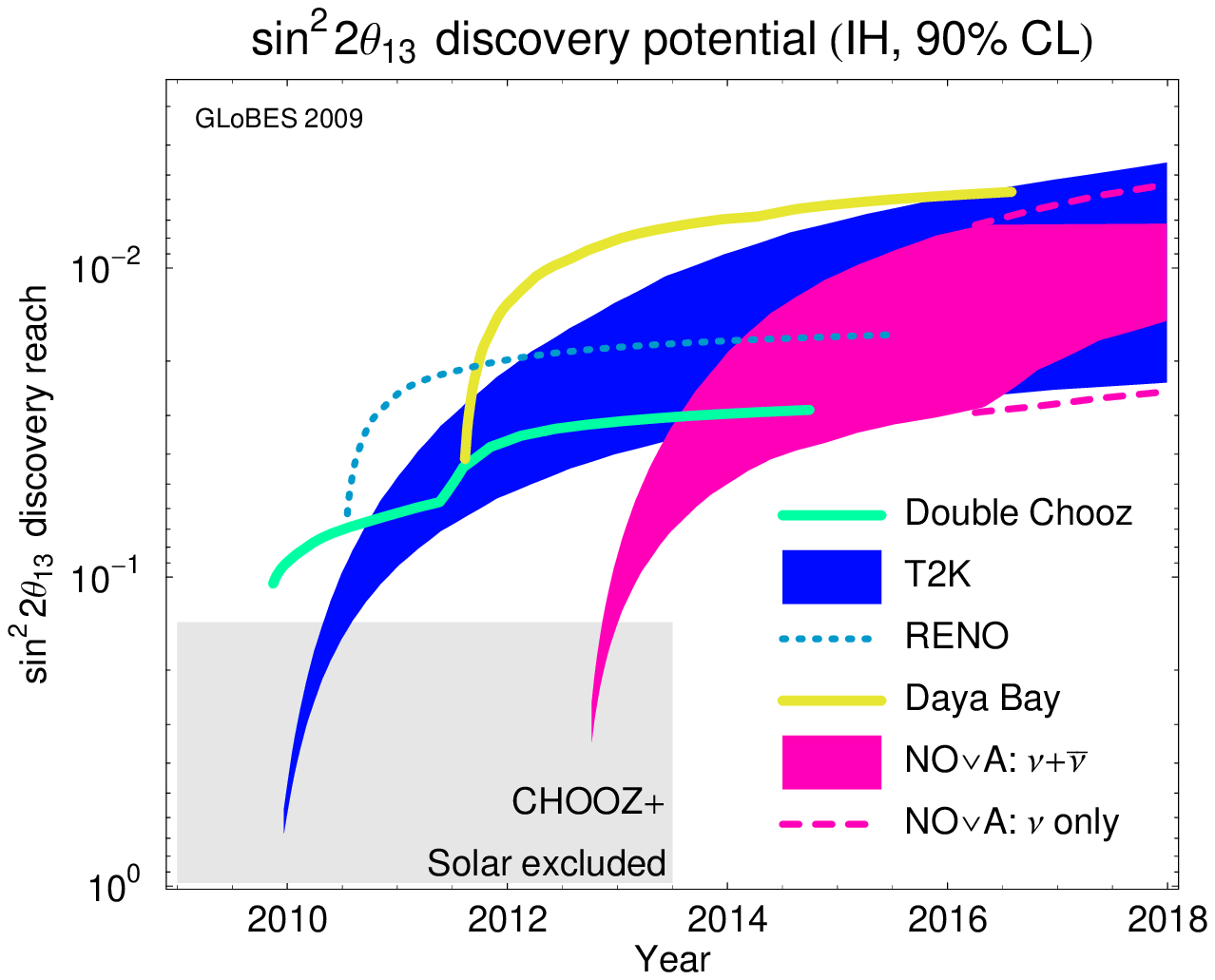}
\end{tabular}
\caption{Top: Time evolution of the sensitivity of $\theta_{13}$ at 90 $\%$ C.L., defined as the limit which is obtained if the true value of $\theta_{13}$ is zero. Center and bottom: Discovery potential of $\theta_{13}$ at 3$\sigma$ as a function of time (given as the smallest value of $\theta_{13}$, which can be distinguished from zero), for a NH (center) and an IH (bottom) neutrino mass spectrum. From~\cite{Huber:2009cw}.}
\label{t13discov}
\end{figure}

The next round of reactor (Double CHOOZ, Daya Bay, and RENO) and accelerator (NO$\nu$A and T2K) neutrino experiments are mainly targeted to the measurement of the neutrino mixing angle $\theta_{13}$, which, if large, could also be on the reach of MINOS and OPERA.  However, it is also interesting to investigate how sensitive these experiments are to CPV and the neutrino mass hierarchy (NMH). This question was recently addressed in~\cite{Huber:2009cw}, where the physics potential of the upcoming reactor and accelerator neutrino oscillation experiments has been analyzed.

In Fig.~\ref{t13discov}, the sensitivity limit and discovery potential of $\theta_{13}$  is given as a function of time for the reactor and superbeam experiments mentioned above. From the top panel of this figure, one concludes that the sensitivity will be dominated by the reactor neutrino experiments and, in particular, by Daya Bay as soon as it becomes operational. The same plot also shows that accelerator experiments are not competitive with the reactor ones.  The discovery potential of $\theta_{13}$ is shown at the center and bottom of the same figure for the NH and IH spectrum, respectively. For the beam experiments, the dependence of the results on the CP phase $\delta$ is reflected by the corresponding shaded regions. Note that there is no dependence on $\delta$ for the reactor experiments since this phase does not appear in the $P_{ee}$ disappearance probability. The comparison of the NH and IH results shows that the discovery potential of $\theta_{13}$ does not depend much on the type of neutrino mass hierarchy. In general grounds, one concludes that we can measure $\theta_{13}$ in the next generation of neutrino experiments, if $\theta_{13}\gtrsim 3$\textdegree.

The analysis of~\cite{Huber:2009cw} shows that NO$\nu$A is required for NMH discovery, due to its long baseline and significant matter effects. If $\sin^2\theta_{13}\simeq 0.1$, the NMH can be established at 90 $\%$ C.L. for about $40\%-50\%$ of all values of $\delta$. Adding other experiments to NO$\nu$A slightly improves the situation in some cases. By themselves, NO$\nu$A and T2K do not have a significant CPV discovery potential. Yet, when combined, these two experiments can be sensitive to CPV for $30\,\%$ of all values of $\delta$, if $\dmatm < 0$. On the other hand, the same two experiments combined have no CPV discovery potential for the NH case~\cite{Huber:2009cw}. Nevertheless, the inclusion of reactor neutrino data significantly improves the situation to a point in which CPV can be established at 90 $\%$ C.L. for about $20\%-30\%$ of all values of $\delta$ if $\sin^2\theta_{13} \gtrsim 0.04$. In conclusion, one can say that the CPV discovery potential in future reactor and superbeam experiments is rather marginal. If $\theta_{13}$ is close to its upper bound, the sensitivity of these setups to CPV and the NMH can be greatly improved with upgraded versions of NO$\nu$A and T2K. In any case, although these experiments may give some indications about the value of $\theta_{13}$, CPV, and the NMH, the confirmation of such hints will require a new generation of experiments like $\beta$ beams or neutrino factories. One should also keep in mind that, even if $\theta_{13}$, CPV, or sgn($\dmatm$) are not measured, the upcoming beam experiments will increase the precision of the atmospheric neutrino parameters through the study of the $\nu_\mu \rightarrow \nu_\mu$ disappearance channel. In particular, deviations from maximal atmospheric mixing can be established at $3\sigma$ for $|\sin^2\theta_{23}-0.5|\gtrsim 0.07$~\cite{Huber:2009cw}.

Recently, the T2K Collaboration reported the results of the first two physics runs (January to June 2010 and November to March 2011)~\cite{Abe:2011sj}. The analysis of the events in the far detector with a single electronlike ring indicates electron-neutrino appearance from a muon-neutrino beam. T2K observed six of such events, which can hardly be explained if $\theta_{13}=0$. Indeed, the probability to observe six or more events for vanishing $\theta_{13}$ is less than 1$\%$. The $90\%$~C.L. interval obtained from the T2K oscillation analysis is $0.03(0.04) < \sin^2 (2\theta_{13}) < 0.28 (0.34)$ with a best fit 0.11(0.14), where the numbers in parenthesis correspond to the results in the case $\dmatm<0$. Further data from T2K and reactor neutrino experiments will help to confirm these results and increase the precision on the determination of $\theta_{13}$. Taking as a reference the best-fit value of the T2K analysis, then we can say that the prospects for determining the NMH and CP violation in the near future are very good.

Examples of second-generation superbeam experiments are the CERN superbeam project~\cite{Mezzetto:2003mm,GomezCadenas:2001eu} based on a super proton linear particle accelerator (SPL), and the upgrade of T2K and T2HK~\cite{Itow:2001ee}. In the former case, the MEMPHYS detector at Fr\'{e}jus in France would detect the CERN SPL neutrinos located at a distance of 130 km. The T2HK beam would be produced at J-PARC in Tokai and sent to the Hyper-Kamiokande detector located at the Kamioka mine, 295 km far from the source. An alternative setup with a second detector placed in Korea (T2KK) at a distance of 1050 km has also been considered~\cite{Ishitsuka:2005qi}. The discovery potential of $\theta_{13}$, CPV, and NMH in those second-generation superbeam experiments has been investigated in~\cite{Campagne:2006yx}.

\begin{figure}[!t]
\begin{tabular}{c}
 \includegraphics[width=7.2cm]{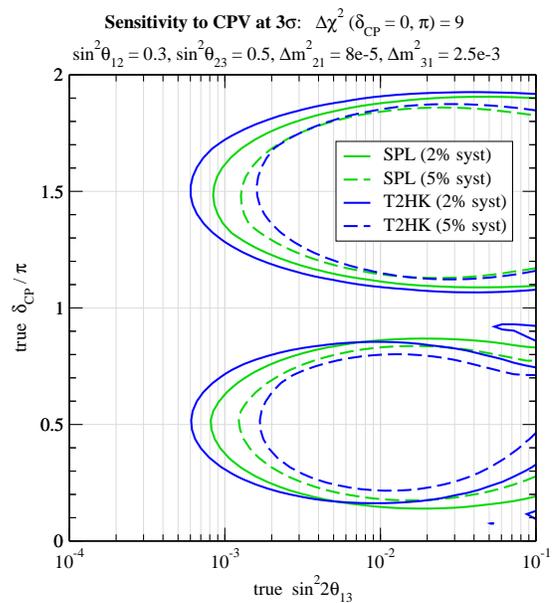}
\end{tabular}
\caption{SPL and T2HK CPV discovery potential. Inside the regions delimited by the solid (dashed) lines, CP conservation can be excluded at the $3\sigma$ level, considering systematic errors of $2\%\,(5\%)$. From~\cite{Campagne:2006yx,Bandyopadhyay:2007kx}.}
\label{SPLT2HK}
\end{figure}

The CPV discovery potential of T2HK and SPL is shown in Fig.~\ref{SPLT2HK}, where the performance of the two experiments is compared. The results show that for maximal leptonic CPV, i.e., for $\delta=\pi/2$ or $3\pi/2$, CPV could be discovered at $3\sigma$ for $\sin^22\theta_{13}\gtrsim 10^{-3}$. Concerning the discovery potential of the mixing angle $\theta_{13}$, the performance of T2HK and SPL is similar, and a measurement down to $\sin^2 2\theta_{13}\simeq 4\times 10^{-3}$ is within their reach for all possible values of $\delta$. Because of the short baseline of the upgraded superbeam experiments, the determination of the NMH at T2HK and SPL is rather limited. The combination of superbeam and $\beta$-beam experiments would also result in an increased $\theta_{13}$ sensitivity. For instance, the 5-year data set of SPL combined with a $\beta$-beam experiment would have a better sensitivity than a 10-year running of T2HK~\cite{Huber:2009cw}. The SPL superbeam combined with a neutrino factory could also help in solving the eightfold degeneracy~\cite{BurguetCastell:2002qx} described in Sec.~\ref{sec.3.2.3}.

\paragraph{${\bm \beta}$-beam experiments}
\noindent\\

One of the main limitations of superbeam experiments is the $\nu_e$ contamination of the initial neutrino beam. A flavor-pure neutrino beam could be obtained using the $\beta$-beam concept~\cite{Zucchelli:2002sa} in which highly boosted $\nu_e$s are obtained from the decay of accelerated unstable ions circulating in a storage ring. Pure electron-neutrino and antineutrino beams can be produced using $^{18}$Ne and $^6$He through the reactions $^{18}{\rm Ne} \rightarrow\, ^{18}{\rm Fe} + e^+ + \nu_e$ and $^{6}{\rm He} \rightarrow\, ^6{\rm Li} + e^- + \bar{\nu}_e$ respectively. The neutrino energy can be accurately set by choosing the required Lorentz factor $\gamma$ of the accelerated mother nuclei. $\beta$-beam experiments aim at studying the $\nu_e \rightarrow \nu_\mu$ and $\bar{\nu}_e \rightarrow \bar{\nu}_\mu$ appearance channels, which can be used to probe $\theta_{13}$ and CP violation. In principle, the $\nu_e \rightarrow \nu_e$ and $\bar{\nu}_e \rightarrow \bar{\nu}_e$ disappearance can also be measured at a $\beta$-beam experiment, although in this case the performance is comparable with the one of reactor neutrino experiments. Although at present there are no concrete $\beta$-beam experiments planned, there has been a great effort to develop this kind of experimental setup~\cite{Lindroos:2010zza}.

A standard low-energy experiment with sub-GeV neutrinos and a baseline of $L=130$~km (distance from CERN to Fr\'{e}jus) has been considered as a possible $\beta$-beam configuration (LE$\beta\beta$)~\cite{Mezzetto:2003ub,Bouchez:2003fy}. Possible candidate isotopes are $^6$He and $^{18}$Ne~\cite{Zucchelli:2002sa} accelerated to a standard Lorentz factor $\gamma_{\rm He,Ne}=100$ at the CERN Super Proton Synchrotron (SPS)~\cite{BurguetCastell:2003vv,Mezzetto:2005ae}. High-energy $\beta$ beams (HE$\beta\beta$) with $E=1-1.5$ GeV and $L \simeq 700$~km (CERN-Canfranc, CERN-Gran Sasso, or Fermilab-Soudan) could also be an alternative. For such cases, the appropriate Lorentz factor $\gamma_{\rm He,Ne}=350$ is achievable at an upgraded SPS or the Tevatron~\cite{BurguetCastell:2003vv}. Alternatively, moderate values of $\gamma \sim 100$ could be appropriate if ions with higher endpoint kinetic energy like $^8$Li or $^8$B are used. Because of its larger baseline, the HE$\beta \beta$ setup would be sensitive to sgn$(\dmatm)$~\cite{Meloni:2008it,Coloma:2007nn,Huber:2005jk,Donini:2005qg,Agarwalla:2006vf}.

\begin{figure}[!t]
\begin{tabular}{c}
\includegraphics[width=7.5cm]{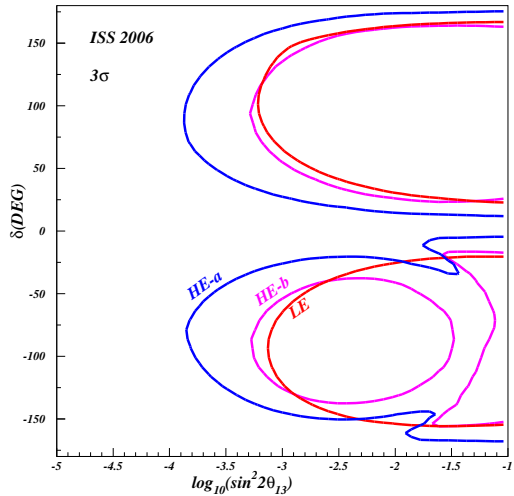}
\end{tabular}
\caption{3$\sigma$ sensitivity to CPV for the three $\beta$-beam configurations: LE$\beta \beta$, HE$\beta \beta$-a and HE$\beta \beta$-b (see text for more details). From~\cite{Bandyopadhyay:2007kx}.}
\label{betab}
\end{figure}

The 3$\sigma$ sensitivity to CPV is shown in Fig.~\ref{betab} for three $\beta$-beam configurations, namely, LE$\beta \beta$ with a 500 Mton water \v{C}erenkov detector, HE$\beta \beta$ with a 500 Mton water \v{C}erenkov detector (HE$\beta \beta$-a), and HE$\beta \beta$ with a liquid-scintillator detector (HE$\beta \beta$-b). From these results, one can see that the HE$\beta \beta$-a provides the best CPV sensitivity, with slightly worse results for negative values of $\delta$ due to the sgn$(\dmatm)$ ambiguity. The potential of these $\beta$-beam setups to sgn$(\dmatm)$ is limited to relatively high values of $\theta_{13}$, namely, $\sin^2\theta_{13}\gtrsim 0.03$. The extraction of $\theta_{13}$ and $\delta$ from the data is also more difficult for the LE$\beta \beta$ setup since the uncertainties are significantly larger and the eightfold degeneracy is present. The situation is improved for the HE$\beta \beta$-a case for which the intrinsic degeneracy is resolved.

The combination of superbeam and $\beta$-beam experiments has also been considered and, in particular, it has been shown that a 5-year run of SPL and $\beta$ beam would result in a better sensitivity to $\theta_{13}$ than 10 years of T2HK~\cite{Huber:2009cw}. Using distinct ions~\cite{Donini:2006dx} with a $\gamma$ reachable at the CERN SPS could also help in resolving the degeneracies due to the different values of $L/\langle E \rangle$.

\paragraph{Electron-capture beams}
\noindent\\

\begin{figure}[!t]
\begin{tabular}{c}
\includegraphics[width=8.5cm]{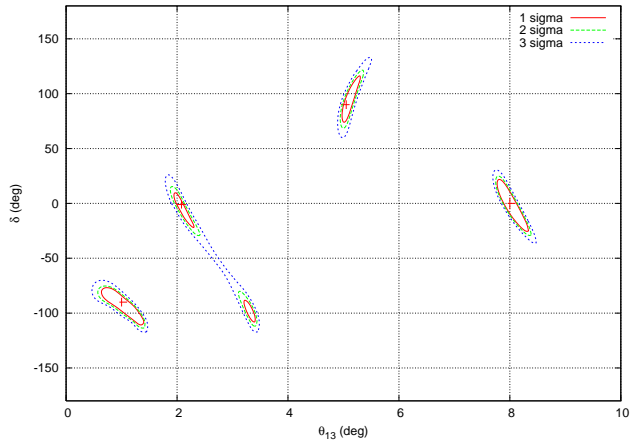}
\end{tabular}
\caption{Fits for $\theta_{13}$ and $\delta$ in an electron-capture experiment with a baseline $L=130$~km (CERN-Fr\'{e}jus) and a 440 kton water \v{C}erenkov detector. Two 5 yr-running periods with $\gamma=195$ and $\gamma=90$ have been considered. From~\cite{Bernabeu:2005jh}.}
\label{ecexp}
\end{figure}

In these experiments, neutrinos are obtained from electron-capture processes~\cite{Bernabeu:2005jh,Sato:2005ma,Bernabeu:2005kq,Orme:2009ak}, in which an atomic electron is captured by a proton of the nucleus leading to a nuclear state of the same mass number $A$. The proton is replaced by a neutron, and an electron neutrino is emitted ($p e^- \rightarrow n \nu_e$) with fixed energy, since this is a two-body decay. Consequently, a flavor-pure and monochromatic neutrino beam can be obtained. The electron-capture beam concept is feasible if the ions decay fast enough. Recent discovery of nuclei far from the stability line having superallowed spin-isospin transitions to Gamow-Teller resonances turn out to be very good candidates. A particular choice is $^{150}$Dy, with a neutrino energy at rest given by 1.4 MeV due to a unique nuclear transition from 100$\%$ electron capture in going to neutrinos. The oscillation channel to study is once more $\nu_e \rightarrow \nu_\mu$, being the prospects for the measurement of $\theta_{13}$ and CP violation quite impressive. Since only a neutrino beam is available, sensitivity to CPV is reached by performing runs at different values of $\gamma$. The attainable precision in such kind of experiments~\cite{Bernabeu:2005jh} is illustrated in Fig.~\ref{ecexp}, where several values for $\theta_{13}$ and $\delta$ have been assumed. The contour lines correspond to the determination of the oscillation parameters at different confidence levels. It has also been shown that the combination of $\beta$ and electron-capture beam experiments using boosted ytterbium could achieve remarkable results in what concerns the determination of the neutrino mass hierarchy, CP violation, and $\theta_{13}$~\cite{Bernabeu:2009np}.

\paragraph{Neutrino factories}
\noindent\\

If $\theta_{13}$ happens to be very small, then its measurement will be only possible at a neutrino factory (NF). This idea was first discussed almost 15 years ago~\cite{Geer:1997iz} and, since then, a great deal of effort has been made in order to plan and optimize the concept. In this type of experiment, muons are accelerated and stored in a storage ring. A boosted and collimated neutrino beam is obtained from the decays of the muons in the straight sections of the ring. Contrary to the $\beta$-beam and electron-capture experiments, at NFs the neutrino beam contains both electron and muon (anti)neutrinos since $\mu^- \rightarrow e^- + \nu_\mu + \bar{\nu}_e$ (or $\mu^+ \rightarrow e^+ + \bar{\nu}_\mu + \nu_e$, if $\mu^+$ are stored). The neutrino beam at a NF can be used to study the leading atmospheric neutrino parameters $\dmatm$ and $\theta_{23}$ through the study of the disappearance channels $\nu_\mu \rightarrow \nu_\mu$ and $\bar{\nu}_\mu \rightarrow \bar{\nu}_\mu$. Nevertheless, the ultimate purpose of a NF is the measurement of subleading effects in the golden appearance channel $\nu_e \rightarrow \nu_\mu$ and its CP conjugated~\cite{Cervera:2000kp}.  The detection of golden channel events requires an effective charge separation of the muons produced in charged-current processes, due to the presence of wrong-sign muons originated from the disappearance channel. This could be achieved with a magnetized iron detector (MIND), which appears as the most straightforward solution for a high-fidelity muon charge measurement. Since the neutrino energy is typically very high (up to 25 or 50 GeV), the detector has to be placed at a distance of several thousand of kilometers in order for oscillations to occur. A very active research and development program is currently undergoing in the framework of the International Design Study for the Neutrino Factory (IDS-NF)~\cite{Bandyopadhyay:2007kx}, to which the reader is referred for more details about the possible NF configurations and performance comparison. Here we limit ourselves to give a general idea about the $\theta_{13}$, CPV, and NMH sensitivities at neutrino factories.

\begin{figure*}[!t]
\begin{tabular}{c}
\includegraphics[scale=0.6]{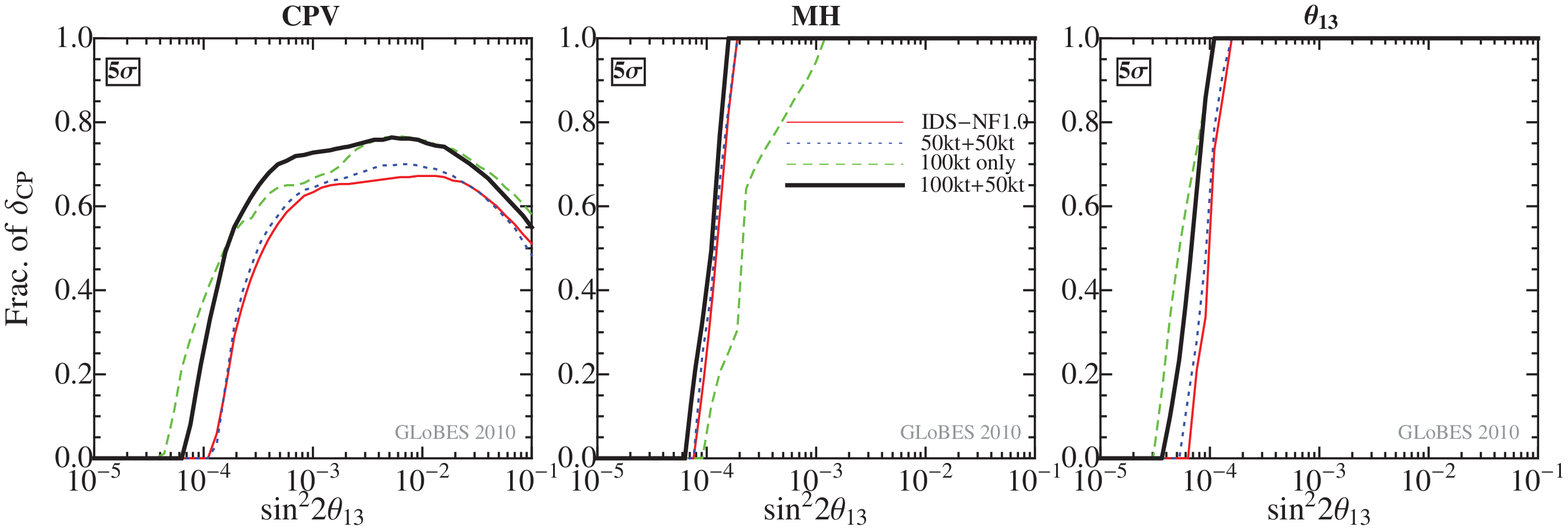}
\end{tabular}
\caption{5$\sigma$ discovery reach of CPV (left), NMH (center) and $\theta_{13}$ (right) for several
NF setups: ``50 kt+50kt'' refers to a combination of two 50 kton MINDs  at
$L_1=4000$~km and $L_2=7500$~km, ``100 kt only'' to a 100~kton MIND at $L=4000$~km,
``100 kt+50 kt'' to a 100~kton MIND at $L_1=4000$~km and a 50~kton MIND at $L_2=7500$~km, and ``IDS-NF 1.0''
to the IDS-NF setup. From~\cite{Agarwalla:2010hk}.}
\label{NFpot}
\end{figure*}

As mentioned, the determination of $\theta_{13}$ and $\delta$ at a NF suffers from several ambiguities. A possible solution to this problem is to combine golden measurements at different baselines or, if an efficient $\tau$ detector is available, to use the silver $\nu_e \rightarrow \nu_\tau$ oscillation channel~\cite{Donini:2002rm,Autiero:2003fu}. The original IDS-NF setup considers a double-baseline NF with $L_1\simeq 3000-5000$~km, $L_2\simeq L_{\rm magic }\simeq 7500$~km, and a muon energy $E_\mu=25$~GeV. Such a standard configuration is advantageous for several reasons: the sensitivity to very small values of $\theta_{13}$ and thus to several three-flavor effects~\cite{Huber:2006wb}, and the robustness against new physics effects like nonstandard interactions in the lepton sector~\cite{Kopp:2008ds} and systematic errors~\cite{Tang:2009na}. An alternative setup with a lower muon energy $E_\mu=5$ GeV, a totally active scintillator detector (TASD) and a baseline of $L\simeq 1300$~km has also been considered as a possible low-energy neutrino factory (LENF) configuration~\cite{Geer:2007kn,Bross:2007ts,FernandezMartinez:2010zza,Tang:2009wp}. This kind of alternative
is particularly suitable for large $\sin^2 2\theta_{13}$.

Since it is unlikely that the accelerator part of a NF will be specially built for this experiment, one has to assume that the neutrino beam will be produced at existing facilities. In such a case, the options are CERN, J-PARC, the Rutherford Appleton Laboratory (RAL) and the Fermi National Accelerator Laboratory (FNAL)~\cite{Apollonio:2008xx}. As for the possible detector locations, a list of candidate sites in the United States~\cite{USAlist} and Europe~\cite{Rubbia:2010zz} has been recently compiled. In Asia, possible detector sites are the Kamioka mine in Japan, the proposed Chinese underground laboratory at CPJL, YangYang in Korea, and the India-based Neutrino Observatory (INO) in India. The possibility of a green-field scenario in which neither the baseline nor the muon energy are constrained has also been considered in NF optimization studies~\cite{Huber:2006wb,Bueno:2001jd,Agarwalla:2010hk}.

As a representative analysis, we show in Fig.~\ref{NFpot}~\cite{Agarwalla:2010hk} the CPV, $\theta_{13}$, and NMH discovery potential for several NF setups. The results indicate that the $\theta_{13}$ sensitivity is comparable for all the cases considered, namely, $\sin^2 2\theta_{13}$ will be measurable at neutrino factories down to $\sim 10^{-4}$, corresponding to $\theta_{13}\sim 0.3$\textdegree. The ``100 kt+50 kt'' setup is the one which performs better on the CPV discovery potential, while the NMH sensitivity is comparable to the one of the remaining two double-baseline options (see the figure caption for more details on the curve labels). The single-baseline configuration ``100 kt only'' has a rather worse NMH discovery reach than the other setups. In general, one can say that for $\sin^22\theta_{13}\gtrsim 10^{-2}$ a LENF is quite effective. On the other hand, a double-baseline high-energy NF will be necessary for smaller values of $\theta_{13}$. As mentioned, the next generation of reactor and superbeam experiments will be able to tell us if $\sin^2 2\theta_{13}\gtrsim 10^{-2}$, allowing for an optimization of a large $\theta_{13}$ scenario at neutrino factories.

To conclude, one can say that the $\theta_{13}$, CPV and NMH discovery potential of future experiments depends mainly on the true value of $\theta_{13}$. If $\sin^22\theta_{13}\gtrsim 10^{-2}$, then the discovery potential of all the above considered experiments is comparable (although NFs will be able to perform more precise measurements). If the value of $\sin^22\theta_{13}$ is in the intermediate range $5\times 10^{-4}\lesssim \sin^22\theta_{13} \lesssim 10^{-2}$, only $\beta$-beam experiments and neutrino factories will be able to probe on CPV and the NMH. In the worst case, in which $\sin^22\theta_{13} \lesssim5\times 10^{-4}$, neutrino factories seem to be the only hope to establish leptonic CP violation and identify the neutrino mass hierarchy. However, since the recent T2K and MINOS data indicate that $\sin^22\theta_{13}$ is not so small, most probably we will not have to wait for neutrino factories to discover LCPV and find out whether the neutrino mass spectrum is normal or inverted.

We conclude this section with a comment on the potential of measuring $\theta_{13}$ and NMH from supernova (SN) neutrinos. The time-dependent energy spectra of $\nu_e$ and $\bar{\nu}_e$ from a future SN can be valuable to obtain information on the neutrino mass and mixing pattern~\cite{Dighe:1999bi}. In fact, identifying the neutrino mass hierarchy is possible for $\theta_{13}$ as small as $10^{-10}$~\cite{Dasgupta:2008my}. For such small values of $\theta_{13}$, the sensitivity of supernova neutrino oscillations to the mass hierarchy stems from collective neutrino oscillations that take place near the supernova core. Therefore, a future galactic SN may become extremely important for the understanding of neutrino mixing and SN astrophysics. Of course, the occurrence of a SN is a rare happening, and to take the most from SN neutrinos one must be prepared with the best detectors.

\subsection{Neutrinoless double beta decay}
\label{sec3.3}

An important process which may unveil crucial aspects about the fundamental nature of neutrinos is neutrinoless double beta decay ($\ndbd$)~\cite{Avignone:2007fu,Tomoda:1990rs,Vergados:2002pv}, where even-even nuclei undergo the transition $(A,Z)\rightarrow (A,Z+2)+2e^-$. This process obviously violates lepton number by two units and therefore the mechanism responsible for $\ndbd$ can also induce Majorana neutrino masses. In short, the observation of $\ndbd$ implies that neutrinos are Majorana particles~\cite{Schechter:1981bd}. Several scenarios beyond the SM predict the occurrence of $\ndbd$ decay like, for instance, supersymmetric theories that violate lepton number and/or R parity~\cite{Mohapatra:1986su,Hirsch:1997dm,Hirsch:1995zi}. The $\ndbd$-decay width is usually factorized as $\Gamma_{\ndbd}=G_{\rm kin} |\mathcal{M}_{0\nu}|^2 F_{\rm part}$, where $G_{\rm kin}$ is a known phase space factor, $\mathcal{M}_{0\nu}$ is the nuclear matrix element (NME), and $F_{\rm part}$ encodes the particle physics part of the process. In the simplest case, when $\ndbd$ is driven by light Majorana neutrino exchange, $F_{\rm part} \propto m_{ee}^2$, where $m_{ee}$ is an effective electron-neutrino mass simply given by $m_{ee}=|(\bmm_\nu)_{11}|$ [see, e.g.,~\cite{Bilenky:2010zz,Rodejohann:2011mu}].

Several experiments have been searching for $\ndbd$ using different nuclei. Up to now, no indications in favor of this process have been obtained, although some members of the Heidelberg-Moscow collaboration claim to have observed $\ndbd$ with a lifetime which corresponds to $m_{ee} \simeq 0.4~\eV$~\cite{KlapdorKleingrothaus:2006ff}. This result will be soon checked by an independent experiment. From the most precise $\ndbd$ experiments, the upper bounds
\begin{eqnarray}
 \label{ndbdlimit}
m_{ee}&<&(0.20-0.32)\,\eV,\quad {\rm Heidelberg-Moscow\,(^{76}Ge)}\nonumber\\
&<& (0.30-0.71)\,\eV,\quad {\rm CUORICINO\, (^{130}Te)}\nonumber\\
&<& (0.50-0.96)\,\eV,\quad {\rm NEMO \,(^{130}Mo)}\,
\end{eqnarray}
obtained by the Heidelberg-Moscow~\cite{Baudis:1999xd}, CUORICINO~\cite{Andreotti:2010vj} and NEMO~\cite{Arnold:2005rz} Collaborations, have been inferred. In the future, $\ndbd$ experiments like GERDA~\cite{Jochum:2010zz}, CUORE~\cite{Andreotti:2010vj}, EXO~\cite{Gornea:2010zz}, MAJORANA~\cite{Gehman:2008zz}, SuperNEMO~\cite{Arnold:2010tu}, SNO+~\cite{Kraus:2010zzb}, KamLAND-ZEN~\cite{Terashima:2008zz}, and others, will be able to probe the value of $m_{ee}$ down to a few $10^{-2}$~eV.

If the dominant contribution to $\ndbd$ is due to the exchange of light active Majorana neutrinos, then $m_{ee}$ depends exclusively on neutrino mass and mixing parameters which enter the definition of the neutrino mass matrix $\bmm_\nu$. Using the parametrization for the leptonic mixing matrix $\bU$ given in Eq.~\eqref{Uparam2}, one has
\begin{equation}
m_{ee}=|\,c_{13}^2\,(m_1c_{12}^2+m_2e^{-i\alpha_1}s_{12}^2)+m_3e^{-i\alpha_2}s_{13}^2\,|\,.
\label{mee}
\end{equation}
This shows that the relation between the particle physics part of ${\ndbd}$ decay and neutrino masses and mixing is direct in the sense that $m_{ee}$ depends on parameters which define the neutrino mass matrix. Therefore, the observation of $\ndbd$ decay can in principle provide valuable information about the type of neutrino mass spectrum~\cite{Pascoli:2002xq,Bilenky:2001rz,Murayama:2003ci}, the absolute neutrino mass scale~\cite{Joaquim:2003pn,Pascoli:2001by,Matsuda:2000iw,Choubey:2005rq}, and the Majorana CP-violating phases~\cite{Pascoli:2005zb,Barger:1999na,Czakon:2000vz,Branco:2002ie}.

The presently available neutrino oscillation data already impose some constraints on the value of $m_{ee}$. In the case of a hierarchical neutrino mass spectrum ($m_1 \ll m_2 \simeq \sqrt{\dmsol}\ll m_3\simeq \sqrt{\dmatm}$) one has
\begin{equation}
 \label{meeHI}
\frac{m_{ee}^{\rm HI}}{\sqrt{\dmatm}}\simeq \left| s_{13}^4+r\, c_{13}^4
s_{12}^4 + \frac{1}{2}
\sqrt{r}\,s_{12}^2 \,\cos\alpha \sin^2(2\theta_{13})\right|^{\frac{1}{2}}\,,
\end{equation}
where $\alpha$ is a Majorana-phase difference. If $\alpha=\pi$, cancellations in
$m_{ee}^{\rm HI}$ may occur for
\begin{equation}
\label{s13canc}
s_{13}^2 = \frac{r s_{12}^2}{1+r s_{12}^2}\sim 0.01\,,
\end{equation}
where in the numerical estimate we have used the STV best-fit values for the neutrino parameters given in Table~\ref{Tabnudata}. Such values of $s_{13}^2$ are close to the best-fit points shown in Table~\ref{Tabnudata}, and will be probed by future neutrino experiments as discussed in Sec.~\ref{sec3.2.4}.

In the case of an IH neutrino mass spectrum, the effective neutrino mass parameter is simply given by
\begin{equation}
 \label{meeIH}
m_{ee}^{\rm IH}\simeq\sqrt{\dmatm}
\sqrt{1-\sin^2(2\theta_{12})\sin^2\frac{\alpha}{2}}\,.
\end{equation}
It is straightforward to conclude that $m_{ee}^{\rm IH}$ is constrained to the range
\begin{equation}
 \label{meeIHlim}
\sqrt{|\dmatm|}\,(1-2s_{12}^2) \lesssim m_{ee}^{\rm IH}\lesssim \sqrt{|\dmatm|}\,,
\end{equation}
which, taking into account the $3\sigma$ allowed ranges for the neutrino parameters given by the STV global analysis (Table~\ref{Tabnudata}), leads to
\begin{equation}
 \label{meeIHrang}
0.013 \lesssim m_{ee}^{\rm IH}\lesssim
0.05\,.
\end{equation}

Therefore, near future $\ndbd$ decay experiments will be able to test the IH neutrino mass spectrum when this process is dominated by neutrino exchange.

\begin{figure}[t]
\begin{tabular}{cc}
\includegraphics[width=8.3cm]{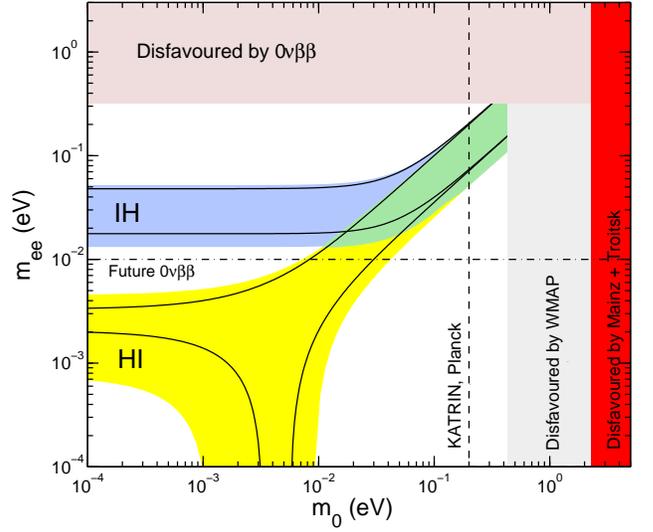}
\end{tabular}
\caption{Dependence of $m_{ee}$ on the lightest neutrino mass $m_0$ for a normal (HI) and inverted (IH) neutrino mass spectrum. The yellow (light blue) region corresponds to $3\sigma$ intervals of the STV for a normal (inverted) neutrino mass spectrum. The Majorana phases $\alpha_{1,2}$ are varied in the interval $[0,2\pi]$. The regions disfavored by kinematical searches and cosmology are delimited by the vertical shaded bands. The future  Katrin and Planck satellite sensitivities are indicated by the vertical dashed line. The horizontal purple band refers to the $m_{ee}$ region disfavored by the $\ndbd$ Heidelberg-Moscow experiment. In turn, the dash-dotted horizontal line at $m_{ee}=0.01~\eV$ illustrates the sensitivity of future $\ndbd$ experiments. The $m_{ee}$ allowed region delimited by the solid black lines is obtained when the best-fit values of the STV global neutrino data analysis are considered (see Table~\ref{Tabnudata}).}
\label{ndbd}
\end{figure}

In Fig.~\ref{ndbd}, we show the dependence of $m_{ee}$ on the lightest neutrino mass $m_0$ for both types of neutrino mass spectra, i.e., normal and inverted hierarchy~\cite{Vissani:1999tu}. The ranges of $m_0$ disfavored by kinematical neutrino mass searches (Mainz and Troitsk) and by cosmology are also shown (see the discussion at the end of Sec.~\ref{sec3}). The $m_{ee}$ allowed region is shown in yellow (light blue) for a normal (inverted) neutrino mass spectrum, taking the $3\sigma$ STV neutrino data of Table~\ref{Tabnudata}, and varying the Majorana phases in the range $[0,2\pi]$. These two regions overlap for $m_0 \gg \dmatm$, where neutrinos are quasidegenerate. The same regions would be delimited by the solid black lines if the best-fit values are considered. In this particular case, one can see that, even if the neutrino mixing angles are fixed, the Majorana phases have a strong impact on $m_{ee}$. It is also clear from this figure that the nonobservation of $\ndbd$ in future experiments sensitive to $m_{ee}$ down to $0.01~\eV$ would exclude the IH and QD neutrino mass spectra. One should, however, keep in mind that the latter conclusion is valid under the assumption that the only contribution to $\ndbd$ is the one mediated by the exchange of light active neutrinos.

If $\ndbd$ decay is observed by future experiments, then one would a priori expect to learn something about Majorana-type CP violation in the lepton sector. In particular, a question which has been often addressed in the literature is whether one can extract the value of the phases $\alpha_{1}$ and $\alpha_{2}$ from a measurement of the $\ndbd$ lifetime of a nucleus. Although this may seem an easy task from the mathematical point of view, the truth is that such a Majorana-phase determination is plagued by uncertainties in the determination of the NMEs $\mathcal{M}_{0\nu}$. Indeed, the computation of these quantities is a highly nontrivial many body problem~\cite{Menendez:2008jp}. It has also been claimed that CP violation is not detectable via $\ndbd$~\cite{Barger:2002vy}. The argument they present is based on the fact that if one considers $x$ as being the sum of the uncertainty in the NME calculation and the experimental error, then the necessary condition for the discovery of CP violation requires that
\begin{equation}
\label{nogo}
\sin^2(2\theta_{12}) > 1-\left(\frac{1-x}{1+x}\right)^2\,.
\end{equation}
Taking the best-fit value of $\sin^2\theta_{12}$, one has $x < 0.46$. This is far beyond what seems reasonable to consider in view of the difficulties in calculating the NME, which presently suffers from an uncertainty factor of 2-3. More refined numerical studies confirmed the above general conclusion that, most probably, Majorana CP violation cannot be established in the near future $\ndbd$ experiments. This could not be the case if the errors in the determination of $m_{ee}$ and the sum of neutrino masses would not exceed 10$\%$. In addition, the corresponding NME should be known within a factor of 1.5~\cite{Pascoli:2005zb}, which seems to be a challenging target to reach.

Note that, although $\ndbd$ decay depends on the Majorana phases $\alpha_{1,2}$, there is no distinction between the  $\ndbd$ rate of a nucleus and that of the corresponding antinucleus. In other words, $\ndbd$ processes do not manifestly exhibit the violation of CP. Still, processes like neutrino $\leftrightarrow$ antineutrino oscillation and rare leptonic decays of $K$ and $B$ mesons (e.g., $ K^{\pm} \rightarrow \pi^{\mp} l^{\pm} l^{\pm}$ and similar modes for the $B$ meson) can actually be sensitive to Majorana-type CPV~\cite{deGouvea:2002gf}.

\subsection{Lepton flavor violation and seesaw neutrino masses}
\label{sec3.4}

In the quark sector, the only source of flavor and CP violation is the CKM mixing matrix. A large number of observables, mainly involving $K$ and $B$ meson sectors, have been crucial to constrain the mixing angles and the CP-violating phase of this matrix, and to test the consistency of the CKM framework. In general, if there is new physics beyond the standard model (BSM), new sources of flavor and CP violation are present. Their contributions to flavor and CP-violating processes may induce deviations from the SM predictions. The situation in the lepton sector is very different, since the only experimental evidence for flavor violation comes from neutrino oscillations, which require the existence of a nontrivial lepton mixing matrix $\bU$, which is the analog of the CKM matrix for leptons. This mixing matrix leads to LFV processes like, for instance, radiative charged-lepton decays $l_i \rightarrow l_j \gamma$~\cite{Cheng:1976uq,Petcov:1976ff,Marciano:1977wx}. Moreover, if CP is violated in the lepton sector, charged-lepton electric dipole moments get also a nonzero contribution~\cite{Ng:1995cs}. However, due to the smallness of the neutrino masses, the corresponding observables are negligibly small and unaccessible to experiments.

The observation of any lepton flavor-violating process other than neutrino oscillations or the measurement of charged-lepton electric dipole moments would then be a direct signature of new physics. This is in clear contrast with what happens in the quark sector, in which new physics effects are subdominant to the SM ones. Up to now, none of these LFV processes have been observed and therefore only upper bounds on their rates are available. The present experimental limits for several charged-lepton LFV decays are shown in Table~\ref{Tabrare}. Several experiments aim at improving these bounds in the near future, namely, the MEG Collaboration plans to reach a sensitivity of ${\rm BR}(\mu \rightarrow e \gamma)\sim 10^{-13}$~\cite{Cavoto:2010mc} until the end of 2012, while a Super B factory would be able to probe LFV $\tau$ decays to a level of $10^{-9}$. As for $\mu\rightarrow 3e$, the rather optimistic projected sensitivity is around $10^{-14}$~\cite{Aysto:2001zs}, while $\mu$-$e$ conversion in titanium could be tested at $10^{-18}$ by the J-PARC experiment PRISM/PRIME~\cite{Yoshimura:2003ai}.

If small neutrino masses are the only source of LFV, then the branching ratios (BR) for the radiative LFV charged-lepton decays are simply given by
\begin{align}
\frac{{\rm BR}(l_i \rightarrow l_j \gamma)}
{{\rm BR}(l_i \rightarrow l_j \bar{\nu}_j \nu_i)}&=\frac{3\alpha}{32\pi}
\left|\sum_{k=2,3}\bU_{ik}^\ast \bU_{jk} \frac{\Delta m^2_{k1}}{m_W^2}
\right|^2 \nonumber\\
& \lesssim  \frac{3\alpha}{32\pi} \left|\frac{\dmatm}{m_W^2}\right|^2 \sim
\mathcal{O}(10^{-53})\,,
\label{ljlinu}
\end{align}
where the unitarity of $\bU$ and the present value for $|\dmatm|$ have been taken into account for the numerical estimate. The above result shows that, if neutrino masses are added to the SM in order to explain the neutrino oscillation data, the rates of LFV processes turn out to be far beyond the sensitivity reach of future experiments. This is due to an extremely strong Glashow-Iliopoulos-Maiani (GIM) suppression mechanism~\cite{Glashow:1970gm} in the lepton sector. Therefore, it is of extreme importance to explore BSM scenarios where this suppression is somehow alleviated.

Particularly interesting scenarios in which LFV is enhanced to observable levels are those when the new LFV sources are in some way related to those responsible for neutrino masses and mixing. For instance, if neutrino masses arise through the seesaw mechanism, then the seesaw mediators may induce LFV at either tree or one-loop level by participating directly in the decays. In such cases, the masses of these new states are required to be not too far from the electroweak scale. In the case of the type I seesaw (see Sec.~\ref{sec2.5}), the flavor dependence of the one-loop amplitudes of the processes $l_i \rightarrow l_j \gamma$ is roughly encoded in the coefficients $F_{ij}=(\bY^{\nu\dag} \mathbf{d}_M^{-2} \bY^\nu)_{ij}$, where $\bY^\nu$  is the Dirac-neutrino Yukawa coupling matrix and $\mathbf{d}_M={\rm diag}(M_1,M_2,M_3)$; $M_i$ are the heavy Majorana neutrino masses. Instead, it follows from Eq.~\eqref{mnutypeIseesaw} that the effective neutrino mass matrix is proportional to the combination $\bY^{\nu} \mathbf{d}_M^{-1}\bY^{\nu T}$. From this simple (but effective) argument, one can see that there is no direct model-independent way of relating the neutrino data with LFV searches in this simple framework. This is mainly due to the fact that one cannot reconstruct the couplings $\bY^\nu$ and masses $M_i$, even if we know the effective neutrino mass matrix.

The situation is somehow different in the type II seesaw mechanism in which neutrino masses are generated by the tree-level exchange of scalar triplets. In this case, $l_i \rightarrow l_j \gamma$ is induced at one loop~\cite{Bilenky:1987ty,Mohapatra:1992uu,Pich:1985uv}, while three-body charged-lepton LFV decays appear already at tree level~\cite{Pal:1983bf,Barger:1982cy}. The BRs for both cases are given by
\begin{align}
\label{LFVlig}
\frac{{\rm BR}(l_i \rightarrow l_j \gamma)}
{{\rm BR}(l_i \rightarrow l_j \bar{\nu}_j \nu_i)}&=\frac{25\alpha}{768\,G_F^2\pi}
\frac{\left|(\bY^{\Delta\dag}\bY^\Delta)_{ij}\right|^2}{M_\Delta^4}\,,\nonumber\\
\frac{{\rm BR}(l_i^{-} \rightarrow l_j^{+} l_k^{-} l_m^{-})}
{{\rm BR}(l_i \rightarrow l_j \bar{\nu}_i \nu_j)}&=(1+\delta_{km})
\frac{|\bY^\Delta_{ij}|^2|\bY^\Delta_{km}|^2}{G_F^2 M_\Delta^4}\,.
\end{align}

Taking into account the bounds in Table~\ref{Tabrare}, one can use the above expressions to constrain combinations of the couplings $\bY^\Delta$, namely,
\begin{align}
&\left|(\bY^{\Delta\dag}\bY^\Delta)_{ij}\right| \simeq  1.9 \times 10^3 \Bigl(\frac{M_\Delta}
{1\,{\rm TeV}}\Bigr)^2 \sqrt{\frac{{\rm BR}(l_i \rightarrow l_j \gamma)}
{{\rm BR}(l_i \rightarrow l_j \bar{\nu}_j \nu_i)}}
\,,\nonumber\\
&|\bY^\Delta_{ij}||\bY^\Delta_{km}|\simeq \frac{16.6}{\sqrt{1+\delta_{km}}} \Bigl(\frac{M_\Delta}
{1\,{\rm TeV}}\Bigr)^2\sqrt{\frac{{\rm BR}(l_i^{-} \rightarrow l_j^{+} l_k^{-} l_m^{-})}
{{\rm BR}(l_i \rightarrow l_j \bar{\nu}_i \nu_j)}}\,.\nonumber\\
\label{limits}
\end{align}

\begin{table}
\caption{\label{Tabrare}Present upper bounds for the branching ratios of flavor-violating charged-lepton decays $l_j\rightarrow l_i \gamma$ and $l_i\rightarrow l_j l_k l_k$ ($j,k\neq i$) and the $\mu$-$e$ conversion rate in titanium (Ti).}
\begin{ruledtabular}
\begin{tabular}{llll}
$\mu \rightarrow e \gamma$ & $2.4\times 10^{-12}$ &\cite{Adam:2011ch} &\\
$\tau \rightarrow \mu \gamma$ &$4.4\times 10^{-8}$ &\cite{Bernard:2009tk}&\\
$\tau \rightarrow e \gamma$ & $3.3\times 10^{-8}$ &\cite{Bernard:2009tk}&\\
$\mu^- \rightarrow e^+e^-e^- $ &$1.0\times 10^{-12}$ &\cite{Bellgardt:1987du}&\\
$\tau^- \rightarrow \mu^+\mu^-\mu^- $ &$3.2\times 10^{-8}$
&\multirow{6}*{$\left.\begin{array}{l} \\ \\ \\ \\ \medskip \\  \end{array}\hspace*{-7mm}\right\}$} &
\hspace*{-37mm}\multirow{6}*{\cite{Hayasaka:2010np}}\\
$\tau^- \rightarrow e^+e^-e^- $ &$3.6\times 10^{-8}$& &\\
$\tau^- \rightarrow e^+\mu^-\mu^-$ &$2.3\times 10^{-8}$ & &\\
$\tau^- \rightarrow e^-\mu^+\mu^-$ &$4.1\times 10^{-8}$ & &\\
$\tau^- \rightarrow \mu^+e^-e^- $ &$2.0\times 10^{-8}$ & &\\
$\tau^- \rightarrow \mu^- e^+e^- $ &$2.7\times 10^{-8}$ & &\\
$\mu\rightarrow e$ in Ti & $4.3\times 10^{-12}$ &\cite{Dohmen:1993mp}\\
\end{tabular}
\end{ruledtabular}
\end{table}

\begin{figure*}[t]
\begin{tabular}{cc}
\includegraphics[width=6.2cm]{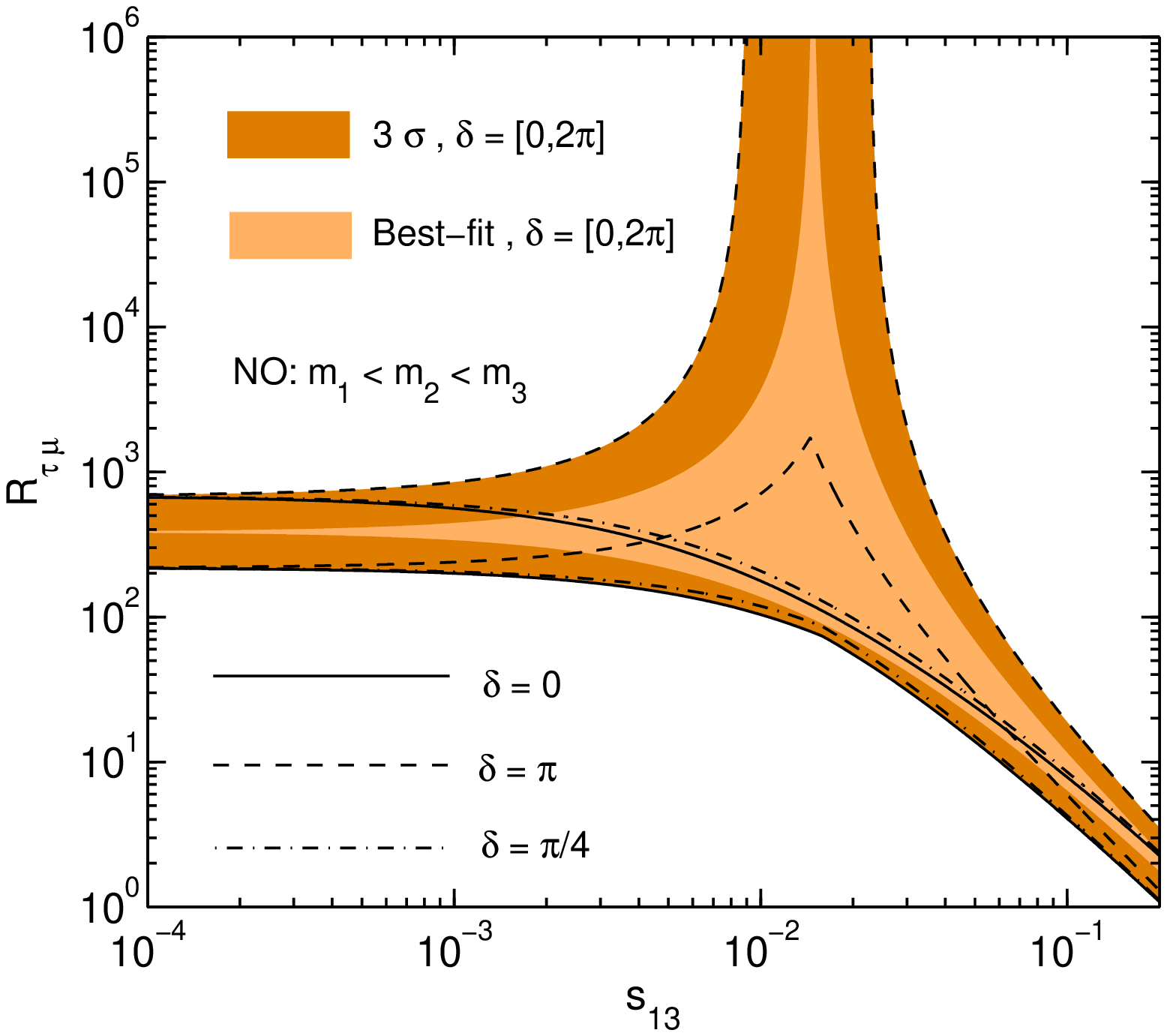}
& \includegraphics[width=6.2cm]{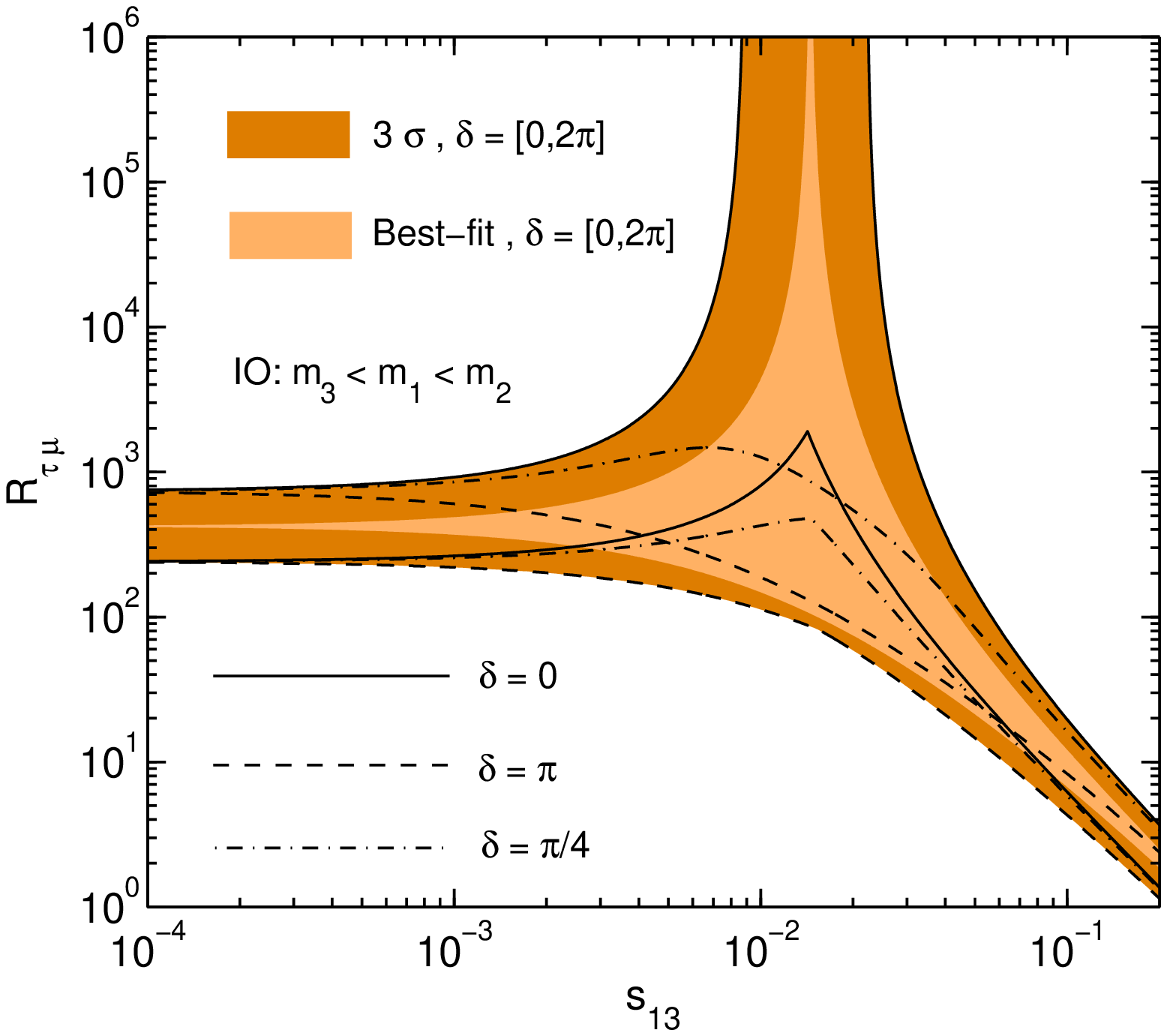}
\\ \includegraphics[width=6.2cm]{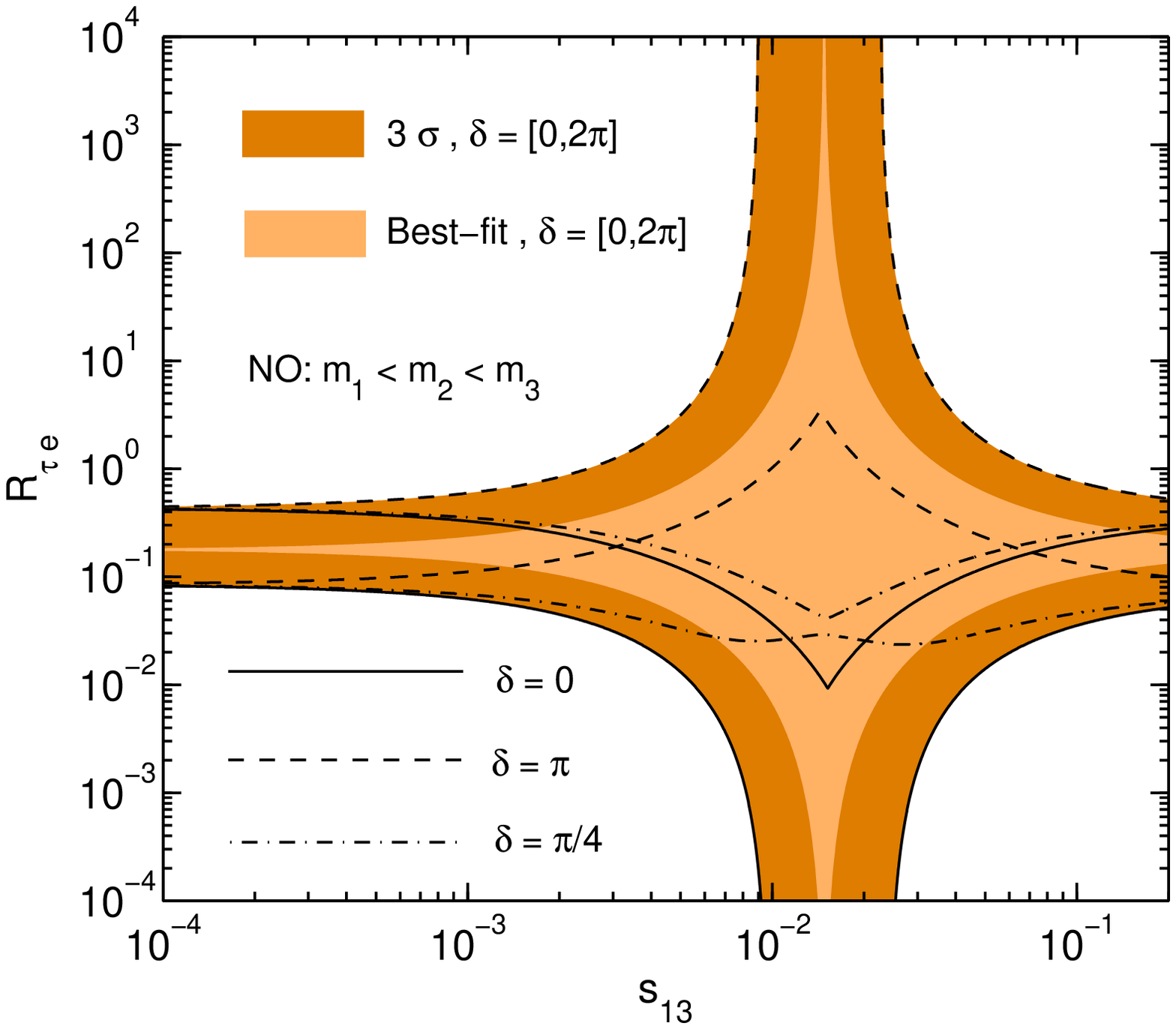}
& \includegraphics[width=6.2cm]{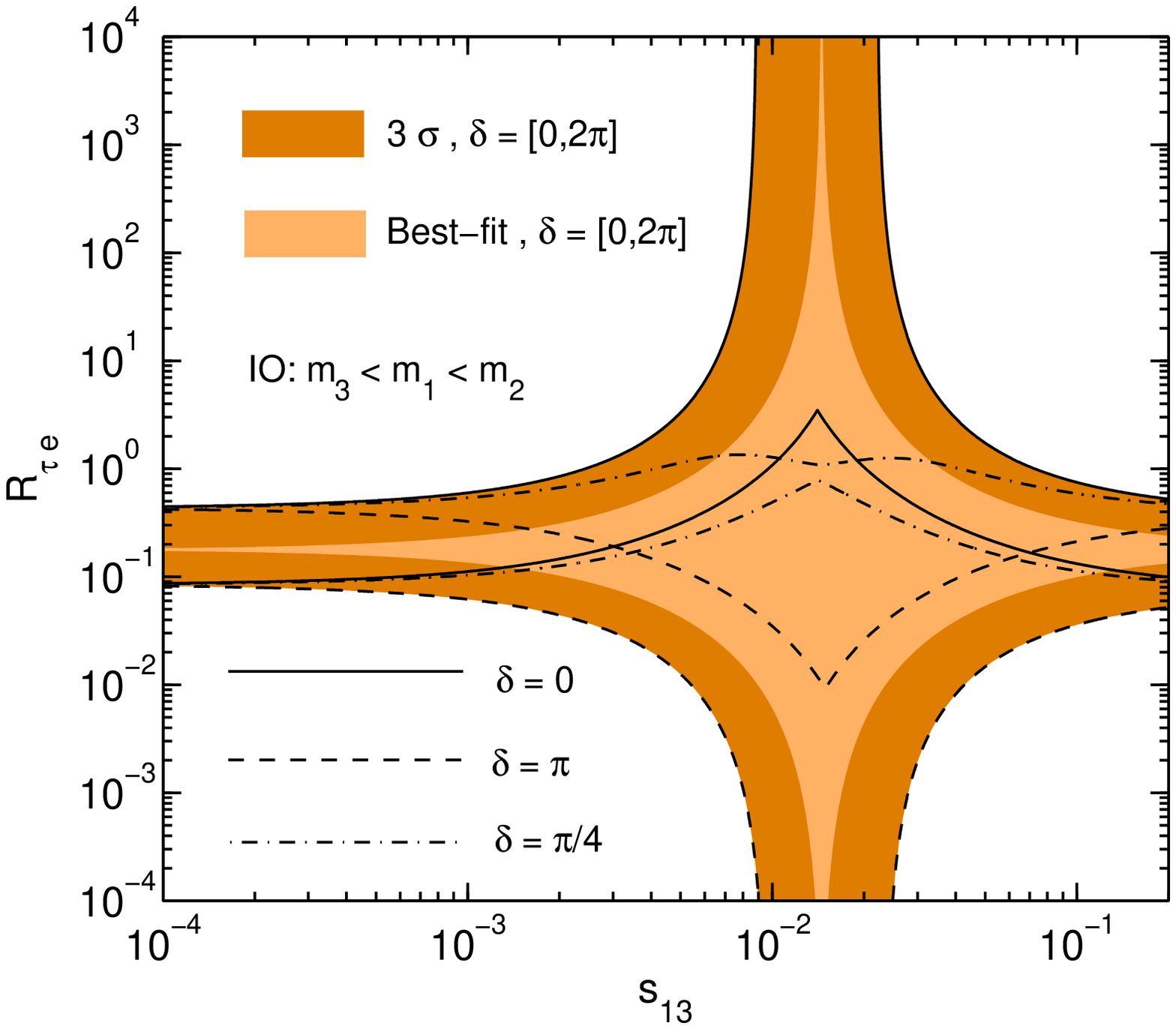}
\end{tabular}
\caption{Allowed regions for $R_{\tau\mu}$ (upper plots) and $R_{\tau e}$ (lower plots) defined in Eqs.~\eqref{Rtauj} and \eqref{m2lmue} as a function of $s_{13}$ and $\delta$, for both the NO (left plots) and the IO (right plots) neutrino mass spectra. In dark (light) shading we show the $3\sigma$ (best-fit) allowed regions obtained by varying the CP-violating phase $\delta$ in the interval $[0,2\pi]$ and using the neutrino data displayed in Table~\ref{Tabnudata}. The solid, dashed, and dash-dotted lines delimit the $3\sigma$ region for $\delta=0$, $\delta=\pi$, and $\delta=\pi/4$, respectively. Adapted from~\cite{Joaquim:2009vp}.} \label{LFVs13}
\end{figure*}

Note also that the flavor dependence of the BRs on the neutrino mass and mixing parameters is direct in the sense that $\bY^\Delta =M_\Delta {\rm \bf m}_\nu/(\mu v^2)$ [see Eq.~\eqref{mnutypeIIseesaw}]. Therefore the way in which the rates of LFV decays depend on the neutrino parameters is model independent. In order to eliminate the dependence on $v$, $\mu$, and $M_\Delta$, it is convenient to define ratios of BRs such as
\begin{equation}
\label{Rtauj}
R_{\tau j}\equiv \frac{{\rm BR}(\tau \rightarrow l_j \gamma)}
{{\rm BR}(\mu \rightarrow e \gamma)}=
\left|\frac{({\rm \bf m}_\nu^{\dag}{\rm \bf m}_\nu)_{\tau j}}
{({\rm \bf m}_\nu^{\dag}{\rm \bf m}_\nu)_{\mu e}}\right|^2
{\rm BR}(\tau \rightarrow l_j \bar{\nu}_j \nu_\tau)\,.
\end{equation}
and
\begin{align}
\label{Rtaujki}
\hat{R}_{\tau jki} &\equiv \frac{{\rm BR}(\tau^{-} \rightarrow l_j^{+} l_k^{-} l_i^{-})}
{{\rm BR}(\mu \rightarrow 3e)}\nonumber\\
&=\frac{2}{1+\delta_{ki}}
\left|\frac{({\rm \bf m}_\nu)_{\tau j} ({\rm \bf m}_\nu)_{ki}}
{({\rm \bf m}_\nu)_{\mu e} ({\rm \bf m}_\nu)_{ee}}\right|^2 {\rm BR}(l_i \rightarrow l_j \bar{\nu}_j \nu_i)\,.
\end{align}
Using now the parametrization for $\mathbf{m}_\nu$ shown in Eq.~\eqref{mnutypeIdiag}, and taking into account the definitions~\eqref{NOIOms}, one can see that the quantities $R_{ij}$ do not depend on the
Majorana phases
$\alpha_{1,2}$ and the lightest neutrino mass~\cite{Rossi:2002zb,Joaquim:2006mn}.
In contrast, the ratios $\hat{R}_{\tau jki}$ may depend on all neutrino parameters~\cite{Chun:2003ej}.

The dependence of ${\rm BR}(l_i \rightarrow l_j \gamma)$ on the neutrino parameters
is~\cite{Joaquim:2009vp,Joaquim:2009zz}
\begin{widetext}
\begin{eqnarray}
\label{m2lmue}
{\rm BR}(\mu \rightarrow e \gamma) &\propto& c_{13}^2\left[r^2c_{23}^2
\sin^2(2\theta_{12})+a^2s_{13}^2s_{23}^2 +
a|r|s_{13}\cos\delta\sin(2\theta_{12})\sin(2\theta_{23})
\right]\,,\nonumber\\
{\rm BR}(\tau \rightarrow e \gamma) &\propto& c_{13}^2\left[r^2s_{23}^2\sin^2(2
\theta_{12})+a^2s_{13}^2c_{23}^2-
a|r|s_{13}\cos\delta\sin(2\theta_{12})\sin(2\theta_{23})\right]
\,,\nonumber\\
{\rm BR}(\tau \rightarrow \mu \gamma)& \propto & \{\,4|r|s_{13}\,\cos\delta
\sin(2\theta_{12})\cos(2\theta_{23})+[2\,b\,c_{13}^2-|r|(\cos(2\theta_{23})-3)\cos(2
\theta_{12})]\sin(2\theta_{23})\,\}^2\nonumber \\
& &+16\, r^2 s_{13}^2\cos\delta\sin(2\theta_{12})
\sin(2\theta_{23})\,.
\end{eqnarray}
\end{widetext}

These expressions are valid for both the NO and IO neutrino mass spectra with $a$ and $b$ defined as
\begin{align}
\label{ab}
{\rm NO:} &\,\,a=2\,(1-|r|s_{12}^2)\simeq 2\,,\quad b=-2+|r|\simeq -2\,,\nonumber\\
{\rm IO:} &\,\,a=-2\,(1+|r|s_{12}^2)\simeq -2\,,\quad b=2+|r|\simeq 2\,,
\end{align}
and the parameter $r$ given in Eq.~\eqref{rdef}.

From the above equations, one can immediately conclude that the ratios $R_{\tau j}$ depend on the lepton mixing angles, the Dirac CP phase $\delta$, and the ratio $r$. Taking into account the present neutrino data summarized in Table~\ref{Tabnudata}, one can study the dependence of $R_{\tau j}$ on $\theta_{13}$ and $\delta$. This is shown in Fig.~\ref{LFVs13} where $R_{\tau \mu}$ (top panels) and $R_{\tau e}$ (bottom panels) are shown for the NO (left panels) and IO (right panels) neutrino mass spectra. From this figure, it is evident that the impact of $\delta$ on $R_{\tau j}$ can be significant for $s_{13}\sim 10^{-2}$. In particular, a flavor suppression may occur in the $\tau \mu$ and $\tau e$ channels in the CP-conserving cases. This may be have profound impact on the LFV predictions of the type II seesaw~\cite{Joaquim:2006mn,Joaquim:2009zz}.

\begin{figure}[t]
\includegraphics[width=8.5cm]{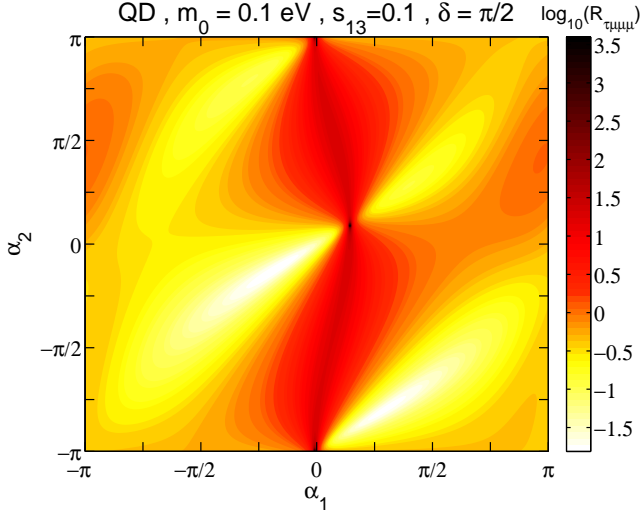}
\caption{Density plot of the ratio $R_{\tau\mu\mu\mu}$ defined in Eq.~\eqref{Rtaujki} as a function of the Majorana phases $\alpha_{1,2}$ for a QD neutrino mass spectrum with $m_0=0.1$~eV, $s_{13}=0.1$, and $\delta=\pi/2$.}
\label{RQD}
\end{figure}

As already mentioned, in the type II seesaw framework, the three-body LFV charged-lepton decay rates may also depend on the neutrino mass scale and the Majorana CP phases. In some cases, this is not true though. For instance, for a HI neutrino mass spectrum and $\theta_{13}=\delta=0$, the BRs of the decays $\mu^- \rightarrow e^+e^-e^-$ and $\tau^- \rightarrow e^+e^-e^-$ depend on the neutrino mixing parameters as
\begin{eqnarray}
\label{BRdep3}
{\rm BR}(\mu^- \rightarrow e^+e^-e^-) &\propto& r^2
c_{12}^2\,c_{23}^2\,s_{12}^6\,, \nonumber\\
{\rm BR}(\tau^- \rightarrow e^+e^-e^-) &\propto& r^2
c_{12}^2\,s_{23}^2\,s_{12}^6\,,
\end{eqnarray}
leading to $R_{\tau e e e}\simeq \tan^2\theta_{23}\,{\rm BR}(\tau\rightarrow e \nu_\tau \bar{\nu}_e)\simeq 0.17$. Therefore, in this specific case, the observation of the $\tau^- \rightarrow e^+e^-e^-$ decay in the near future would exclude a scenario where these decays occur due to the exchange of the scalar triplet which gives rise to neutrino masses, for any value of the Majorana phases. This is not the case for the IH neutrino spectrum for which
\begin{align}
\label{BRdepIH}
{\rm BR}(\mu^- \rightarrow e^+e^-e^-) &\propto  c_{12}^2 s_{12}^2 c_{23}^2\sin^2(\alpha_1/2)
\nonumber\\
&\times\left[1-\sin^2(2\theta_{12})\sin^2(\alpha_1/2)\right],
\end{align}
for $\theta_{13}=\delta=0$ and at zero order in $r$. This expression exhibits a strong dependence on the (only) Majorana phase $\alpha_1$. In Fig.~\ref{RQD}, we show the dependence of the ratio $R_{\tau\mu\mu\mu}$ on the Majorana phases $\alpha_{1,2}$ for $s_{13}=0.1,\delta=\pi/2$ (large Dirac CP violation), and a QD neutrino mass spectrum ($m_0\simeq0.1$~eV). The density plot of ${\rm log}_{10}(R_{\tau\mu\mu\mu})$ shows that, depending on the values of $\alpha_{1}$ and $\alpha_{2}$, $R_{\tau\mu\mu\mu}$ can change by several orders of magnitude.

If the neutrino mass mediators are very heavy, their direct effect on LFV processes becomes irrelevant. Still, they can participate indirectly on the generation of new LFV terms, as may happen in supersymmetric versions of the seesaw mechanism~\cite{Borzumati:1986qx,Rossi:2002zb}, when renormalizable Yukawa interactions involving the heavy and SM fields induce, through renormalization, LFV soft supersymmetry (SUSY) breaking. This scenario has been the subject of a large number of studies~\cite{Raidal:2008jk}. In the SUSY type I seesaw, singlet neutrino superfields $N_i$ with masses $M_i$ are added to the minimal supersymmetric standard model (MSSM) superfield content, in such a way that the superpotential $W$ is just $W=W_{\rm MSSM}+ \bY^\nu N L H_2+\frac{1}{2}\, M_i N_iN_i$, where $L$ and $H_2$ are the lepton and Higgs superfields, respectively. Considering (flavor-blind) universal boundary conditions for the soft SUSY-breaking terms at a scale $\Lambda > M_i$, LFV terms may be generated at lower scales due to renormalization group effects induced by the presence of $\bY^\nu$~\cite{Borzumati:1986qx}. In particular, in the simplest case in which only the LFV effects induced in the left-handed scalar sector are relevant, the soft SUSY-breaking terms $\tilde{L} \bmm_{\tilde{L}}^2\tilde{L}$ are such that
\begin{equation}
\label{m2LI}
(\bmm_{\tilde{L}}^2)_{ij} \simeq -\frac{3m_0^2+A_0^2}{8\pi^2}\,
(\bY^{\nu\dagger}\bY^\nu)_{ij}\ln \frac{\Lambda}{M}\;,\;(i \neq j)\,,
\end{equation}
where $m_0$ and $A_0$ are the universal SUSY-breaking soft mass and trilinear parameters at the scale $\Lambda$. For simplicity, we have taken in the above expression a common mass $M$ for all the heavy Majorana neutrinos. The existence of LFV entries in the slepton masses $(\bmm_{\tilde{L}}^2)_{ij}$ opens the window for the LFV processes discussed above at the loop level. For the specific case of radiative charged-lepton decays,
\begin{equation}
\label{BRap}
{\rm BR}(l_i \rightarrow l_j\gamma) \simeq \frac{48\pi^3 \alpha}{G_F^2}\,
|C_{ij}|^2 \tan^2\!\beta\,\, {\rm BR}(l_i \rightarrow
l_j\nu_i\bar{\nu}_j)\,,
\end{equation}
where the coefficients $C_{ij}$ encode the LFV dependence of the rates. Taking a common mass $m_S$ for the SUSY particles in the loops, one has
\begin{equation}
\label{Cij}
C_{ij} \sim
\frac{g_2^2}{16\pi^2}\frac{(\bmm_{\tilde{L}}^2)_{ij}}{m_S^4}\;,\; (i\neq
j=e,\mu,\tau)\,.
\end{equation}

It is straightforward to see that the rates of the LFV processes depend on a combination of couplings which is different from the one which appears in the neutrino mass matrix. Therefore, as in the case for the low-energy seesaw discussed above, a model-independent reconstruction of $(\bY^{\nu^\dagger}\bY^\nu)_{ij}$ is not possible from low-energy data. For instance, it has been recently shown that the $C_{ij}$ coefficients are not as sensitive to the unknown mixing angle $\theta_{13}$ as previously advocated~\cite{Casas:2010wm}. In other words, the way that the SUSY LFV terms depend on the neutrino parameters in the SUSY type I seesaw mechanism is not model independent. Nevertheless, it can be shown that the phases entering in the neutrino mixing matrix may have a strong impact in LFV processes~\cite{Petcov:2006pc} and the electric dipole moments of charged leptons~\cite{Ellis:2001xt,Joaquim:2007sm,Farzan:2004qu,Masina:2003wt}.

In the case of the SUSY type II seesaw, the left-handed LFV soft scalar masses are given by~\cite{Rossi:2002zb}
\begin{equation}
\label{m2LII}
(\bmm_{\tilde{L}}^2)_{ij}-\frac{9m_0^2+3A_0^2}{8\pi^2}\, (\bY^{\Delta\dagger}
\bY^\Delta)_{ij} \, \ln \frac{\Lambda}{M_\Delta}\,,
\end{equation}
where $\bY^\Delta$ are the couplings of the triplet with the lepton superfields and $M_\Delta$ the triplet mass. Note that, since the effective neutrino mass matrix $\bmm_{\nu}$ is again proportional to $\bY^\Delta$, the ratios of BRs defined in Eq.~\eqref{Rtauj} are still valid in the SUSY case. In particular, the predictions shown in Fig.~\ref{LFVs13} also hold in the present case. The same is not true for three-body decays and $\mu$-$e$ conversion in nuclei which, in the MSSM, are induced at one loop due to the presence of LFV soft SUSY-breaking terms like $(\bmm_{\tilde{L}}^2)_{ij}$. Consequently, the rates for these processes will be also independent from the Majorana phases and the lightest neutrino mass~\cite{Joaquim:2006mn}. In general, this is valid in all cases with LFV in the soft SUSY-breaking sector induced by the couplings $\bY^\Delta$, as in the universal boundary condition limit~\cite{Rossi:2002zb,Joaquim:2009vp,Joaquim:2009zz}, or in the gauge-Yukawa SUSY-breaking mediation scenario~\cite{Joaquim:2006uz,Joaquim:2006rg}. It has also been shown that, in a type II seesaw scenario with neutrino masses generated from K\"{a}hler effective terms, the same relation of LFV processes and neutrino data is obtained~\cite{Brignole:2010bj,Brignole:2010nh}.

In the previous examples, the CP phases affecting the LFV rates are those that can be potentially measured in neutrino experiments. However, it is well known that it is possible to probe on CPV in the leptonic sector by adopting an effective Lagrangian approach to extract some information on the CP-violating structure of the LFV effective operators~\cite{Zee:1985pt,Treiman:1977dj,Okada:1999zk,deGouvea:2000cf}. For instance, this can be achieved by measuring the polarization of the final-state particles in $\mu\rightarrow e \gamma$~\cite{Farzan:2007us,Ayazi:2008gk} and $\mu$-$e$ conversion in nuclei~\cite{Davidson:2008ui}. Similar conclusions can be drawn if one performs a spin measurement of the more energetic positron in the final state of $\mu^+ \rightarrow e^+ e^- e^+$. Although such studies could shed some light on the CP-violating structure of the effective Lagrangian, the origin of such effects would be hardly identifiable, since their connection with CP violation in neutrino oscillations is difficult to establish without further theoretical assumptions. Still, it is undeniable that detecting such CPV effects in LFV processes could be a powerful tool for discriminating BSM scenarios in which the LFV effective operators arise.

\subsection{Impact of LCPV at colliders}
\label{sec3.5}

High-energy accelerators like the LHC may also provide valuable information about the neutrino mass generation mechanism. In particular, if the neutrino mass mechanism operates at scales not far from the electroweak scale, then new phenomena can manifest in colliders. Most of the research performed in this direction concerns the study of new signals, which result from decays of the seesaw mediators~\cite{Akeroyd:2007zv,Kadastik:2007yd,delAguila:2008cj,Han:2005nk,Bajc:2007zf}. Although these decays do not lead to explicit CPV effects, the presence of CPV phases affects the decay rates, since the couplings of the SM particles to the seesaw mediators depend on the phases $\alpha_{1,2}$ and $\delta$ of the lepton mixing matrix $\bU$.

The connection between LCPV, collider processes, and neutrino oscillation experiments is not straightforward to establish. In particular, in the case of the type I (III) seesaw, it is not possible to reconstruct in a model-independent way the couplings of the fermion singlets (triplets) with the Higgs and charged-lepton fields. However, the situation changes in the type II seesaw since, as mentioned, the couplings of the scalar triplet $\Delta$ with the lepton doublets have the same flavor structure as the effective neutrino mass matrix. In this framework, if the triplet mass is close to the electroweak scale, $\Delta$ may be produced in high-energy collisions. More specifically, the production of its doubly-charged Higgs component occurs via the Drell-Yan process $q\bar{q}\rightarrow \gamma^\ast Z^\ast \rightarrow\Delta^{++}\Delta^{--}$, and also (subdominantly) by photon-photon fusion $\gamma\gamma\rightarrow\Delta^{++}\Delta^{--}$. Provided the triplet VEV is small enough, the decays of $\Delta^{\pm\pm}\rightarrow l^\pm l^\pm$ are dominant over $\Delta^{++}\rightarrow \Delta^+\Delta^+$, $\Delta^{++}\rightarrow \Delta^+ W^+$ and $\Delta^{++}\rightarrow W^+ W^+$. In this case, the decay of the $\Delta^{\pm\pm}$ pair into four charged leptons gives a very clear signature, which is almost free of any SM background~\cite{delAguila:2008cj, Han:2005nk}.

Assuming that neutrino masses are generated through the exchange of $\Delta$, the decay rate of $\Delta^{\pm\pm}\rightarrow l_i^\pm l_j^\pm$ is proportional to $|(\bmm_\nu)_{ij}|^2$, which is sensitive to the LCPV phases. The branching ratios $\BR_{\Delta ij}\equiv\BR(\Delta^{\pm\pm}\rightarrow l_i^\pm l_j^\pm)$ are simply given by
\begin{equation}
\label{BR1}
\BR_{\Delta ij}=\frac{2}{1+\delta_{ij}}
\frac{\sum_k|m_k\bU_{ik}\bU_{jk}|^2}{\sum_{n}m_n^2}\,,
\end{equation}
where $\delta_{ij}$ is the Kronecker symbol, introduced to account for the decays into charged leptons of the same flavor. The term in the denominator is $\sum_{p}m_p^2= 3m_0^2+\dmsol +\dmatm$ for a NO neutrino mass spectrum, and $\sum_{p}m_p^2=3m_0^2+\dmsol+2|\dmatm|$ for an IO one. The above BRs depend exclusively on the lepton mixing angles, CPV phases and the neutrino masses. In some specific limits, very simple relations can be obtained. In particular, in the HI case (NO with $m_0=0$), and taking $\theta_{13}=0$ one has
\begin{align}
\label{BRHI}
\BR_{\Delta ee}^{\rm HI}&=\dfrac{rs_{12}^4}{1+r}\,,\nonumber\\
\BR_{\Delta \mu e}^{\rm HI}&=\dfrac{r c_{23}^2\sin^2(2\theta_{12})}
{2\,(1+r)}\,,\nonumber\\
\BR_{\Delta \mu\mu}^{\rm HI}&=\dfrac{rc_{12}^4c_{23}^4+s_{23}^4+
2\sqrt{r}\,c_{12}^2c_{23}^2s_{23}^2 \cos {\alpha_{21}}}{1+r}\,.
\end{align}

Note that, in this particular case, the $e^\pm e^\pm$ and $\mu^\pm e^\pm$ decays are suppressed by the parameter $r\ll 1$. Moreover, only the $\mu^\pm\mu^\pm$ channel is sensitive to leptonic CPV effects associated with the Majorana-phase difference $\alpha_{21}=\alpha_2-\alpha_1$ (the decays into $\mu^\pm\tau^\pm$ are also sensitive to $\alpha_{21}$). In the IH limit (IO with $m_0=0$) the above BRs are instead approximately given by
\begin{align}
\label{BRIH}
\BR_{\Delta ee}^{\rm IH} &\simeq \dfrac{1}{2}\left(s_{12}^4+c_{12}^4+2\,c_{12}^2s_{12}^2\cos\alpha_1
\right)\,,\nonumber\\
\BR_{\Delta \mu e}^{\rm IH} &=\sin^2(2\theta_{12})\,c_{23}^2\sin^2\frac{\alpha_1}{2}\,,\nonumber\\
\BR_{\Delta \mu \mu}^{\rm IH} &\simeq c_{23}^4 \,\BR_{\Delta ee}^{\rm IH}.
\end{align}
As for the QD case ($m_0 \gg \dmatm$), the following relations hold:
\begin{align}
\label{BRQD}
\BR_{\Delta ee(\mu)}^{\rm QD} &\simeq \frac{2}{3} \BR_{\Delta ee(\mu)}^{\rm IH}\,, \nonumber\\
12\,\BR_{\Delta \mu\mu}^{\rm QD}&= c_{23}^4 \left[ 3+\cos(4\theta_{12})+2\sin^2(2\theta_{12})\cos\alpha_1  \right] \nonumber \\&+
4s_{23}^4+2(c_{12}^2\cos\alpha_{21}+s_{12}^2\cos\alpha_2)\sin^2(2\theta_{23})            \,.
\end{align}

\begin{figure}[]
\begin{tabular}{c}
\includegraphics[width=7.7cm]{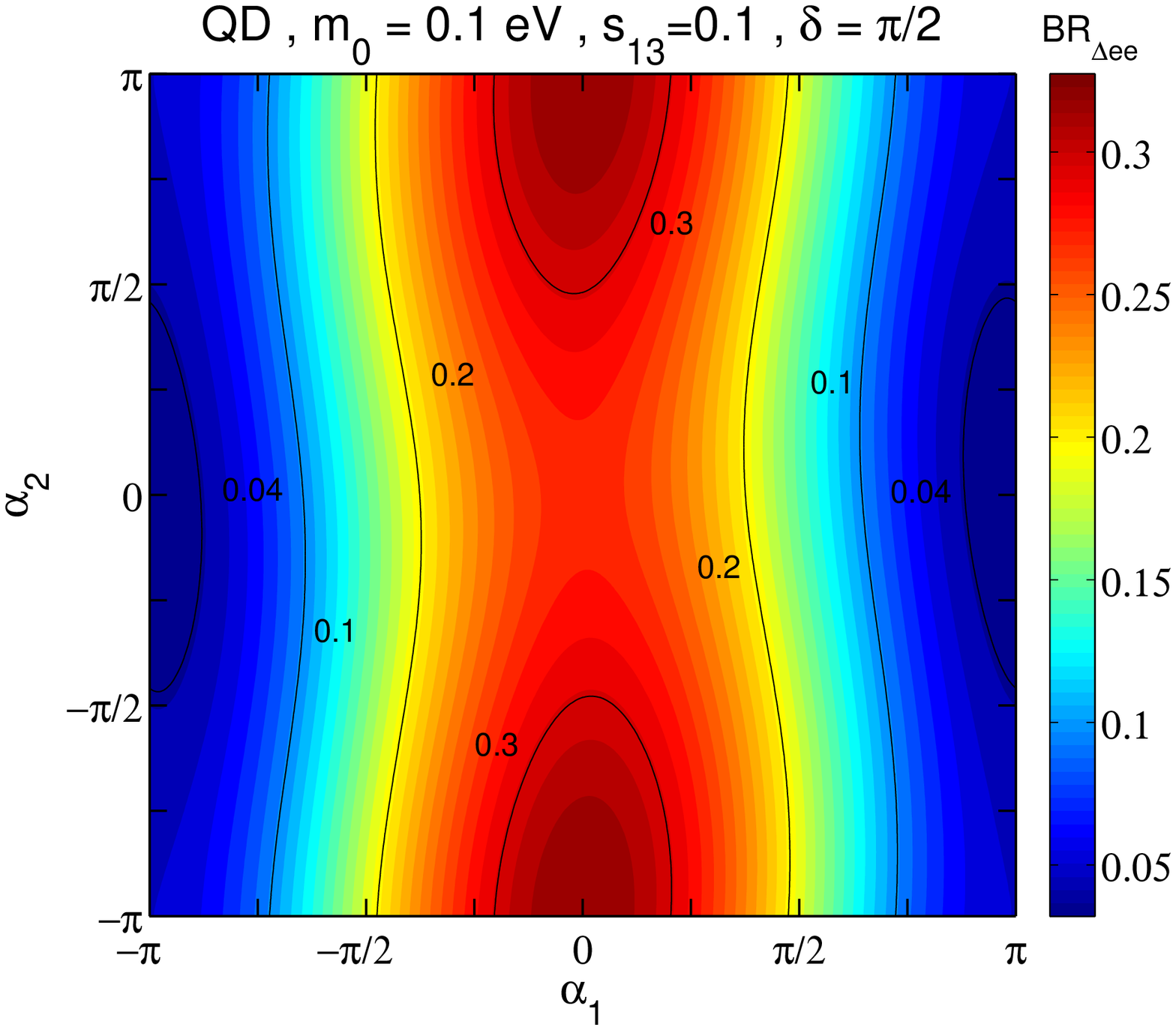} \\
\includegraphics[width=7.7cm]{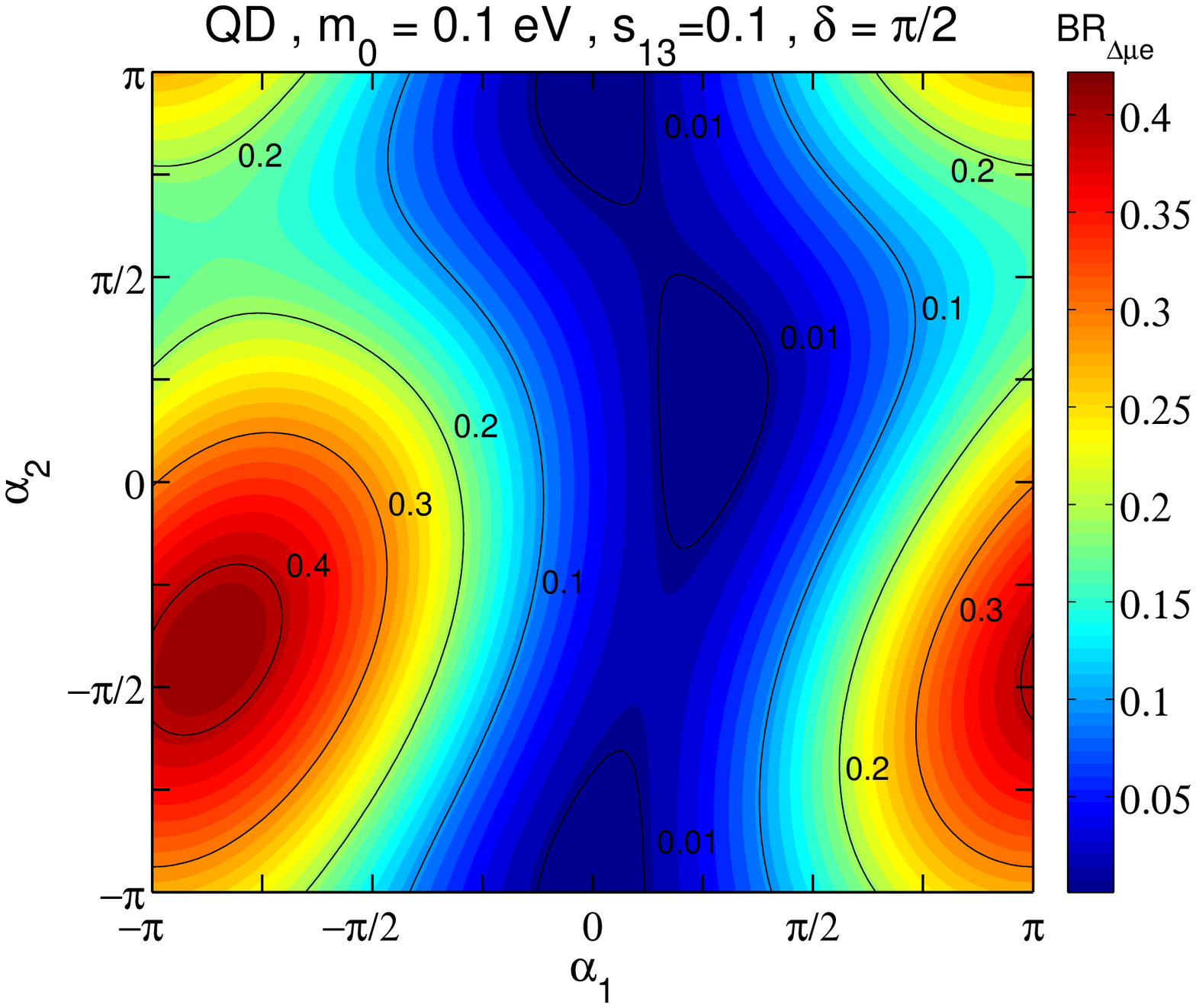}\\
\includegraphics[width=7.7cm]{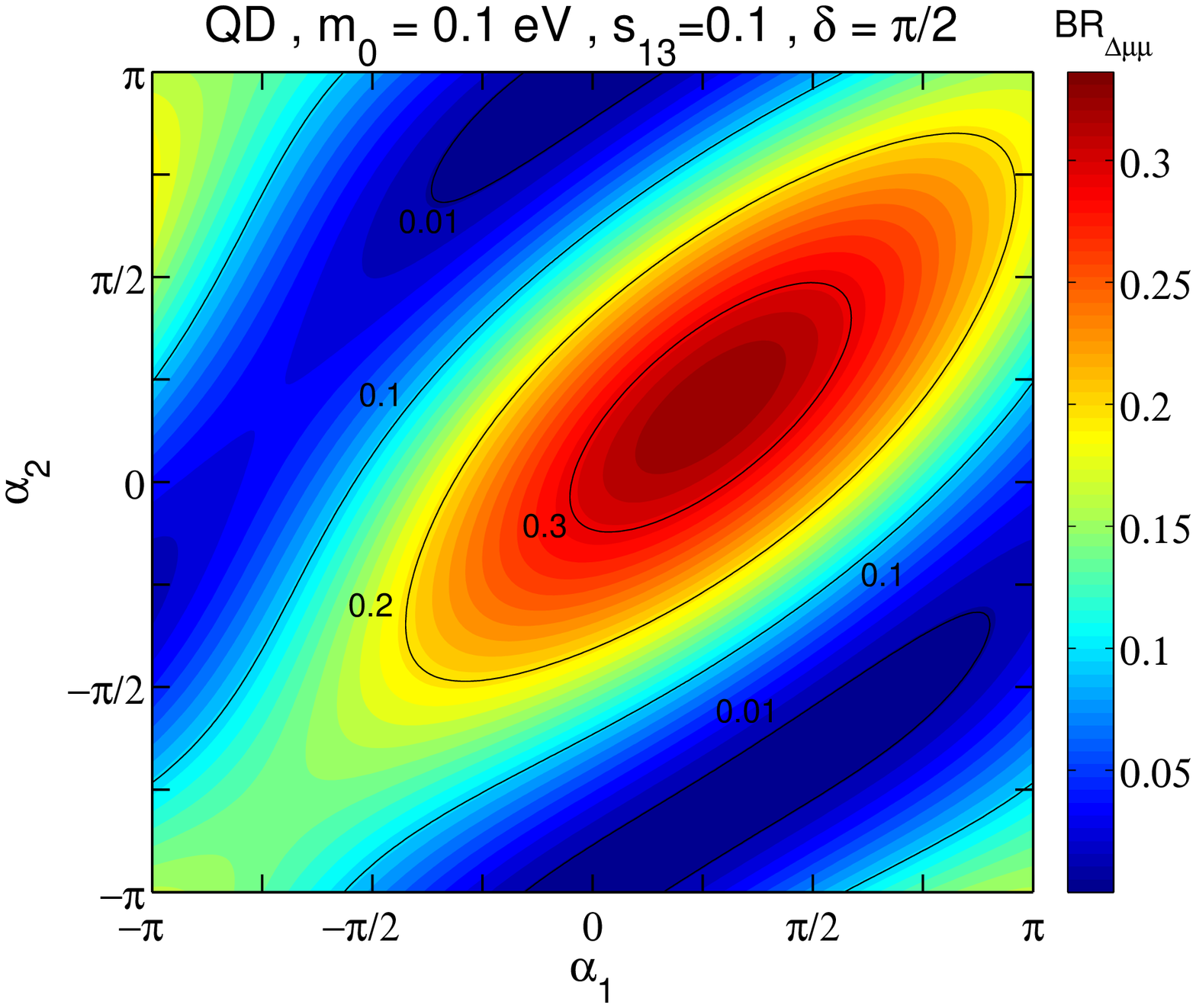}
\end{tabular}
\caption{Variation of $\BR_{\Delta ee}$ (top), $\BR_{\Delta \mu e}$ (center), and $\BR_{\Delta \mu\mu}$ (bottom) in the $\alpha_{1}$-$\alpha_2$ parameter space for a quasidegenerate neutrino mass spectrum with $m_0=0.1~\eV$, $s_{13}=0.1$, and $\delta=\pi/2$. The remaining neutrino parameters are taken at the best-fit values of the STV
analysis (see Table~\ref{Tabnudata}).}
\label{BRsT}
\end{figure}

The above results hold in the simple limits of HI, IH and QD neutrino masses with $\theta_{13}=0$. A complete study including the dependence on the lightest neutrino mass and CPV phases can be found in~\cite{Garayoa:2007fw}. The possibility of extracting information on the Majorana phases from the doubly-charged Higgs decays into leptons has been addressed in~\cite{Akeroyd:2007zv} and the connection with neutrinoless double beta decay in~\cite{Petcov:2009zr}. In particular, it has been shown that it is possible to extract some information about $m_0$ and $\alpha_{1,2}$ from $\BR_{ee}$, $\BR_{\mu\mu}$, and $\BR_{e\mu}$.

In Fig.~\ref{BRsT}, we show how $\BR_{\Delta ee}$ (top panel), $\BR_{\Delta \mu e}$ (center panel), and $\BR_{\Delta \mu \mu}$ (bottom panel) depend on the Majorana phases $\alpha_{1,2}$, for the specific case $\delta=\pi/2$, $s_{13}=0.1$ and a QD neutrino mass spectrum with $m_0=0.1~\eV$. The results show that the rates for the decays of the triplets into leptons are considerably affected by the Majorana phases $\alpha_{1,2}$. In particular, one can see from these plots that $\BR_{\Delta \mu e}$ tends to be suppressed when $\BR_{\Delta ee}$ and $\BR_{\Delta \mu \mu}$ are larger.

\subsection{Nonunitarity effects in the lepton sector}
\label{sec3.6}

Searches for deviations from unitary mixing are a sensitive probe of physics beyond the SM. In the quark sector, several studies have been carried out in the direction of finding possible deviations from the unitarity of the CKM matrix. Similarly, nonunitarity (NU) effects may occur in the lepton sector in the presence of BSM physics. This is the case if, for instance, new states with mass far above the electroweak scale are added to the SM particle content. Probably the best example of such a framework is the seesaw mechanism described in Sec.~\ref{sec3.5}. In the type I version, the mass matrix is extended to a $(3+n_R)\times (3+n_R)$ form, where $n_R$ is the number of heavy right-handed neutrinos with typical mass $M \gg v$. In this case, the NU of the lepton mixing matrix stems from the fact that this matrix is now a sub-block of a larger unitary one, since the complete theory has to respect probability conservation~\cite{Schechter:1980gr}. After the decoupling of these states, an effective dimension-six operator of the type $(\bar{\ell} \tilde{\phi})i\partial\!\!\!/\, (\tilde{\phi}^\dag\ell)/M^2$ is generated~\cite{Broncano:2002rw} which induces a contribution to the neutrino kinetic energy, suppressed by $v^2/M^2$, upon electroweak symmetry breaking.  Therefore, a field redefinition is demanded to bring back the kinetic term to its canonical form. This, in turn, introduces NU mixing in the charged and neutral current Lagrangian terms.

In the conventional type I seesaw, the NU effects are too small to be observed. Nevertheless, this may not be the case in alternative realizations like the inverse seesaw~\cite{Mohapatra:1986bd,GonzalezGarcia:1988rw}, in which the effect of the mass suppression can be alleviated without prejudice of the smallness of neutrino masses. In other words, in this scenario the effective dimension-five operator responsible for the suppression of neutrino masses can be somehow decoupled from the dimension-six one, allowing at the same time not too small NU effects so that interesting new phenomenology may appear~\cite{Malinsky:2009gw,Malinsky:2009df,Deppisch:2005zm,Deppisch:2004fa,Dev:2009aw}. Similar effects arise in other models with large light-heavy neutrino mixing~\cite{Tommasini:1995ii,Nardi:1994nw}, and in scenarios with extra dimensions where the mixing of Kaluza-Klein modes with the light neutrinos may induce NU effects~\cite{DeGouvea:2001mz,Branco:2002hc,Bhattacharya:2009nu}. Another possible source of nonunitarity arises from loop corrections to the charged-lepton or neutrino self-energies~\cite{Bellazzini:2010gn} which modify the corresponding kinetic terms, thus inducing NU effects. There can also be direct corrections to the lepton mixing matrix $\bU$.

In studying NU effects in the lepton sector, a model-independent approach can be adopted such that the sources of NU are not specified. In particular, we focus here on a framework called minimal unitarity violation (MUV), in which NU sources are allowed only in neutrino Lagrangian terms and three light neutrinos are considered~\cite{Antusch:2006vwa}. Under these assumptions, the mass and flavor neutrino eigenstates are related by a nonunitary $3\times 3$ matrix $\bN$ such that $\nu_\alpha=\bN_{\alpha k}\nu_k$. In the corresponding mass basis, the charged and neutral current Lagrangian terms become~\cite{Schechter:1980gr}
\begin{align}
\label{NULCC}
\mathcal{L}_{CC}&=-\frac{g}{\sqrt{2}}\left(W_\mu^+\bar{l}_\alpha\gamma_\mu P_L \bN_{\alpha k}
\nu_k+\Hc\right)\,,\\
\label{NULNC}
\mathcal{L}_{NC}&=-\frac{g}{\cos\theta_W}\left[Z_\mu\bar{\nu}_kP_L(\bN^\dag\bN)_{kj}\nu_j+
\Hc\right]\,.
\end{align}

These modifications give rise to new effects in several physical phenomena such as neutrino oscillations, universality tests and electroweak decays, which can be used to test unitarity in lepton mixing. In this direction, detailed analysis have been performed in the literature with the goal of quantifying the deviations from unitarity of $\bN$, taking into account several physical processes. In the following,  we briefly review the main conclusions of those studies.

\subsubsection{Neutrino oscillations with NU}
\label{sec3.6.1}

In the presence of NU, the neutrino flavor and mass eigenstates cannot be simultaneously orthogonal. As a consequence, the oscillation probabilities $\nu_\alpha\rightarrow\nu_\beta$, as a function of the distance $L$ travelled by neutrinos, now read ~\cite{Czakon:2001em}
\begin{equation}
\label{NUOSCP}
P_{\alpha\beta}=\frac{\left|\sum_{k}\bN_{\beta k}\,e^{-iE_k L}
\bN_{\alpha k}^\ast\right|^2}{(\bN\bN^\dag)_{\alpha\alpha}(\bN\bN^\dag)_{\beta\beta}}\,,
\end{equation}
which reduces to Eq.~\eqref{tramp2} in the limit of a unitary $\bN$. An immediate consequence of the above result is that a flavor transition is possible at zero distance $(L=0)$ before oscillations~\cite{Langacker:1988up}, with a transition probability
\begin{equation}
\label{zeroDp}
P_{\alpha\beta}(L=0)=\frac{\left|(\bN\bN^\dag)_{\beta\alpha}\right|^2}
{(\bN\bN^\dag)_{\alpha\alpha}(\bN\bN^\dag)_{\beta\beta}}\neq\delta_{\alpha\beta}\,.
\end{equation}
This result can be probed at neutrino oscillation experiments with near detectors. In particular, the data from NOMAD~\cite{Astier:2001yj}, Bugey~\cite{Declais:1994su}, KARMEN~\cite{Declais:1994su}, and the MINOS~\cite{Adamson:2008zt} near detector impose the following constraints on $\bN \bN^\dag$:
\begin{eqnarray}
\label{limNNNO}
|(\bN \bN^\dag)_{e\alpha}|&\simeq& (1.00\pm 0.04, <0.05, <0.09)\,,\nonumber\\
|(\bN \bN^\dag)_{\mu\alpha}|&\simeq& (<0.05, 1.00\pm 0.04, <0.013)\,,\nonumber\\
|(\bN \bN^\dag)_{\tau\alpha}|&\simeq& (<0.09, <0.013, ?)\,,
\end{eqnarray}
at $90\%$ C.L.~\cite{Antusch:2006vwa}.

In vacuum, the disappearance oscillation probability is then given by
\begin{align}
\label{disPN}
P_{\alpha\alpha}=\sum_{k=1}^3|\bN_{\alpha k}|^4+
\sum_{k\neq j=1}^3|\bN_{\alpha k}|^2|\bN_{\alpha j}|^2\cos\frac{\Delta m^2_{kj}L}{2E}.
\end{align}
Instead, the oscillation probabilities in matter are modified with respect to the unitary case since the effective potential felt by neutrinos is no longer diagonal~\cite{delAguila:2002sx,Bekman:2002zk,Holeczek:2007kk,FernandezMartinez:2007ms}. In addition, the NC contribution to the matter potential contributes to the evolution equation once it cannot be interpreted as a global phase.

Depending on the range of $L/E$, the above equation can be simplified and used to constrain the elements of $\bN$ (or combinations of them), considering the experimental neutrino oscillation data suitable for each case. The combined fit of the KamLAND~\cite{Araki:2004mb}, CHOOZ\cite{Apollonio:2002gd}, SNO~\cite{Ahmad:2002ka} and K2K~\cite{Ahn:2002up} data allow for the following determination of $|\bN|$ at $90\%$ C.L.~\cite{Antusch:2006vwa}
\begin{eqnarray}
\label{limN1}
|\bN_{ej}|&\simeq& (0.75-0.89,0.45-0.66, <0.34)\,,\nonumber\\
|\bN_{\mu 1}|^2&+&|\bN_{\mu 2}|^2=0.57-0.86\,,\nonumber\\
|\bN_{\mu 3}|&\simeq& 0.57-0.86\,,
\end{eqnarray}
where $|\bN_{e 2}|$ and $|\bN_{e 1}|$ are determined by the SNO and KamLAND data (combined with the others), respectively, and $|\bN_{e 3}|$ is constrained by CHOOZ. On the other hand, atmospheric and accelerator experiments do not allow for a discrimination between $|\bN_{\mu 1}|^2$ and $|\bN_{\mu 2}|^2$. Nevertheless, these two quantities can be disentangled taking into account the constraints shown in Eq.~(\ref{limNNNO}), leading to the final result
\begin{align}
\label{limNNO}
|\bN_{e j}|&\simeq (0.75-0.89,0.45-0.66, <0.27)\,,\nonumber\\
|\bN_{\mu j}|&\simeq (0.00-0.69, 0.22-0.81, 0.57-0.85)\,.
\end{align}
The absence of constraints for the elements in the third row of $\bN$ is due to the lack of $\nu_\tau$ oscillation signals.

\subsubsection{NU constraints from electroweak decays}
\label{sec3.6.2}

It has been known for quite a long time that nonunitarity of the leptonic mixing matrix induced by light-heavy neutrino mixing can manifest itself in tree-level processes like $\pi$, $W$, and $Z$ decays~\cite{Nardi:1991rg,Langacker:1988ur,Nardi:1994iv,Korner:1992an}, in rare charged-lepton decays $l_j\rightarrow l_i\gamma$, $l_j\rightarrow 3l_j$, $l_j\rightarrow l_il_il_k$, and $\mu$-$e$ conversion in nuclei~\cite{Tommasini:1995ii,Ilakovac:1994kj,Langacker:1988up}. The interest on this subject has been recently revived in a series of works, where the constraints on NU effects in the lepton sector have been analyzed, considering the above electroweak processes in view of the most recent experimental data~\cite{Antusch:2008tz,Antusch:2006vwa,Abada:2007ux,Abada:2008ea}.

In the MUV framework, $W \rightarrow l_\alpha \nu_\alpha$ and invisible $Z$ decays lead to the conditions
\begin{eqnarray}
\label{ZWNU}
\frac{(\bN\bN^\dag)_{\alpha\alpha}}{\sqrt{(\bN\bN^\dag)_{ee}
(\bN\bN^\dag)_{\mu\mu}}}&=&f_\alpha\,,\\
\frac{\sum_{\alpha\beta}|(\bN\bN^\dag)_{\alpha\beta}|^2}{\sqrt{(\bN\bN^\dag)_{ee}
(\bN\bN^\dag)_{\mu\mu}}}&=&2.984\pm 0.009\,,
\end{eqnarray}
respectively, with $f_{e,\mu,\tau}=(1.000\pm 0.024,0.986\pm 0.028,1.002\pm 0.032)$. On the other hand, from charged-lepton decays $l_\alpha \rightarrow l_\beta\, \gamma$, one can write
\begin{equation}
\label{LVFNU}
\frac{|(\bN\bN^\dag)_{\alpha\beta}|^2}{\sqrt{(\bN\bN^\dag)_{\alpha\alpha}
(\bN\bN^\dag)_{\beta\beta}}}=\frac{96\pi}{100\alpha_{\rm em}}\frac{\BR(l_\alpha \rightarrow l_\beta \gamma)}
{\BR(l_\alpha \rightarrow \nu_\alpha l_\beta \bar{\nu}_\alpha)}\,.
\end{equation}

The present experimental limits on the branching ratios entering the above expression are shown in Table~\ref{Tabrare}. The combination of constraints coming from electroweak decays leads then to the following limits\footnote{We report here the result obtained in~\cite{Antusch:2006vwa}, improved by considering the most recent BABAR bounds on the radiative $\tau$ decays shown in Table~\ref{Tabrare}. In practice, this only affects the limits on $|(\bN\bN^\dag)_{\tau \mu}|$ and $|(\bN\bN^\dag)_{\tau e}|$ [see Eq.~\eqref{LVFNU}].} for $|\bN\bN^\dag|$:
\begin{equation}
\label{NNEW}
|\bN\bN^\dag| \approx {\small \left(\begin{array}{ccc}
 1.002\pm 0.005   & <7.2\times 10^{-5}  & <8.8\times 10^{-3} \\
 <7.2\times 10^{-5}   & 1.003\pm 0.005       & < 10^{-2} \\
 <8.8\times 10^{-3}   & < 10^{-2}       & 1.003\pm 0.005
\end{array}
\right)}\,.
\end{equation}

In conclusion, data from weak decays provide strong constraints on the unitarity of the lepton mixing matrix, which is satisfied at the percent level. The improvement of the limits on the rare charged-lepton decays will further improve the bounds on leptonic NU effects. Moreover, future precision measurements performed in neutrino oscillation facilities will certainly play a crucial role in testing unitarity in the lepton mixing. It is also worth emphasizing that the above conclusions were drawn taking MUV as a reference framework in the analysis of lepton NU. If one goes beyond this simple scenario and considers particular cases with NU effects due to new physics, then other constraints may arise. For instance, if fermion triplets are added to the SM particle content, as in the type III seesaw mechanism, decay processes like $l_j\rightarrow l_il_kl_k$ (cf. Table~\ref{Tabrare}) or $\mu$-$e$ conversion in nuclei are possible at tree level. Consequently, the constraints imposed on the NU of the lepton mixing matrix become stronger in this case when compared with the MUV ones. In particular, from the present bound on the $\mu$-$e$ conversion rate, one obtains $|(\bN\bN^\dag)_{e\mu}|<1.7\times 10^{-7}$. Furthermore, the $|(\bN\bN^\dag)_{e\tau}|$ and $|(\bN\bN^\dag)_{\mu\tau}|$ bounds are also improved down to the level of $\sim 10^{-3}$ when considering the experimental bounds on the $\tau\rightarrow 3l$ rates~\cite{Abada:2008ea}.

\subsubsection{Nonunitarity and leptonic CPV}
\label{sec3.6.3}

In analogy with the quark sector, the observation of LCPV would automatically raise the question on whether this signal can be explained within a minimal framework in which the only source of CPV in neutrino oscillations is the Dirac phase $\delta$. This could not be the case if lepton mixing is nonunitary. For instance, in the previously discussed MUV framework, three extra phases in the leptonic mixing matrix $\bN$ act as new sources of LCPV. At present, these phases are not bounded by the available neutrino oscillation and electroweak data. Although the MUV is a representative scenario of NU in the lepton sector, it has been shown that there is room for considerable new CPV effects even in such a limited framework~\cite{Altarelli:2008yr,FernandezMartinez:2007ms}.

Following the notation of~\cite{FernandezMartinez:2007ms}, one can parametrize deviations from unitarity by writing $\bN = (\openone + \bm{\eta})\bU$, where $\bm{\eta}$ is a Hermitian matrix containing nine new parameters (six moduli and three phases). The bounds on $\bm{\eta}_{\alpha\beta}$ can be easily obtained from the ones on $\bN \bN^\dag$ considering that $(\bN \bN^\dag)_{\alpha\beta}\simeq \delta_{\alpha\beta}+2\bm{\eta}_{\alpha\beta}$~\cite{FernandezMartinez:2007ms}. The main question is then how much room do these possible deviations from unitarity leave for the observation of nonstandard CP violation in neutrino oscillations. In order to understand this, one has to write the transition probabilities $P_{\alpha\beta}$ and CP asymmetries $A_{\alpha\beta}\equiv (P_{\alpha\beta}-\bar{P}_{\alpha\beta})/(P_{\alpha\beta}+\bar{P}_{\alpha\beta})$ in the MUV framework~\cite{FernandezMartinez:2007ms,Altarelli:2008yr,Goswami:2008mi}, which will receive new contributions from $\bm{\eta}_{\alpha\beta}\equiv \eta_{\alpha\beta}e^{i\theta_{\alpha\beta}}$, where $\theta_{\alpha\beta}$ are the new CP-violating phases.

\begin{figure*}[]
\begin{tabular}{cc}
{\scriptsize (a)} & \hspace*{0.7cm}{\scriptsize(b)}\\
\includegraphics[width=6.6cm]{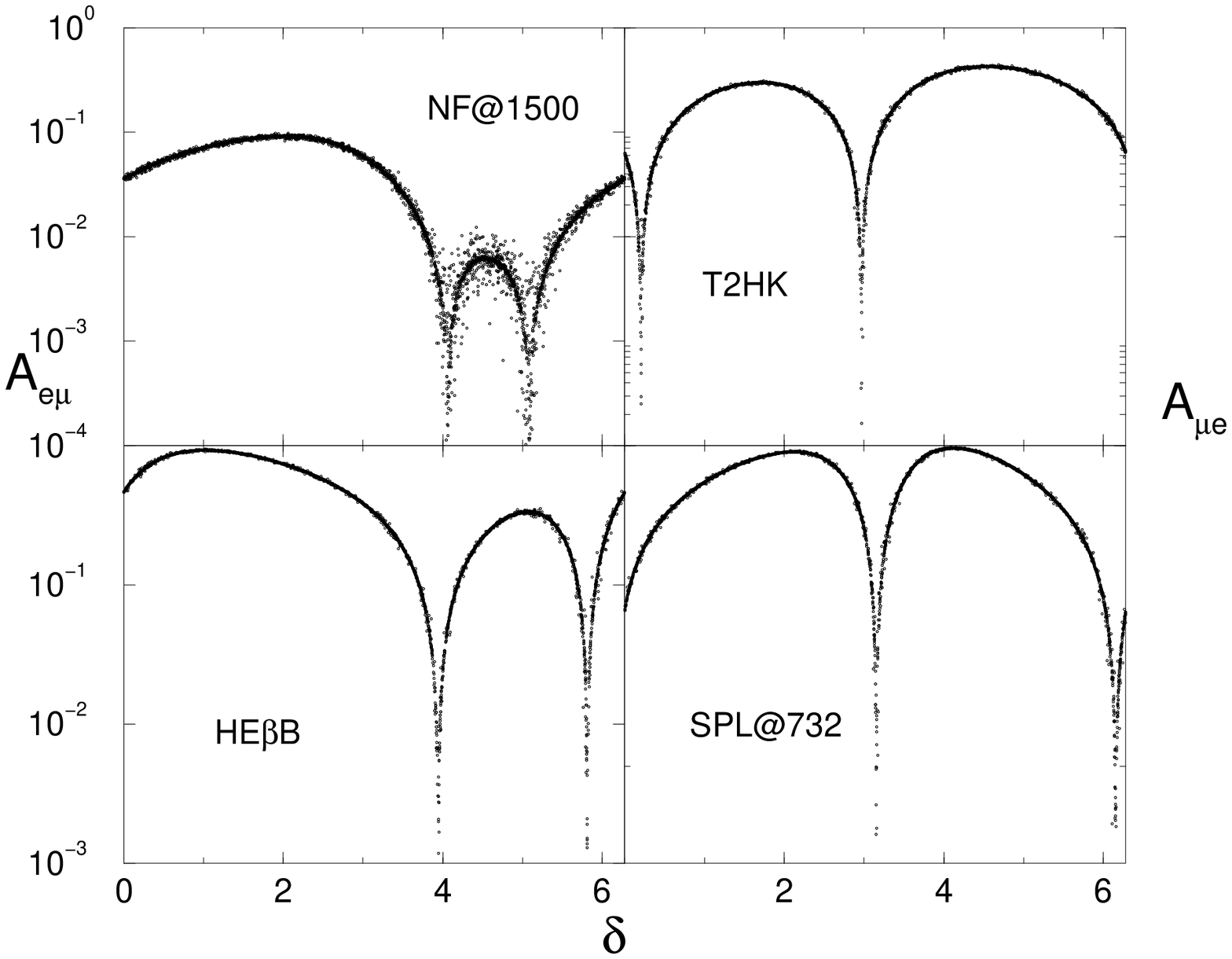}
&\includegraphics[width=6.1cm]{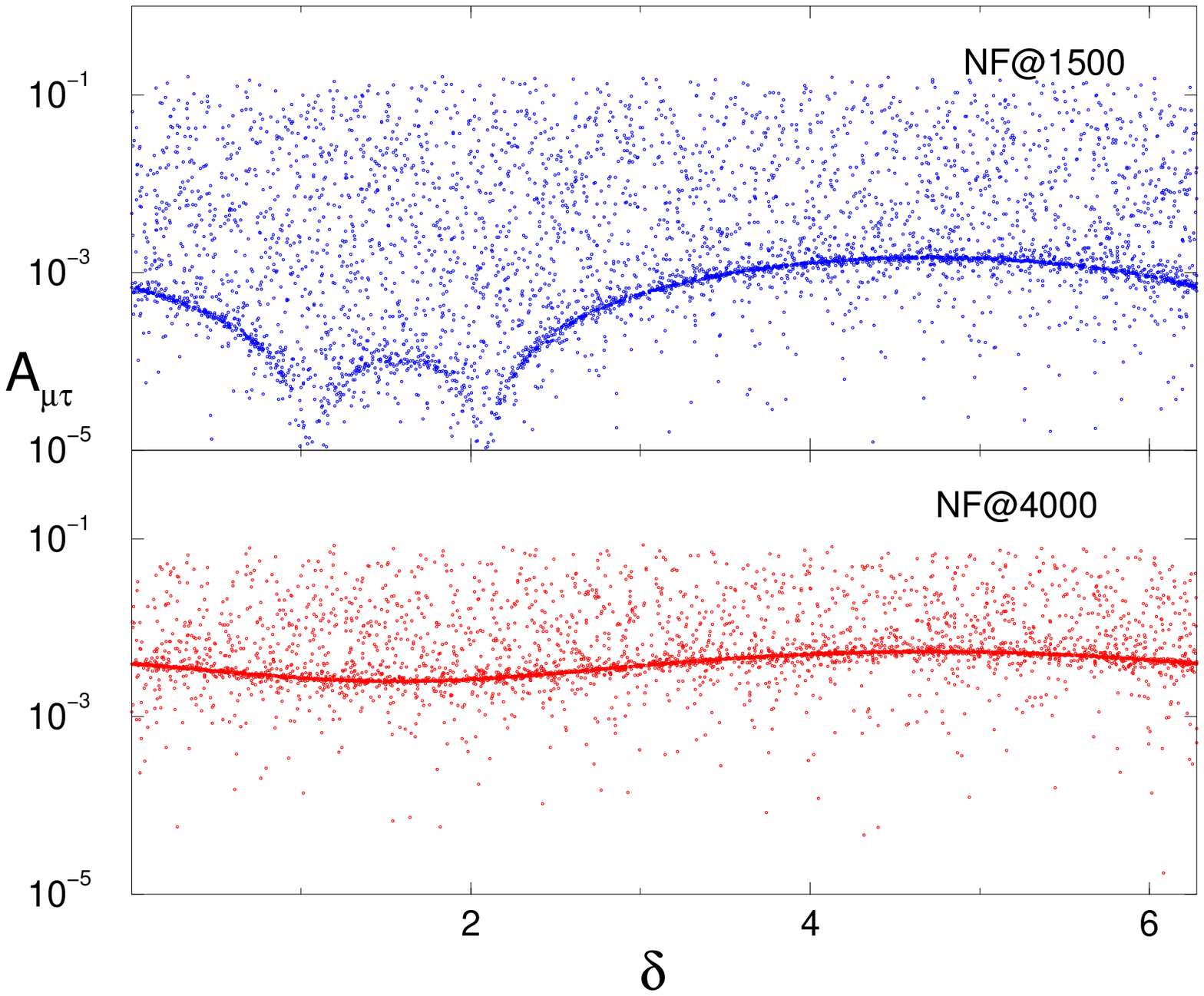}     \\
\hspace*{0.7cm}{\scriptsize (c)}
&\hspace*{0.7cm} {\scriptsize(d)}         \\
\includegraphics[width=6.2cm,height=4.8cm]{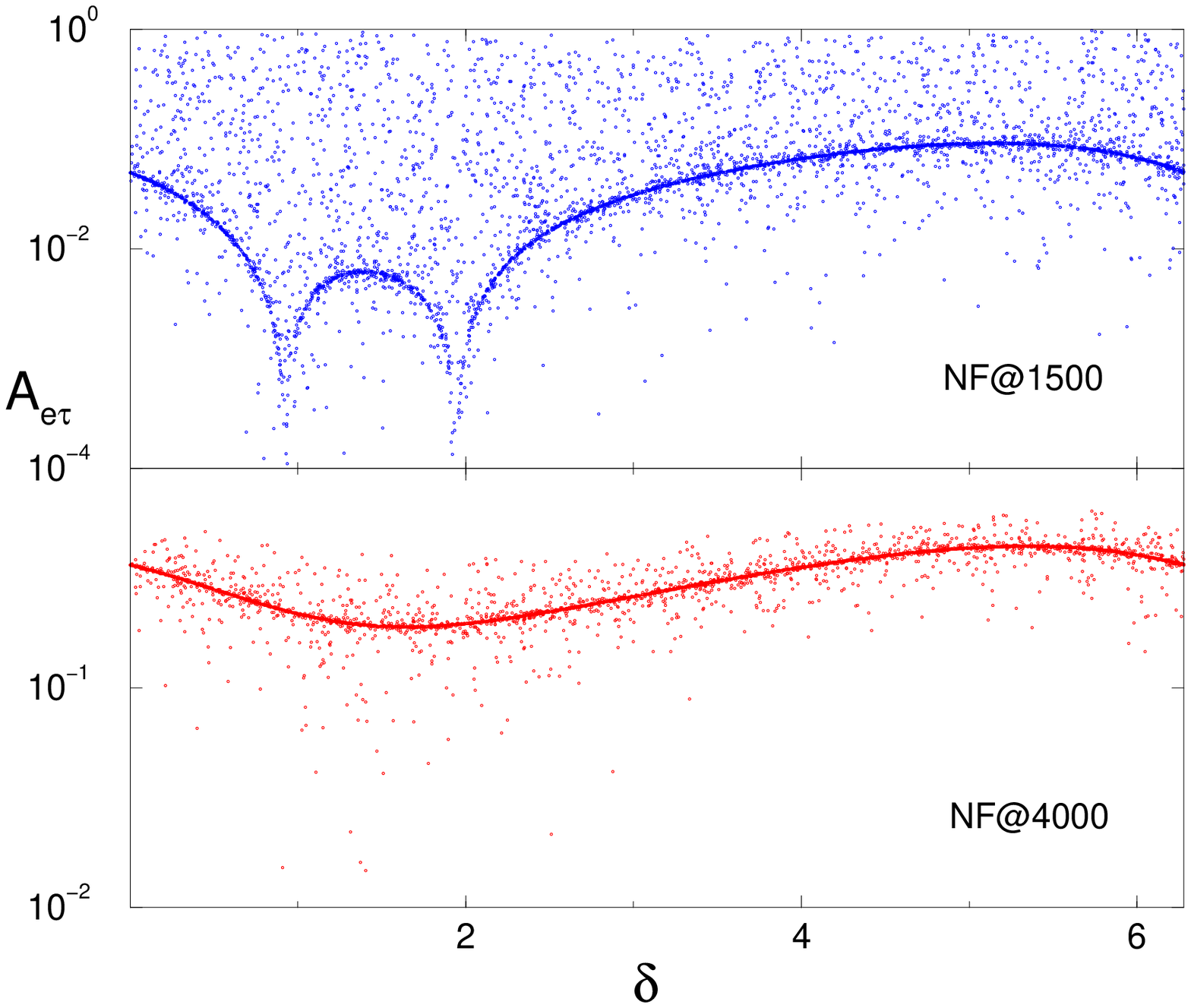}
& \includegraphics[width=6.7cm,height=5.0cm]{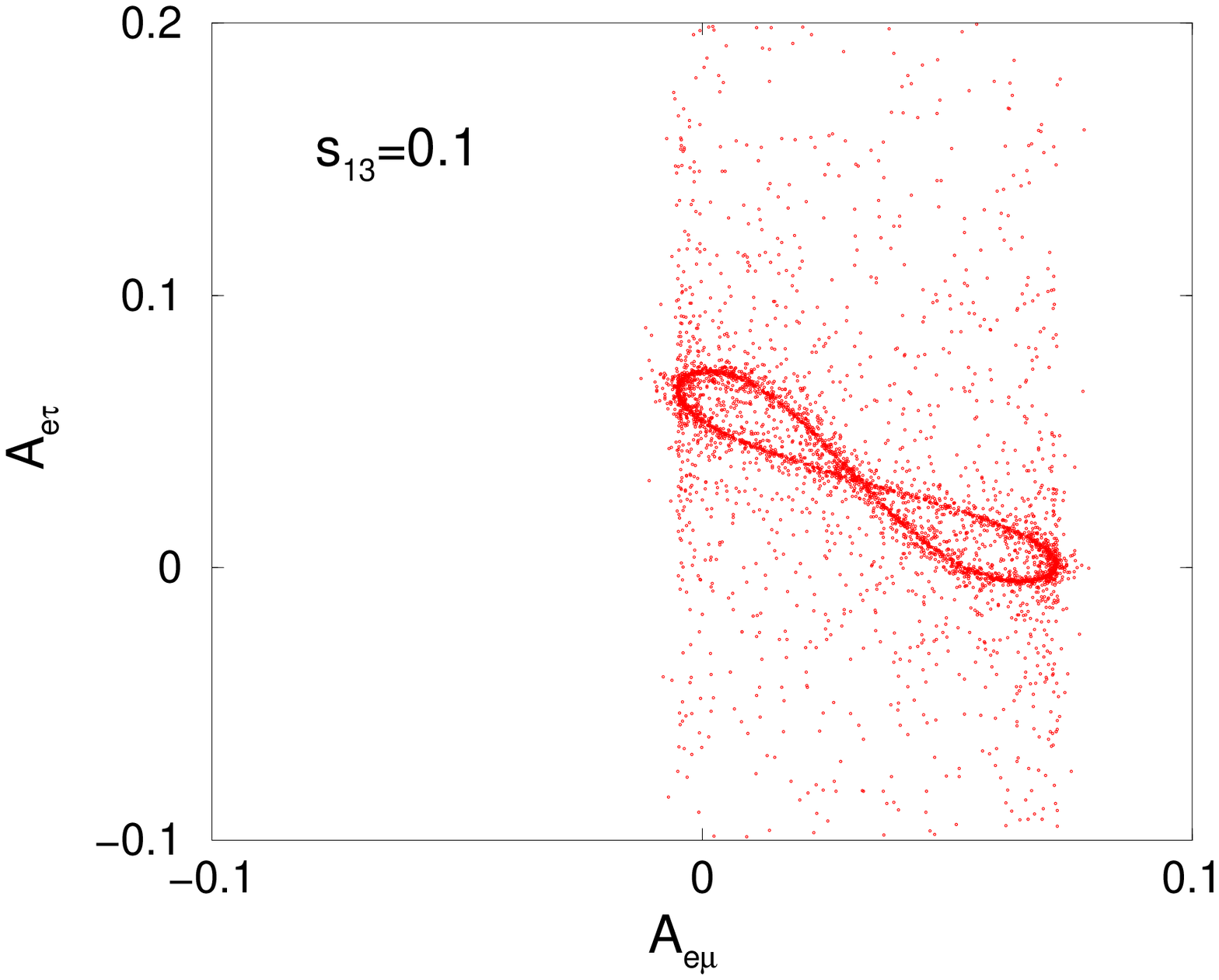}
\end{tabular}
\caption{(a) to (c) Scatter plots for $|A_{e\mu}|$, $|A_{\mu\tau}|$ and $|A_{e\tau}|$, respectively, as a function of the Dirac phase $\delta$. The neutrino parameters are fixed at $s_{13}=0.1$, $s_{12}^2=1/3$, $\theta_{23}=\pi/4$, $\dmsol=8\times 10^{-5}\, \eV^2$, and $\dmatm=2.4\times 10^{-3}\,\eV^2$. The results are presented considering several experimental setups, namely, ${\rm HE}\beta {\rm B}$ (high-energy beta beam with $E=1 \,\GeV$ and $L=732\,\km$), the upgraded T2K, T2HK ($E=0.75\,\GeV$ and $L=295\,\km$), the CERN superbeam project SPL ($E=0.3\,\GeV$ and $L=130\,\,{\rm or}\,\,732\,\km$), and neutrino factories (NF@$L$) with $E=35$ and $30\,\GeV$ in panels (b) and (c), respectively. (d) $A_{e\tau}$ as a function of $A_{e\mu}$ considering a baseline $L=1500\,\km$ and $E=30\,\GeV$. The neutrino parameters are the same as in the previous panels. In all cases, the MUV parameters are varied in their allowed ranges and the solid lines correspond to the standard unitary limit. From~\cite{Altarelli:2008yr}.}
\label{Aalphamu}
\end{figure*}

In the MUV framework, the golden channel asymmetries $A_{e\mu}$ do not deviate significantly from the standard unitary case due to the strong bounds on $\eta_{e\mu}$. Since the new physics effects are already constrained to be small in this case, the above channel is probably the most appropriate for a clean determination of lepton mixing parameters. On the other hand, the transition probabilities and their corresponding asymmetries for the remaining oscillation channels may be considerably affected by new physics effects. For instance, for $\mu\tau$ oscillations~\cite{Altarelli:2008yr,FernandezMartinez:2007ms}
\begin{equation}
\label{Amutau}
A_{\mu\tau}\simeq A_{\mu\tau}^{SM}-4\eta_{\mu\tau}\cot\Delta_{31}\sin\delta_{\mu\tau}\,,
\end{equation}
where $A_{\mu\tau}^{SM}$ is the CP asymmetry in the standard unitary scenario, which is typically $\mathcal{O}(10^{-3})$, while the new contribution proportional to $\eta_{\tau\mu}$ can be as large as $\sim 10^{-1}$. In Fig.~\ref{Aalphamu}, we show the behavior of the CP asymmetries $A_{\alpha\beta}$ as a function of the Dirac CP phase $\delta$ [panels (a) to (c)] for $s_{13}=0.1$ and several experimental setups (see the figure caption for more details). The parameters $\eta_{\alpha\beta}$ are varied in their allowed intervals and the phases $\delta_{\alpha\beta}$ are kept free. From Fig.~\ref{Aalphamu}a it is apparent the small impact of the new physics effects on $A_{e\mu}$ in the MUV framework. One should, however, keep in mind that in a more general picture with other new physics effects, the deviations with respect to the standard unitary scenario could be more significant. As for the $\mu\tau$ and $e\tau$ asymmetries, the NU effects can be quite dramatic, as illustrated in Figs.~\ref{Aalphamu}b and \ref{Aalphamu}c, where the solid lines indicate the result in the unitary case for which $\eta_{\alpha\beta}=0$. This analysis shows that the new physics effects are more pronounced for the facilities with the smallest $L/E$, which makes neutrino factories with small baselines and large $E$ more appropriate for the detection of new physics effects in $\nu_\mu \rightarrow \nu_\tau$~\cite{Goswami:2008mi}.

The standard unitary picture for LCPV would be automatically disproved in case one or more asymmetries are not compatible with their bounds. If indeed the neutrino mixing and LCPV patterns are described by a unitary matrix, then the trajectory spanned by a pair of asymmetries is a well-defined line which is obtained by varying the value of $\delta$. Therefore, in the standard unitary scenarios, any pair of measured asymmetries should fall in the corresponding line. Once one considers the MUV framework, the allowed space is enlarged outside these lines. This is shown in Fig.~\ref{Aalphamu}d, where $A_{e\tau}$ is plotted against $A_{e\mu}$ (the least affected asymmetry), varying the MUV parameters in their allowed ranges. From this plot, one clearly distinguishes the closed line which corresponds to the case in which $\eta_{\alpha\beta}=0$. Moreover, it is clear that the deviations to the standard unitary limit allowed by the present bounds on the MUV parameters are quite significant. One should also keep in mind that these results have been obtained in the MUV scenario, in which the new physics effects are pretty much constrained. Larger deviations to the standard unitary case could be observed in other frameworks with a wider allowed range for $A_{e\mu}$. Moreover, one should also take into account the experimental accuracy in the determination of the asymmetries, and the impact of the degeneracies discussed in Sec.~\ref{sec.3.2.3}, which can make the task of testing the standard LCPV framework more difficult~\cite{FernandezMartinez:2007ms,Altarelli:2008yr,Goswami:2008mi}. In particular, it has been shown that deviations from the standard picture of LCPV could be established with a modest precision, when considering the uncertainties on the $A_{\alpha\beta}$ asymmetries. This has been confirmed for a particular NF setup with detectors at $L=1500\,\km$~\cite{Altarelli:2008yr} and $E=50\,\GeV$.

\section{LEPTONIC CP VIOLATION AND THE ORIGIN OF MATTER}
\label{sec4}

If we take for granted that inflation~\cite{Linde:2007fr} took place in the early Universe, any primordial cosmological charge asymmetry would have been exponentially wiped out during the inflationary period. Thus, rather than being an initial accidental state, the observed dominance of matter over antimatter should be dynamically generated. In 1967, more than a decade before inflation was put forward and just three years after the discovery of CP violation in the $K_L \rightarrow 2\pi$ decays, Sakharov realized the need for generating the baryon asymmetry through a dynamical mechanism. Three necessary ingredients to create a baryon asymmetry from an initial state with a baryon number equal to zero were formulated in his work~\cite{Sakharov:1967dj}\footnote{Sakharov did not enunciate these conditions as clearly as they are traditionally presented. The three key assumptions in his seminal paper ``Violation of CP-invariance, C asymmetry, and baryon asymmetry of the Universe" are now known as the Sakharov conditions.}: (i) baryon number violation, (ii) C and CP violation, (iii) departure from thermal equilibrium.

The need for B violation is somehow obvious. If B is conserved by the interactions, and our Universe is initially symmetric (${\rm B}=0$), then no baryon production may take place. Indeed, since the baryon number commutes with the Hamiltonian $\mathcal{H}$, i.e. $[B,\mathcal{H}]=0$, at any time one has $B(t)=\int_0^t\,[B,\mathcal{H}]\,dt^\prime=0$. Thus, if B is conserved, the present asymmetry can only reflect asymmetric initial conditions. In grand unified theories, quarks and leptons are unified in the same multiplets, thus baryon number violation mediated by gauge bosons and scalars is natural. In the SM, however, the baryon number and the lepton flavor numbers (${\rm L}_{e,\mu,\tau}$) are accidentally conserved, and it is not possible to violate these symmetries at any perturbative level. Nevertheless, due to the chiral anomaly, nonperturbative instanton effects may give rise to processes that violate $({\rm B} + {\rm L})$ while conserving $({\rm B}-{\rm L})$~\cite{tHooft:1976up,tHooft:1976fv} . Although exponentially suppressed at zero temperature, such configurations, often referred to as sphalerons~\cite{Klinkhamer:1984di}, are frequent in the early Universe, at temperatures above the electroweak phase transition~\cite{Kuzmin:1985mm}.

The second Sakharov condition, namely, the violation of C and CP symmetries, is more
subtle. The baryon number operator,
\begin{align}\label{Boper}
\hat{B} = \frac{1}{3} \sum_i \int d^3x:\psi_i^{\dagger}({\bf x},t)\psi_i({\bf x},t):,
\end{align}
where $\psi_i({\bf x},t)$ denotes the quark field of flavor $i$ and the colons represent the normal ordering, is C odd and CP odd. This can easily be seen by recalling how the $C$, $P$, and $T$ operators act on the  quark fields. Using the standard phase convention,
\begin{align}\label{CPTtransf1}
\begin{split}
P\psi_i({\bf x},t)P^{-1} & =  \gamma^0\psi_i({-\bf x},t), \\
P\psi_i^{\dagger}({\bf x},t)P^{-1} & = \psi_i^{\dagger}({-\bf x},t) \gamma^0, \\
C\psi_i({\bf x},t)C^{-1} & =  i\gamma^2 \psi_i^{\dagger}({\bf x},t), \\
C\psi_i^{\dagger}({\bf x},t)C^{-1} & =  i \psi_i({\bf x},t)\gamma^2, \\
T\psi_i({\bf x},t)T^{-1} & =  -i \psi_i({\bf x},-t)\gamma_5\gamma^0\gamma^2, \\
T\psi_i^{\dagger}({\bf x},t)T^{-1} & =  -i \gamma^2 \gamma^0\gamma_5 \psi_i^{\dagger}({\bf x},-t).
\end{split}
\end{align}
Thus,
\begin{align}\label{CPTtransf2}
\begin{split}
P:\psi_i^{\dagger}({\bf x},t)\psi_i({\bf x},t):P^{-1}= :\psi_i^{\dagger}({-\bf x},t)\psi_i({-\bf x},t):,\\
C:\psi_i^{\dagger}({\bf x},t)\psi_i({\bf x},t):C^{-1} = - :\psi_i^{\dagger}({\bf x},t)\psi_i({\bf x},t):, \\
T:\psi_i^{\dagger}({\bf x},t)\psi_i({\bf x},t):T^{-1}= :\psi_i^{\dagger}({\bf x},-t)\psi_i({\bf x},-t):,
\end{split}
\end{align}
and one obtains
\begin{align} \label{CPtransf1}
& C \hat{B} C^{-1}  =  -\hat{B}, \quad (CP) \hat{B} (CP)^{-1} =  -\hat{B},\nonumber \\
& (CPT) \hat{B} (CPT)^{-1} =  -\hat{B}.
\end{align}
If C is conserved, then $[C,\mathcal{H}]=0$ and from the time evolution of $\hat{B}$ and
Eq.~\eqref{CPtransf1} one concludes
\begin{align}
    \langle\hat{B}(t)\rangle & =\langle e^{i\mathcal{H}t} \hat{B}(0) e^{-i\mathcal{H}t}\rangle = \langle C^{-1} e^{i\mathcal{H}t} C \hat{B}(0) C^{-1} e^{-i\mathcal{H}t} C \rangle \nonumber\\
    & =- \langle e^{i\mathcal{H}t} \hat{B}(0) e^{-i\mathcal{H}t}\rangle = -\langle\hat{B}(t)\rangle.
\end{align}
Therefore, a nonzero expectation value $\langle\hat{B}\rangle$ requires that the Hamiltonian violates C. The same arguments apply to the CP symmetry.

Finally, the third Sakharov requirement can be understood as follows. In thermal equilibrium, thermal averages are described by the density operator $\rho=\exp(-\beta \mathcal{H})$, with $\beta=1/T$. If the Hamiltonian is CPT invariant, using Eq.~\eqref{CPtransf1} it then follows
\begin{align}
 \langle \hat{B} \rangle _T & = \mathop{\rm Tr}\nolimits ( e^{-\beta \mathcal{H}} \hat{B} ) = \mathop{\rm Tr}\nolimits [(CPT)(CPT)^{-1} e^{-\beta \mathcal{H}} \hat{B}] \nonumber\\
  & = \mathop{\rm Tr}\nolimits [e^{-\beta \mathcal{H}}(CPT)^{-1}\hat{B}(CPT)] =  - \mathop{\rm Tr}\nolimits ( e^{-\beta H} \hat{B} ) \nonumber\\ &= - \langle \hat{B} \rangle _T,
 \end{align}
i.e. $\langle \hat{B} \rangle _T=0$ in thermal equilibrium. In other words, in thermal equilibrium the rate for a given process that produces an excess of baryons is equal to the rate of its corresponding inverse process, so that no net asymmetry can be generated since the inverse process destroys the baryon excess as fast as the direct process creates it. Departure from thermal equilibrium is very common in the early Universe, when interaction rates cannot keep up with the expansion rate. A simple example is provided by the out-of-equilibrium decay of a heavy particle $X$ with a mass $M_X>T$ at time of decay. In this case, the rate of the direct process is of order $T$, while the inverse decay rate is Boltzmann suppressed $\sim \exp(-M_X/T)$.

The present value of the baryon asymmetry of the Universe inferred from WMAP 7-year data combined with baryon acoustic oscillations is~\cite{Komatsu:2010fb}
\begin{align}\label{etaobs}
    \eta_B \equiv \frac{n_B-n_{\bar{B}}}{n_\gamma}=(6.20 \pm 0.15) \times 10^{-10},
\end{align}
where $n_B, n_{\bar{B}}$ and $n_\gamma$ are the number densities of baryons, antibaryons, and photons at present time, respectively\footnote{An equivalent definition of the baryon asymmetry is the baryon-to-entropy ratio $Y_B = (n_B-n_{\bar{B}})/s$. The two measures are related as $Y_B \approx \eta_B/7.04$.}. The explanation of such a small but nonzero number poses a challenge to both particle physics and cosmology. It is remarkable that the SM contains the three Sakharov ingredients. Yet not all of them are available in a sufficient amount. The baryon number is violated by the electroweak sphaleron processes, which are fast and unsuppressed in the early Universe. The C symmetry is maximally violated by the weak interactions, and CP is violated by the CKM phase. Nevertheless, if baryogenesis occurs at the electroweak phase transition scale $T_{\rm ew} \sim \mathcal{O}(100)$~GeV, the strength of CP violation, parametrized in the SM by the invariant $\mathcal{J}^\text{CP}_\text{quark}$ of Eq.~\eqref{ICPquark}, seems insufficient to generate the required value of $\eta_B$. The naive estimate $\mathcal{J}^\text{CP}_\text{quark}/T_{\rm ew}^{12} \sim 10^{-20}$ indicates that at such temperatures electroweak baryogenesis~\cite{Trodden:1998ym} requires new sources of CP violation.\footnote{In the cold electroweak baryogenesis scenarios, where baryogenesis takes place at temperatures well below $T_{\rm ew}$, the strength of CP violation in the SM may be enough to account for the observed $\eta_B$~\cite{Krauss:1999ng,GarciaBellido:1999sv,Tranberg:2009de,Enqvist:2010fd}.} Finally, at the electroweak phase transition departure from thermal equilibrium takes place. However, a successful baryogenesis requires a strongly first order phase transition, which can occur if the Higgs mass is rather light, $m_\text{Higgs} \lesssim 70$~GeV. This value is nevertheless well below the present experimental lower bound $m_H > 114.4$~GeV~\cite{Nakamura:2010zzi}. Thus, the explanation of the baryon asymmetry observed in our Universe requires new physics beyond the SM.

Among the several viable baryogenesis scenarios, leptogenesis~\cite{Fukugita:1986hr} is undoubtedly one of the simplest, most attractive, and well-motivated mechanisms. Many aspects of leptogenesis have been widely discussed in the literature and there are excellent reviews on the subject [see, for instance,~\cite{Buchmuller:2004nz,Buchmuller:2005eh,Davidson:2008bu}]. In its simplest realization, new heavy (bosonic or fermionic) particles are introduced in the theory in such a way that the interactions relevant for leptogenesis are simultaneously responsible for the nonvanishing and smallness of the neutrinos masses via the seesaw mechanism. The three Sakharov conditions are naturally fulfilled in this framework: the seesaw mechanism requires lepton-number violation and sphalerons partially convert the lepton asymmetry into a baryon asymmetry; neutrino complex Yukawa couplings provide the necessary source of CP violation; and last, departure from thermal equilibrium is guaranteed by the out-of-equilibrium decays of the new heavy particles. It is precisely on these simple thermal leptogenesis scenarios that this section of the review focuses. We do not aim at covering all the theoretical ideas on leptogenesis extensively developed over the last years. It is our goal, instead, to describe the role that leptonic CP violation may have played in the origin of matter.

\subsection{Leptogenesis mechanisms}
\label{sec4.1}

In this section, we briefly review the simplest nonsupersymmetric leptogenesis scenarios based on the seesaw mechanism for neutrino masses. As discussed in Sec.~\ref{sec2.5}, seesaw models are characterized by the properties of the exchanged heavy particles. In particular, in type I, type II and type III seesaw mechanisms, these particles are $SU(3)\times SU(2)\times U(1)$-singlet fermions, $SU(2)$-triplet scalars, and $SU(2)$-triplet fermions, respectively. As it turns out, thermal leptogenesis can be successfully implemented in each framework. Yet, in general, specific constraints must be satisfied in order to generate the required value of the baryon asymmetry.

The baryon asymmetry $\eta_B$ produced by thermal leptogenesis can be obtained by taking into account the suppression factors given by the Sakharov conditions. The final asymmetry is the result of the rivalry between the processes that produce it and the washout processes that tend to erase it. Assuming that after inflation the Universe reheats to a thermal bath composed of particles with gauge interactions, the asymmetry can be estimated as the product of three factors: (the leptonic CP asymmetry $\epsilon$ produced in heavy particle decays)$\times$ (an efficiency factor $\eta$ due to washout processes in scattering, decays, and inverse decays) $\times$ (a reduction factor due to chemical equilibrium, charge conservation, and the redistribution of the asymmetry among different particle species by fast processes). The computation of each of these factors is model dependent. In particular, the calculation of the efficiency factor $\eta\, (0 \leq \eta \leq 1)$ requires the solution of a full set of Boltzmann equations which describe the out-of-equilibrium dynamics of the processes involving the heavy particles responsible for leptogenesis. Simple analytical estimates can also be obtained in some specific regimes~\cite{Buchmuller:2004nz,Giudice:2003jh,Abada:2006ea}.

Departure from thermal equilibrium is provided by the expansion of the Universe, characterized by the Hubble expansion rate $H(T) \sim 1.66 g_\ast^{1/2} T^2/M_P$, where $g_\ast$ is the number of relativistic degrees of freedom in the thermal bath ($g_\ast =106.75$ within the SM) and $M_P=1.22\times10^{19}$~GeV is the Planck mass. Nonequilibrium takes place whenever a crucial interaction rate becomes smaller that $H$ so that it is not fast enough to equilibrate particle distributions. Furthermore, flavor effects can play a significant role in this process. As first discussed in~\cite{Barbieri:1999ma,Endoh:2003mz} and more recently emphasized in~\cite{Pilaftsis:2005rv,Abada:2006fw,Nardi:2006fx,Abada:2006ea}, when the interactions mediated by the charged-lepton Yukawa couplings are in thermal equilibrium, the flavored leptonic asymmetries and the Boltzmann equations for individual flavor asymmetries must be properly taken into account. Since the time scale for leptogenesis is $H^{-1}$ and the typical interaction rates for the charged-lepton Yukawa couplings $y_\alpha$ are $\Gamma_\alpha \simeq 10^{-2} y^2_\alpha T$~\cite{Cline:1993bd}, interactions involving the $\tau$ and $\mu$ Yukawa couplings are in equilibrium for $T \lesssim 10^{12}$~GeV  and $T \lesssim 10^{9}$~GeV, respectively. Below these temperature scales the corresponding lepton doublets are distinguishable mass eigenstates and, as such, should be properly introduced into the Boltzmann equations.

Since the leptonic CP asymmetries are the relevant quantities in establishing a link between leptonic CP violation and the matter-antimatter asymmetry, in what follows we discuss these quantities in more detail within each seesaw framework\footnote{The main conclusions of this section are expected to remain valid also in the minimal supersymmetric extension of each framework. Although new decay channels will enhance the generated CP asymmetry, these additional contributions tend to be compensated by the washout processes which are typically stronger than in the nonsupersymmetric case.}. Readers interested in a more complete understanding of the mechanism of leptogenesis are referred, \textit{e.g.}, to the recent pedagogical review~\cite{Davidson:2008bu} and the extensive list of references quoted therein.

\subsubsection{Type I seesaw leptogenesis}
\label{sec4.1.1}

\begin{figure*}[t]
\includegraphics{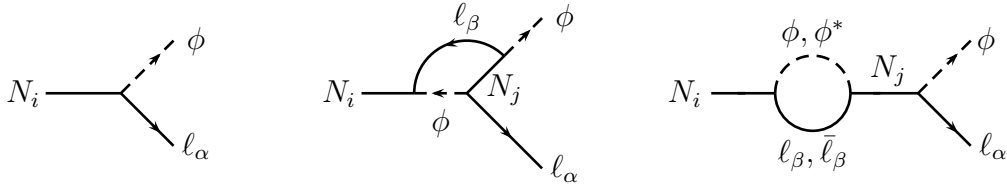}
\caption{\label{fig51} Diagrams contributing to the CP asymmetry $\epsilon_i^{\alpha}$ in
type I seesaw leptogenesis. The last diagram corresponds to the wave function
corrections: the one with an internal $\ell_\beta$ is lepton flavor and lepton-number
violating, while the one with an internal $\bar{\ell}_\beta$ is lepton flavor violating
but lepton-number conserving, thus giving no contribution to the unflavored CP
asymmetry.}
\end{figure*}

In the type I seesaw framework, at least two singlet fermions must be added to the SM particle content to correctly reproduce the observed neutrino mass square differences. The existence of more than one singlet fermion also turns out to be crucial for the mechanism of thermal leptogenesis. We consider the SM extended by three singlet fermions $N_i\, ($i=1,2,3$)$ with large Majorana masses $M_i$. In this case, the relevant Lagrangian interactions terms are given by Eq.~\eqref{LtypeIseesaw}. Working in the mass eigenbasis of the heavy neutrinos $N_i$ and the charged leptons $\ell_\alpha$, the CP asymmetry $\epsilon_i^\alpha$ in the lepton flavor $\alpha$ produced in the $N_i$ decays is given by
\begin{widetext}
\begin{align}\label{cpasymI}
          \epsilon_i^\alpha & \equiv \frac{\Gamma(N_i\rightarrow \phi \ell_\alpha)-\Gamma(N_i\rightarrow \phi^\dagger \bar{\ell}_\alpha)}{\sum_\beta \left[\Gamma(N_i\rightarrow \phi \ell_\beta)+\Gamma(N_i\rightarrow \phi^\dagger \bar{\ell}_\beta)\right]}\nonumber \\
         & = \frac{1}{8\pi}\frac{1}{\mathbf{H}_{ii}^\nu}\sum_{j \neq i} \left\{ \text{Im}\left[\mathbf{Y}^{\nu\ast}_{\alpha i} \mathbf{H}^\nu_{ij} \mathbf{Y}^\nu_{\alpha j}\right]\,\left(f(x_{j})
         + g(x_{j})\right) + \text{Im}\left[\mathbf{Y}^{\nu\ast}_{\alpha i}\mathbf{H}^\nu_{ji} \mathbf{Y}^\nu_{\alpha j}\right] g^\prime(x_{j})\right\},
    \end{align}
\end{widetext}
where $\mathbf{H}^\nu \equiv \mathbf{Y}^{\nu\dagger}\mathbf{Y}^\nu$, $x_{j} =
M_j^2/M_i^2$, and
\begin{align}\label{loopfunc}
    \begin{split}
       f(x) & =\sqrt{x}\left[1-(1+x)\ln\left(1+x^{-1}\right)\right], \\
         g(x) &= \sqrt{x}\, g^\prime(x) = \frac{\sqrt{x}\,(1-x)}{(x-1)^2+a_j^2},\quad a_j = \frac{\Gamma_{N_j}}{M_i},
     \end{split}
\end{align}
are the vertex and self-energy one-loop functions, respectively. The quantity
$\Gamma_{N_j}$ denotes the $N_j$ total tree-level decay rate,
\begin{align}\label{GammaDj}
    \Gamma_{N_j} = \frac{\mathbf{H}^\nu_{jj}M_j}{8\pi}.
\end{align}
The CP asymmetry given in Eq.~\eqref{cpasymI} arises from the interference of the tree-level and one-loop diagrams depicted in Fig.~\ref{fig51}~\cite{Covi:1996wh}. The presence of complex phases in the Yukawa couplings involved as well as nonzero absorptive parts in the loop diagrams are necessary conditions to have a nonvanishing asymmetry. The last diagram in Fig.~\ref{fig51} corresponds to the wave-function corrections. The diagram with an internal $\ell_\beta$ is lepton flavor and lepton number violating. On the other hand, the diagram with an internal $\bar{\ell}_\beta$ is lepton flavor violating but lepton number conserving. Thus it vanishes when summed over the lepton flavors~\cite{Covi:1996wh}.

We note that in the self-energy loop functions $g$ and $g^\prime$ of Eq.~\eqref{loopfunc} the corrections due to the mixing of nearly degenerate heavy Majorana neutrinos have been included. They are parametrized here through the quantities $a_j$~\cite{Pilaftsis:1997jf,Pilaftsis:2003gt,Pilaftsis:2005rv}. In~\cite{Anisimov:2005hr}, a different regulator of the loop functions was obtained in the degenerate limit $M_i \sim M_j$. Instead of $a_j^2$, the term $(\sqrt{x}\,a_j-a_i)^2$ was found. Both results agree when $\mathbf{H}^\nu_{jj} \gg \mathbf{H}^\nu_{ii}$. The above corrections become relevant in the so-called resonant leptogenesis scenario~\cite{Pilaftsis:2003gt}, i.e., in the limit when the mass splitting between $N_i$ and $N_j$ is comparable with their decay widths.

Summing over the lepton flavors one recovers the standard result:
\begin{align}\label{cpasymIa}
       \epsilon_i = \sum_\alpha \epsilon_i^\alpha = \frac{1}{8\pi}\frac{1}{\mathbf{H}^\nu_{ii}}\sum_{j \neq i} \text{Im}\left[(\mathbf{H}_{ij}^{\nu})^2\right] \,\left(f(x_{j}) + g(x_{j})\right).
\end{align}

In the so-called $N_1$-dominated scenario with $M_1 \ll M_j\, (j=2,3)$, one has $x_{j} \gg 1$ and the one-loop functions are approximated by $f(x) \simeq -1/(2\sqrt{x}), g(x) \simeq -1/\sqrt{x}$, and $g^\prime(x) \simeq -1/x$. In this case, the flavored asymmetry in Eq.~\eqref{cpasymI} becomes
\begin{align}\label{cpasymIb}
       \epsilon_1^\alpha \simeq -\frac{3}{16\pi}\frac{1}{\mathbf{H}^\nu_{11}}\sum_{j \neq 1} \frac{M_1}{M_j}\,\text{Im}\left[\mathbf{Y}^{\nu\ast}_{\alpha 1}\mathbf{H}^\nu_{1j} \mathbf{Y}^\nu_{\alpha j}\right],
\end{align}
 while the unflavored asymmetry \eqref{cpasymIa} reads
\begin{align}\label{cpasymIc}
       \epsilon_1 \simeq -\frac{3}{16\pi}\frac{1}{\mathbf{H}^\nu_{11}}\sum_{j \neq 1} \frac{M_1}{M_j}\text{Im}\left[(\mathbf{H}_{1j}^{\nu})^2\right].
\end{align}

A remarkable feature of the unflavored asymmetry \eqref{cpasymIc} is that it has the upper bound~\cite{Hamaguchi:2001gw,Davidson:2002qv}
\begin{align}\label{epsupper}
    |\epsilon_1| &\lesssim \frac{3}{16\pi}\frac{M_1}{v^2}\left(m_\text{max} - m_\text{min}\right) \nonumber\\
    &\simeq 10^{-6}\left(\frac{M_1}{10^{10}\,\text{GeV}}\right)\left(\frac{m_\text{max} - m_\text{min}}{m_\text{atm}}\right),
\end{align}
where $v \approx 175$~GeV is the vacuum expectation value of the neutral component of the Higgs doublet; $m_\text{max}$ and $m_\text{min}$ are the largest and smallest light neutrino masses, respectively; $m_\text{atm}$ is the atmospheric neutrino mass scale. Moreover, this bound gets more stringent for a quasidegenerate light neutrino spectrum ($m_\text{max} \approx m_\text{min}$). On the other hand, the asymmetry in a given flavor~\eqref{cpasymIb} is bounded by~\cite{Abada:2006ea}
\begin{align}\label{epsupperflav}
    |\epsilon_1^\alpha| &\lesssim \frac{3}{16\pi}\frac{M_1m_\text{max}}{v^2}\sqrt{\frac{\mathbf{Y}^{\nu\ast}_{\alpha 1}\mathbf{Y}^\nu_{\alpha 1}}{\sum_\beta |\mathbf{Y}^\nu_{\beta 1}|^2}}\,,
\end{align}
which goes as the square root of the branching ratio to that flavor and is not suppressed
for a degenerate light neutrino spectrum.

From the requirement that leptogenesis successfully reproduces the baryon asymmetry in Eq.~\eqref{etaobs}, the bound in Eq.~\eqref{epsupper} leads to two important consequences~\cite{Buchmuller:2003gz,Giudice:2003jh,Buchmuller:2004nz}:
\begin{itemize}
  \item[(\emph{i})] A lower bound on $M_1$ and the reheating temperature of the
      Universe, $M_1, T_\text{reh} \gtrsim 2\times 10^9$~GeV.
  \item[(\emph{ii})] An upper bound on the light neutrino mass scale, $m \lesssim
      0.15$~eV.
\end{itemize}
While the bound in (\emph{i}) is not relaxed with the inclusion of flavor effects~\cite{Blanchet:2006be,JosseMichaux:2007zj}, the arguments leading to the bound in (\emph{ii}) do not apply in the flavored regime\footnote{In the unflavored regime, the upper bound on the neutrino mass scale can be relaxed if, for instance, the expansion rate of the Universe is modified at the leptogenesis epoch due to brane cosmology~\cite{Bento:2005je,Okada:2005kv}.}. There is presently no consensus on the precise upper bound on the light neutrino mass scale inferred from flavored leptogenesis. Analytical and numerical calculations~\cite{DeSimone:2006dd,JosseMichaux:2007zj} suggest that one can easily saturate the cosmological bound and reach values of $m$ up to 1 eV.

One may wonder whether the bound on $M_1$ (and $T_\text{reh}$) can be evaded without adding new particles or interactions. We recall that this bound applies only for hierarchical heavy neutrinos. For quasidegenerate $N_i$ the leptonic CP asymmetries can be much larger than the upper value of Eq.~\eqref{epsupper}. In particular, if $x_{j}-1 = a_j$ (or, equivalently, $|M_j - M_i| \simeq \frac{1}{2}\, \Gamma_{N_j}$), the asymmetries $\epsilon_i^\alpha$ are resonantly enhanced due to the self-energy contribution. In this case, the loop functions are approximately given by $g^\prime(x) \simeq g(x) \simeq 4\pi/\mathbf{H}^\nu_{jj}$ so that at the resonance
\begin{align}\label{cpasymIres}
           \epsilon_{i,\text{res}}^\alpha
          &\simeq -\frac{1}{2}\sum_{j \neq i} \left\{ \frac{\text{Im}\left[\mathbf{Y}^{\nu\ast}_{\alpha i} \mathbf{H}^\nu_{ij} \mathbf{Y}^\nu_{\alpha j}\right]}{\mathbf{H}^\nu_{ii}\mathbf{H}^\nu_{jj}} + \frac{\text{Im}\left[\mathbf{Y}^{\nu\ast}_{\alpha i}\mathbf{H}^\nu_{ji} \mathbf{Y}^\nu_{\alpha j}\right]}{\mathbf{H}^\nu_{ii}\mathbf{H}^\nu_{jj}} \right\}\nonumber\\
          &=-\sum_{j \neq i}  \frac{\text{Re}\,\left[\mathbf{H}^\nu_{ij}\right]\, \text{Im}\left[\mathbf{Y}^{\nu\ast}_{\alpha i}  \mathbf{Y}^\nu_{\alpha j}\right]}{\mathbf{H}^\nu_{ii}\mathbf{H}^\nu_{jj}}.
\end{align}
After summing over the flavors one finds
\begin{align}\label{cpasymIares}
       \epsilon_{i,\text{res}} = -\frac{1}{2} \sum_{j \neq i} \frac{\text{Im}\left[(\mathbf{H}_{ij}^{\nu})^2\right]} {\mathbf{H}^\nu_{ii}\mathbf{H}^\nu_{jj}}.
\end{align}
Thus, one concludes that the resonantly enhanced CP asymmetry is not suppressed by the light neutrino masses or the heavy Majorana masses; it is just bounded by unitarity, $|\epsilon_{i}|\leq 1/2$. This in turn implies that leptogenesis can occur at a much lower energy scale.

Although theoretically challenging, it is possible to construct models in which the heavy Majorana neutrino mass splitting is naturally as small as the decay width at the leptogenesis scale. For instance, in the so-called radiative resonant leptogenesis scenario~\cite{GonzalezFelipe:2003fi,Turzynski:2004xy,Branco:2005ye}, the required splitting can be generated by the renormalization group running from the GUT scale down to the leptogenesis scale, assuming that the heavy Majorana neutrinos are exactly degenerate at the GUT scale. The assumption of a completely degenerate right-handed neutrino spectrum at the GUT scale is compatible with the solar and atmospheric neutrino oscillation data~\cite{GonzalezFelipe:2001kr}. Such a degeneracy can be achieved, for instance, by imposing some discrete or Abelian symmetries~\cite{Branco:2005ye}, or in models with minimal lepton flavor violation~\cite{Cirigliano:2006nu,Branco:2006hz,Cirigliano:2007hb} as described in Sec.~\ref{sec2.7}.

\subsubsection{Type II seesaw leptogenesis}
\label{sec4.1.2}

\begin{figure*}[t]
\includegraphics{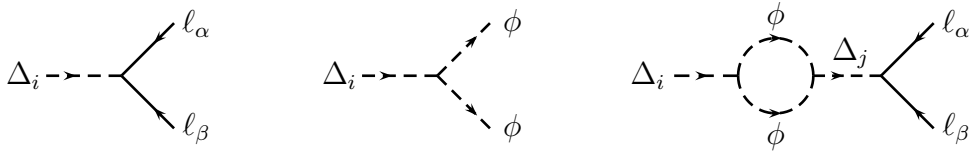}
\caption{\label{fig52} Tree-level diagrams for the scalar triplet decays and one-loop
diagram contributing to the CP asymmetry $\epsilon_{i}^{\alpha\beta}$ in type II seesaw
leptogenesis.}
\end{figure*}

As shown in Sec.~\ref{sec2.5}, the type II seesaw is very economical in the
sense that it has a single source of flavor structure, namely, the symmetric complex
Yukawa coupling matrix $\mathbf{Y}^\Delta$ that couples the $SU(2)_L$ scalar triplet
$\Delta$ to leptons. Furthermore, in its minimal realization, with only one scalar
triplet, the flavor pattern of $\mathbf{Y}^\Delta$ uniquely determines the flavor
structure of the low-energy effective neutrino mass matrix $\mathbf{m}_\nu$ of
Eq.~\eqref{mnutypeIIseesaw}. There is, however, a drawback with leptogenesis in this
minimal setup, namely, the leptonic CP asymmetry that is induced by the triplet decays is
generated only at higher loops and is highly suppressed. Therefore, new sources for
neutrino masses are required to implement thermal leptogenesis in a type II seesaw
framework~\cite{Ma:1998dx,Hambye:2000ui,Hambye:2003ka,Hambye:2003rt,DAmbrosio:2004fz}.
These new sources could come, e.g., from other type I, type II or type III
contributions. For illustration, we describe a simple
nonsupersymmetric leptogenesis scenario with only two scalar triplets, but other mixed
seesaw leptogenesis scenarios are conceivable as
well~\cite{Hambye:2003ka,Antusch:2004xy,Hambye:2005tk}. In particular, renormalizable
left-right symmetric theories and grand unified models based on $SO(10)$ provide a
natural framework for the simultaneous presence of singlet fermions and Higgs triplets.

We consider the SM extended with two scalar triplets $\Delta_i\, (i=1,2)$ of
hypercharge $+1$ (in the normalization with hypercharge $-1/2$ for the lepton doublets)
and masses $M_{\Delta_i}$. In the $SU(2)$ representation we write
\begin{align}
\Delta_i=\begin{pmatrix}
         \Delta^0_i & -\Delta^+_i/\sqrt{2} \\
         -\Delta^+_i/\sqrt{2} & \Delta^{++}_i \\
       \end{pmatrix}.
\end{align}
The relevant Lagrangian terms are given by Eq.~\eqref{LtypeIIseesaw}, which include now
the contributions from both scalar triplets,
\begin{align}
\mathcal{L}_{\Delta} \ni & \sum_{i} \left(- \mathbf{Y}_{\alpha\beta}^{\Delta_i}\, \ell_\alpha^T C \Delta_i \ell_\beta +\mu_i M_{\Delta_i} \tilde{\phi}^T\Delta_i\tilde{\phi} +\text{H.c.}\right)\nonumber\\
&-\sum_i M^2_{\Delta_i}\,\text{Tr}(\Delta_i^\dagger \Delta_i),
\end{align}
where $\mathbf{Y}^{\Delta_i}$ are symmetric $3\times3$ complex Yukawa coupling matrices,
and $\mu_i$ are dimensionless complex couplings.

In the presence of CP-violating interactions, the decay of $\Delta_i$ into two leptons
generates a nonvanishing leptonic asymmetry for each triplet component ($\Delta^0_i,
\Delta^+_i, \Delta^{++}_i$),
\begin{align} \label{cpasymII}
\epsilon_{i}^{\alpha\beta} = \Delta {\rm L} \times \frac{\Gamma (\Delta_i^* \rightarrow \ell_\alpha + \ell_\beta) - \Gamma (\Delta_i \rightarrow \bar{\ell}_\alpha + \bar{\ell}_\beta)}{\Gamma_{\Delta_i} +\Gamma_{\Delta^*_i}}\,,
\end{align}
where $\Gamma_{\Delta_i}$ denotes the total triplet decay width and the overall factor
$\Delta {\rm L}=2$ arises because the triplet decay produces two leptons. It is useful to
define
\begin{align}\label{BlBphi}
    \mathcal{B}^\ell_i\, \Gamma_{\Delta_i} &\equiv \sum_{\alpha,\beta} \Gamma(\Delta_i^* \rightarrow \ell_\alpha+\ell_\beta)=\frac{M_{\Delta_i}}{8\pi} \text{Tr}\, (\mathbf{Y}^{\Delta_i\dagger} \mathbf{Y}^{\Delta_i}),\nonumber\\
    \mathcal{B}^\phi_i \Gamma_{\Delta_i} &\equiv \Gamma(\Delta_i^* \rightarrow \phi+\phi)
    =\frac{M_{\Delta_i}}{8\pi}|\mu_i|^2\,,
\end{align}
where $\mathcal{B}^\ell_i \equiv \text{BR}(\Delta_i^* \rightarrow \ell + \ell)$ and
$\mathcal{B}^\phi_i \equiv \text{BR}(\Delta_i^* \rightarrow \phi+\phi)$ are the
tree-level branching ratios to leptons and Higgs doublets, respectively
($\mathcal{B}^\ell_i+\mathcal{B}^\phi_i=1$). The total triplet decay width is then given
by
\begin{align}\label{tripletdecaywidth}
    \Gamma_{\Delta_i} = \frac{M_{\Delta_i}}{8\pi} \Bigl[ \text{Tr}\, (\mathbf{Y}^{\Delta_i\dagger} \mathbf{Y}^{\Delta_i}) + |\mu_i|^2\Bigr].
\end{align}

When the triplet decays into leptons with given flavors $\ell_\alpha$ and $\ell_\beta$, a
nonvanishing asymmetry $\epsilon_{i}^{\alpha\beta}$ is generated by the interference of
the tree-level decay process with the one-loop self-energy diagram shown in
Fig.~\ref{fig52}. One finds
\begin{align}\label{cpasymIIa}
\epsilon_{i}^{\alpha\beta} \simeq -\frac{g(x_j)}{2\pi} \frac{c_{\alpha\beta}\,\text{Im}\bigl[\mu_i^\ast\mu_j \mathbf{Y}^{\Delta_i}_{\alpha\beta} \mathbf{Y}^{\Delta_j\ast}_{\alpha\beta} \bigr]}{\text{Tr}\left(\mathbf{Y}^{\Delta_i\dagger} \mathbf{Y}^{\Delta_i}\right)+\left|\mu_i\right|^2}, \quad (j \neq i),
\end{align}
where $c_{\alpha\beta}=2-\delta_{\alpha\beta}$ for $\Delta^0_i$ and $\Delta^{++}_i$,
$c_{\alpha\beta}=1$ for $\Delta^+_i$; $x_{j} = M_{\Delta_j}^2/M_{\Delta_i}^2$,  and the
loop function $g(x)$ is defined in Eq.~\eqref{loopfunc}, with the parameter $a_j$ now
given by $a_j=\Gamma_{\Delta_j}/M_{\Delta_i}$.

Recalling that in the type II seesaw framework under discussion the effective light
neutrino mass matrix is
\begin{align} \label{mnuII}
\mathbf{m}_\nu=\mathbf{m}_\nu^{(1)}+\mathbf{m}_\nu^{(2)}, \quad \mathbf{m}_\nu^{(i)}=2\mu_i^*\frac{v^2}{M_{\Delta_i}}\, \mathbf{Y}^{\Delta_i},
\end{align}
and using the relation
\begin{align} \label{GammaPhirel}
  16\pi v^2\, \Gamma_{\Delta_i} (\mathcal{B}^\ell_i\, \mathcal{B}^\phi_i)^{1/2} = M_{\Delta_i}^2\,
  \bigl[\mathrm{Tr}\,\bigl(\mathbf{m}_\nu^{(i)^\dagger} \mathbf{m}_\nu^{(i)}\bigr)\bigr]^{1/2},
\end{align}
Eq.~\eqref{cpasymIIa} can be recast in the more convenient form
\begin{align}\label{cpasymIIb}
   \epsilon_{i}^{\alpha\beta} &\simeq -\frac{g(x_j)}{4\pi} \frac{M_{\Delta_j}(\mathcal{B}^\ell_i\, \mathcal{B}^\phi_i)^{1/2}}{v^2}\,\frac{c_{\alpha\beta} \mathrm{Im}\bigl[\bigl(\mathbf{m}_{\nu}^{(i)}\bigr)_{\alpha\beta} \bigl(\mathbf{m}_{\nu}^{(j)}\bigr)^\ast_{\alpha\beta}\bigr]}{\bigl[\mathrm{Tr}\,\bigl(\mathbf{m}_\nu^{(i)^\dagger} \mathbf{m}_\nu^{(i)}\bigr)\bigr]^{1/2}} \nonumber\\
   &= -\frac{g(x_j)}{4\pi} \frac{M_{\Delta_j}(\mathcal{B}^\ell_i\, \mathcal{B}^\phi_i)^{1/2}}{v^2}\,\frac{c_{\alpha\beta} \mathrm{Im}\bigl[\bigl(\mathbf{m}_{\nu}^{(i)}\bigr)_{\alpha\beta}  \bigl(\mathbf{m}_{\nu}^{\ast}\bigr)_{\alpha\beta}\bigr]}{\bigl[\mathrm{Tr}\,\bigl(\mathbf{m}_\nu^{(i)^\dagger} \mathbf{m}_\nu^{(i)}\bigr)\bigr]^{1/2}}\,.
\end{align}

In the hierarchical limit $M_{\Delta_i} \ll M_{\Delta_j}$, Eq.~\eqref{cpasymIIb} reduces
to
\begin{align}\label{cpasymIIc}
   \epsilon_{i}^{\alpha\beta} \simeq \frac{M_{\Delta_i}(\mathcal{B}^\ell_i\, \mathcal{B}^\phi_i)^{1/2}}{4\pi v^2}\,\frac{c_{\alpha\beta}\, \mathrm{Im}\bigl[\bigl(\mathbf{m}_{\nu}^{(i)}\bigr)_{\alpha\beta}  \bigl(\mathbf{m}_{\nu}^{\ast}\bigr)_{\alpha\beta}\bigr]}{\bigl[\mathrm{Tr}\,\bigl(\mathbf{m}_\nu^{(i)^\dagger} \mathbf{m}_\nu^{(i)}\bigr)\bigr]^{1/2}}\,.
\end{align}
Summing over the final lepton flavors,  Eq.~\eqref{cpasymIIc} leads to the following
expression for the unflavored asymmetry~\cite{Hambye:2005tk,Dorsner:2005ii}:
\begin{align}\label{cpasymIId}
    \epsilon_{i}= \sum_{\alpha,\beta} \epsilon_{i}^{\alpha\beta} = \frac{M_{\Delta_i}(\mathcal{B}^\ell_i\, \mathcal{B}^\phi_i)^{1/2}}{4\pi v^2}\,\frac{\mathrm{Im}\bigl[\mathrm{Tr}\, \bigl(\mathbf{m}_{\nu}^{(i)} \mathbf{m}_{\nu}^{\dagger}\bigr)\bigr]}{\bigl[\mathrm{Tr}\,\bigl(\mathbf{m}_\nu^{(i)^\dagger} \mathbf{m}_\nu^{(i)}\bigr)\bigr]^{1/2}}\,.
\end{align}

It is then straightforward to show that the following upper bound
holds~\cite{Hambye:2005tk}:
\begin{align} \label{epsmaxII}
  |\epsilon_{i}| &\leq \frac{M_{\Delta_i}(\mathcal{B}^\ell_i\, \mathcal{B}^\phi_i)^{1/2}}{4\pi v^2}\,
  \bigl[\mathrm{Tr}\,\bigl(\mathbf{m}_\nu^\dagger \mathbf{m}_\nu\bigr)\bigr]^{1/2}\nonumber\\
  &=\frac{M_{\Delta_i}(\mathcal{B}^\ell_i\, \mathcal{B}^\phi_i)^{1/2}}{4\pi v^2}\,
  \Bigl( \sum_k m_k^2\Bigr)^{1/2}.
\end{align}
Thus, unlike the type I seesaw case, the upper bound on the asymmetry increases as the
light neutrino mass scale increases. For hierarchical light neutrinos one obtains:
\begin{align}\label{epsmaxIIa}
    |\epsilon_i| \lesssim 10^{-6}\bigl(\mathcal{B}^\ell_i\, \mathcal{B}^\phi_i\bigr)^{1/2} \left(\frac{M_{\Delta_i}}{10^{10}\,\text{GeV}}\right)\biggl(\frac{m_\text{atm}}{0.05\,\text{eV}}\biggr).
\end{align}
We remark that, although the absolute maximum in Eqs.~\eqref{epsmaxII} and \eqref{epsmaxIIa} is attained when $\mathcal{B}^\ell_i = \mathcal{B}^\phi_i=1/2$, this situation does not necessarily correspond to a maximal baryon asymmetry. The efficiency of leptogenesis, dictated by the solution of the relevant Boltzmann equations, is not necessarily maximal in such a case. In fact, it turns out that the efficiency is minimal for $\mathcal{B}^\ell_i = \mathcal{B}^\phi_i =1/2$ and maximal when either $\mathcal{B}^\ell_i \ll \mathcal{B}^\phi_i$ or $\mathcal{B}^\ell_i \gg \mathcal{B}^\phi_i$~\cite{Hambye:2005tk}. Consequently, in the limits when the efficiency is maximal the leptonic CP asymmetry is suppressed.

A major difference between type I and type II seesaw leptogenesis scenarios is that, unlike the singlet Majorana neutrinos, the scalar triplets couple to the SM gauge bosons. Since gauge interactions keep the triplets close to thermal equilibrium at temperatures $T \lesssim 10^{15}$~GeV, it may seem difficult to fulfill the third Sakharov condition. Nevertheless, estimates of the thermal leptogenesis efficiency~\cite{Hambye:2000ui,Hambye:2003ka} as well as a more precise calculation of it by solving the full set of Boltzmann equations~\cite{Hambye:2005tk} indicate that leptogenesis is efficient even at a much lower temperature. For hierarchical scalar triplets and in the absence of extra sources of CP violation, leptogenesis is efficient for $M_{\Delta_i} \gtrsim 10^{9}$~GeV.

If the scalar triplets are quasidegenerate in mass, the leptonic asymmetry can be
resonantly enhanced provided that $|M_{\Delta_j} - M_{\Delta_i}| \sim
\frac{1}{2}\,\Gamma_{\Delta_j}$. In this case, from Eq.~\eqref{cpasymIIb} one obtains
\begin{align}\label{cpasymIIres}
   \epsilon_{i}^{\alpha\beta} \simeq \frac{(\mathcal{B}^\ell_i\, \mathcal{B}^\phi_i)^{1/2} c_{\alpha\beta}\, \mathrm{Im}\bigl[\bigl(\mathbf{m}_{\nu}^{(i)}\bigr)_{\alpha\beta} \bigl(\mathbf{m}_{\nu}^{(j)}\bigr)_{\alpha\beta}^{\ast}\bigr]}{\bigl[\mathrm{Tr}\,\bigl(\mathbf{m}_\nu^{(i)^\dagger} \mathbf{m}_\nu^{(i)}\bigr)\bigr]^{1/2}\bigl[\mathrm{Tr}\,\bigl(\mathbf{m}_\nu^{(j)^\dagger} \mathbf{m}_\nu^{(j)}\bigr)\bigr]^{1/2}}\,,
\end{align}
which, after summing over the lepton flavors, yields
\begin{align}\label{epsmaxIIres}
   \epsilon_{i,\text{res}} \simeq \frac{(\mathcal{B}^\ell_i\, \mathcal{B}^\phi_i)^{1/2} \, \mathrm{Im}\bigl[\text{Tr}\,\bigl(\mathbf{m}_{\nu}^{(i)} \mathbf{m}_{\nu}^{(j)\dagger}\bigr)\bigr]}{\bigl[\mathrm{Tr}\,\bigl(\mathbf{m}_\nu^{(i)^\dagger} \mathbf{m}_\nu^{(i)}\bigr)\bigr]^{1/2}\bigl[\mathrm{Tr}\,\bigl(\mathbf{m}_\nu^{(j)^\dagger} \mathbf{m}_\nu^{(j)}\bigr)\bigr]^{1/2}}\,.
\end{align}
This leads to the upper bound $|\epsilon_{i,\text{res}}| \lesssim (\mathcal{B}^\ell_i\, \mathcal{B}^\phi_i)^{1/2}$, which is suppressed by neither the light neutrino masses nor the scalar triplet masses [it is just bounded by the unitarity constraint $|\epsilon_i| < 2 \min(\mathcal{B}^\ell_i, \mathcal{B}^\phi_i)$]. This opens the possibility for type II seesaw leptogenesis scenarios at the TeV scale. We note, however, that in the latter case there is a dependence on $M_{\Delta_i}$ that strongly suppresses the leptogenesis efficiency when $M_{\Delta_i} \sim \mathcal{O}$(TeV). Moreover, the final baryon asymmetry crucially depends on the triplet annihilation rate in the nonrelativistic limit, which is affected by nonperturbative corrections to the $s$-wave coefficient that reduce further the leptogenesis efficiency by about 30\%~\cite{Strumia:2008cf}.  Since after the electroweak symmetry breaking, at temperatures $T \lesssim m_\text{Higgs}$, sphaleron interactions are suppressed and no longer can convert the lepton asymmetry into a baryon asymmetry, a stringent lower bound on the triplet mass is obtained. To successfully reproduce the observed baryon asymmetry, a triplet mass $M_{\Delta_i} \gtrsim 1.6$~TeV is required~\cite{Strumia:2008cf}, which is too heavy to give detectable effects at the LHC~\cite{Nath:2010zj}.

\subsubsection{Type III seesaw leptogenesis}
\label{sec4.1.3}

\begin{figure*}[th]
\includegraphics{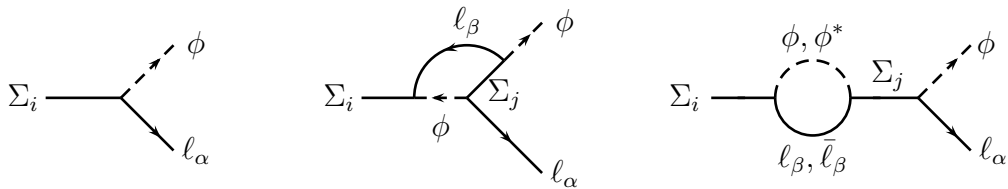}
\caption{\label{fig53} Diagrams contributing to the CP asymmetry $\epsilon_i^{\alpha}$ in type III seesaw leptogenesis. As in the type I seesaw case, the last diagram involves two graphs, one which is lepton flavor and lepton-number violating and another which is lepton flavor violating but does not give contribution to the unflavored CP asymmetry.}
\end{figure*}

As explained in Sec.~\ref{sec2.5}, light neutrino masses can also be mediated by the tree-level exchange of $SU(2)$-triplet fermions with zero hypercharge. Such triplets naturally arise in theories based on grand unification, e.g., when the adjoint $24_F$ fermion representation is introduced in $SU(5)$, and their masses could be low enough to be accessible at the LHC~\cite{Bajc:2006ia}. Apart from the kinetic term, the type III seesaw Lagrangian has the same structure as in the type I seesaw case, but with different contractions of the $SU(2)$ indices in the Yukawa interaction terms [cf. Eq.~\eqref{LtypeIIIseesaw}]. Thus, in what concerns neutrino masses, the type I and type III seesaw mechanisms share the same qualitative features. Yet, there are a few differences in the implementation of leptogenesis that are worth mentioning. First, in the CP asymmetry generated by the triplet fermion decay, the relative sign between the vertex and self-energy contributions is opposite to that of the type I seesaw case. Therefore, for a hierarchical triplet spectrum, the asymmetry turns out to be 3 times smaller than in the singlet fermion case. Nevertheless, this is compensated by the fact that the triplet has three components and, consequently, the final baryon asymmetry is 3 times bigger. Second, fermion triplets have gauge interactions which tend to keep them close to thermal equilibrium and reduce the efficiency of leptogenesis~\cite{Hambye:2003rt}.

Since all of the conclusions previously drawn for type I seesaw leptogenesis essentially remain valid in the present case, we briefly comment next on the main differences. Considering the type III seesaw Lagrangian of Eq.~\eqref{LtypeIIIseesaw} with three fermion triplets $\Sigma_i\, (i=1,2,3)$, the CP asymmetry generated in the decays of $\Sigma_i$ into a lepton $\ell_\alpha$ and the Higgs $\phi$ comes from the interference of the tree-level and one-loop graphs depicted in Fig.~\ref{fig53}. It differs from its analogous of the type I case [cf. Eq.~\eqref{cpasymI}] in the overall sign of the vertex contribution, and the substitutions $M_i \rightarrow M_{\Sigma_i}, \mathbf{Y}^\nu \rightarrow \mathbf{Y}^\Sigma, \mathbf{H}^\nu \rightarrow \mathbf{H}^\Sigma$ and $\Gamma_{N_i}\rightarrow\Gamma_{\Sigma_i}$. Thus, in a $\Sigma_1$-dominated scenario with hierarchical fermion triplets, $M_{\Sigma_1} \ll M_{\Sigma_j}\,(j=2,3)$, instead of the usual $f(x)+g(x) \simeq -3/(2\sqrt{x})$ factor, the factor $g(x)-f(x) \simeq -1/(2\sqrt{x})$ appears. This means, in particular, that the right-hand sides in Eqs.~\eqref{cpasymIb}-\eqref{epsupperflav} get reduced by a factor of 3. On the other hand, the resonant asymmetries in Eqs.~\eqref{cpasymIres} and \eqref{cpasymIares} remain unaltered. As in the type II seesaw case, gauge interactions play a crucial role in the efficiency of the type III leptogenesis scenario. Assuming a hierarchical triplet mass spectrum and neglecting flavor effects, leptogenesis can succeed if $M_{\Sigma_1} \gtrsim 1.5\times 10^{10}$~GeV and the neutrino mass scale is $m \lesssim 0.12$~eV~\cite{Hambye:2003rt}. These bounds are slightly stronger than in type I leptogenesis. On the other hand, if leptogenesis occurs at the TeV scale, the correct amount of baryon asymmetry can only be generated for $M_{\Sigma_i} \gtrsim 1.6$~TeV~\cite{Strumia:2008cf}, which is too large to be within the energy reach of the LHC~\cite{Nath:2010zj}. Accounting for flavor effects does not weaken this bound~\cite{AristizabalSierra:2010mv}.

\subsubsection{Dirac leptogenesis}
\label{sec4.1.4}

All the leptogenesis scenarios discussed in this section are based on the seesaw mechanism, which gives Majorana masses to the light neutrinos. Although well motivated from a theoretical and phenomenological viewpoint, this is not the only possibility to explain neutrino masses. Indeed, neutrinos could be Dirac particles and lepton number may not be violated at the perturbative level. It is therefore pertinent to ask whether leptogenesis can be implemented in such a framework. As it turns out, models with Dirac neutrinos and viable leptogenesis can be constructed as well~\cite{Akhmedov:1998qx,Dick:1999je,Murayama:2002je}. The main idea behind the Dirac leptogenesis scenarios can be understood as follows. Suppose that the CP-violating decay of a heavy particle produces a nonzero lepton number ${\rm L}<0$ ($-{\rm L}>0$) for left-handed (right-handed) particles. Since the Yukawa interactions of the SM are fast enough, they rapidly equilibrate the left-handed and right-handed particles so that L goes to zero. However, this does not apply to Dirac neutrinos, which have Yukawa couplings exceedingly small, $y_\nu \lesssim \mathcal{O}(\text{1 eV})/v \sim 10^{-11}$. For them, the equilibrium between the lepton number stored in each chirality occurs when $\Gamma_\nu/ H \simeq y_\nu^2\, M_P/T \gtrsim 1$, i.e., at a temperature which is far below the electroweak scale. Thus, by the time L-equilibration takes place the left-handed lepton number has already been partially converted into a net baryon number by the sphalerons, leading to a Universe with ${\rm B}={\rm L}>0$. Clearly, one of the consequences of Dirac leptogenesis is the absence of any signal in $0\nu\beta\beta$ decay searches.

\subsection{Leptonic CP violation from high to low energies}
\label{sec4.2}

One of the distinctive features of the leptogenesis mechanisms described in the previous section is the fact that the interactions relevant for leptogenesis can simultaneously be responsible for the nonvanishing and smallness of the neutrinos masses. This raises the question of whether there is a direct link between leptogenesis and low-energy leptonic observables. More specifically, if the strength of CP violation at low energies in neutrino oscillations is measured, what can one infer about the viability or nonviability of leptogenesis? From the sign of the baryon asymmetry, can one predict the sign of the CP asymmetries in neutrino oscillations, namely the sign of the low-energy CP invariant $\mathcal{J}_\text{lepton}^\text{CP}$? Is there any connection between leptogenesis and the low-energy Majorana phases measurable in $0\nu\beta\beta$ decay? The answers to these questions are, however, not straightforward.

In general, the seesaw framework contains many more (unconstrained) parameters than measurable quantities at low energies. We recall that, apart from the three charged-lepton masses, the lepton sector contains nine parameters: the three light neutrino masses plus the three mixing angles and three CP-violating phases contained in the PMNS leptonic mixing matrix $\mathbf{U}$. Only four of these nine parameters have been measured: the mass-squared differences ($\Delta m_{21}^2, \Delta m_{31}^2$) and two mixing angles ($\theta_{12}, \theta_{23}$). The lightest neutrino mass and the Dirac and Majorana phases in $\mathbf{U}$ are unknown. But even if these unknown parameters would be measured, and a partial correspondence with the leptonic sector at high energies could be established, there remain several high-energy free parameters which are not accessible to experiments. Some of the latter are relevant for leptogenesis. Consequently, any connection between leptogenesis and low-energy leptonic observables can only be found in a model-dependent way~\cite{Buchmuller:1996pa,Branco:2001pq,Branco:2002kt}. In particular, thermal leptogenesis can be unsuccessful despite the presence of low-energy leptonic CP violation. Conversely, leptogenesis can take place even without Dirac and/or Majorana phases at low energies~\cite{Branco:2002kt,Rebelo:2002wj}.

In this section we discuss some general aspects of the interplay between the leptonic CP violation responsible for leptogenesis at high energies and the one measurable at low energies, which originates from the leptonic mixing matrix $\mathbf{U}$. Our aim is to analyze some simple cases in which such a link can exist and manifest itself through the leptonic CP asymmetries. We restrict our discussion to the type I seesaw leptogenesis scenario. All the conclusions will be equally valid for the type III seesaw case (with some obvious changes in the notation). Other scenarios, in which the connection can be established taking into account not only the leptonic asymmetry but also the effects that affect the efficiency of leptogenesis (e.g., charged-lepton flavor effects), will be briefly commented on at the end of Sec.~\ref{sec4.2.2}.

In order to address the above questions in a type I seesaw framework, one should keep in mind that, in the mass eigenbasis of the charged leptons and heavy Majorana neutrinos, all the information about the leptonic mixing and CP violation is contained in the Dirac-neutrino Yukawa coupling matrix $\mathbf{Y}^\nu$. It then becomes clear that any bridge between high-energy and low-energy CP violation can only be established for specific choices of this matrix. Below we describe a few possibilities.

\subsubsection{Triangular parametrization}
\label{sec4.2.1}

It can be easily shown that any arbitrary complex matrix can be written as the product of a unitary matrix $\mathbf{V}$ and a lower triangular matrix
$\mathbf{Y}_{\triangle}$~\cite{Morozumi:1997af}. In particular, the Dirac-neutrino Yukawa coupling matrix can be written as
\begin{align}\label{Ytri1}
\mathbf{Y}^\nu = \mathbf{V}\,\mathbf{Y}_{\triangle}, \quad
\mathbf{Y}_{\triangle}= \left(\begin{array}{ccc}
y_{11} & 0 & 0 \\
y_{21}\,e^{i\beta_1} & y_{22} & 0 \\
y_{31}\,e^{i\beta_2} & y_{32}\,e^{i\beta_3} & y_{33}
\end{array}\right),
\end{align}
where $y_{ij}$ are real positive numbers. Since $\mathbf{V}$ is unitary, in general it
contains six phases. However, three of these phases can be rephased away by a
simultaneous phase transformation on the left-handed fields $\ell$, which leaves the
leptonic charged-current invariant. Furthermore, $\mathbf{Y}_{\triangle}$ defined in
Eq.~\eqref{Ytri1} can be rewritten in the form
\begin{align}\label{Ytri2}
\mathbf{Y}_{\triangle}= \mathbf{P}_{\beta}^\dagger \hat{\mathbf{Y}}_{\triangle} \mathbf{P}_{\beta},
\end{align}
where $\mathbf{P}_\beta ={\rm diag} (1, e^{-i\beta_1}, e^{-i\beta_2})$ and
\begin{align}\label{Ytri3}
\hat{\mathbf{Y}}_{\triangle}= \left(\begin{array}{ccc}
y_{11} & 0 & 0 \\
y_{21}  & y_{22} & 0 \\
y_{31}  & y_{32}\,e^{i\sigma} & y_{33}
\end{array}\right) \,,
\end{align}
with $\sigma=\beta_3- \beta_2+ \beta_1$. It follows then from Eqs.~\eqref{Ytri1} and
\eqref{Ytri2} that the matrix $\mathbf{Y}^\nu$ can be decomposed as
\begin{align}
\mathbf{Y}^\nu=\mathbf{U}_{\rho}\,\mathbf{P}_{\alpha}\, {\hat{\mathbf{Y}}_{\triangle}}\,\mathbf{P}_{\beta}\,,
\label{mDdec}
\end{align}
where $\mathbf{P}_\alpha ={\rm diag} (1, e^{i\alpha_1}, e^{i\alpha_2})$ and
$\mathbf{U}_\rho$ is a unitary matrix containing only one phase $\rho$. Therefore, in the
mass eigenbasis of the charged leptons and heavy Majorana neutrinos, the phases $\rho$,
$\alpha_1$, $\alpha_2$, $\sigma$, $\beta_{1}$, and $\beta_{2}$ are the only physical
phases characterizing CP violation in the lepton sector.

The triangular parametrization given in Eq.~\eqref{mDdec} is in general not suitable to disentangle the phases appearing in the flavored leptogenesis asymmetries of Eq.~\eqref{cpasymI}, which depend on the quantities $\text{Im}\left[\mathbf{Y}^{\nu\ast}_{\alpha i} \mathbf{H}^\nu_{ij} \mathbf{Y}^\nu_{\alpha j}\right]$. Nevertheless, for the unflavored leptogenesis asymmetry in
Eq.~\eqref{cpasymIa}, the relevant phases are only those contained in the matrix
$\mathbf{H}^\nu=\mathbf{Y}^{\nu\dagger} \mathbf{Y}^\nu$. From
Eqs.~\eqref{Ytri1}-\eqref{mDdec}, we then conclude that these phases are $\sigma$,
$\beta_{1}$, and $\beta_{2}$. Since the phases $\alpha_1$,
$\alpha_2$, and $\rho$ do not contribute to leptogenesis, and all six phases of $\mathbf{Y}^\nu$ are present in the leptonic mixing matrix $\mathbf{U}$, it is clear that a necessary condition for a direct link between the unflavored leptogenesis asymmetry and low-energy CP violation is the requirement that the matrix $\mathbf{V}$ in Eq.~\eqref{Ytri1} contains no CP-violating phases. We note that, although the above condition was derived in a specific
weak basis, and using the parametrization of Eq.~\eqref{Ytri1}, it can be applied to any
model. A specific class of models which satisfy the above necessary condition in a
trivial way are those for which $\mathbf{V}=\openone$, leading to
$\mathbf{Y}^\nu=\mathbf{Y}_\triangle$~\cite{Branco:2002xf}. This condition is necessary
but not sufficient to allow for a prediction of the sign of the CP asymmetry in neutrino
oscillations, given the observed sign of the baryon asymmetry and the low-energy neutrino
data. A more restrictive class of matrices $\mathbf{Y}^\nu$ should be
considered~\cite{Branco:2002xf}. Below we illustrate the possibility of a direct link
between leptogenesis and low-energy CP violation with a simple example.

We consider an $N_1$-dominated scenario with $M_1 \ll M_{2,3}$. Assuming that
$y_{31}=0$ and $\beta_3=0$, the matrix $\mathbf{Y}^\nu$ in Eq.~\eqref{Ytri1} has the
simple zero-texture structure\footnote{Approximate texture zeros commonly arise in flavor
model constructions based on the Froggatt-Nielsen mechanism~\cite{Froggatt:1978nt}.}
\begin{align}\label{Ytri4}
\mathbf{Y}^\nu = \left(\begin{array}{ccc}
y_{11} & 0 & 0 \\
y_{21}\,e^{i\beta_1} & y_{22} & 0 \\
0 & y_{32} & y_{33}
\end{array}\right),
\end{align}
so that $\text{Im}\left[\mathbf{Y}^{\nu\ast}_{\mu 1}\mathbf{H}^\nu_{12}
\mathbf{Y}^\nu_{\mu 2}\right]$ and $\text{Im}\left[(\mathbf{H}_{12}^{\nu})^2\right]$ are
the only nonvanishing quantities in the flavored and unflavored CP asymmetries of
Eqs.~\eqref{cpasymIb} and \eqref{cpasymIc}, respectively. One obtains
\begin{align}\label{imhij}
       \epsilon_1^\mu &\simeq \frac{3}{16\pi}\frac{M_1}{M_2}\frac{y_{21}^2\,y_{22}^2}{y_{11}^2+y_{21}^2}\times \sin(2\beta_1),\nonumber\\
       \epsilon_1^e &=\epsilon_1^\tau=0,
\end{align}
and summing over the flavors, $\epsilon_1 =\epsilon_1^\mu$. On the other hand, the strength of CP violation at
low energies is controlled by the CP invariant $\mathcal{J}^\text{CP}_\text{lepton}$
defined in Eq.~\eqref{ICPlepton}, with the neutrino mass matrix given by the seesaw
formula~\eqref{mnutypeIseesaw}. In this case,
\begin{align}\label{Jfin}
\mathcal{J}_\text{CP}&=-\frac{\text{Im}\left[(\mathbf{m}_{\nu}\mathbf{m}_{\nu}^{\dagger})_{12} (\mathbf{m}_{\nu}\mathbf{m}_{\nu}^{\dagger})_{23}(\mathbf{m}_{\nu}\mathbf{m}_{\nu}^{\dagger})_{31} \right]}{\Delta m^2_{21}\Delta m^2_{31}\Delta m^2_{32}}\nonumber\\
&= \frac{y_{11}^2\,
y_{21}^2\, y_{32}^2\,y_{22}^2\,v^{12}}{M_1^3 M_2^3\Delta m^2_{21}\Delta m^2_{31}\Delta m^2_{32}}\times \sin(2\beta_1)\nonumber\\ & \times \left[y_{21}^2 y_{32}^2 +y_{11}^2 y_{22}^2+y_{11}^2 y_{32}^2+y_{33}^2 (y_{11}^2+ y_{21}^2)\frac{M_2}{M_3}\right].
\end{align}
Thus, in this toy example not only the relative sign between the low-energy CP invariant $\mathcal{J}^\text{CP}_\text{lepton}$ and the flavored ($\epsilon_1^\mu$) and unflavored ($\epsilon_1$) asymmetries can be predicted (these quantities have the same sign), but also their dependence on the CP-violating phase $\beta_1$ is such that they are simultaneously maximized when
$\beta_1=\pi/4$. We also note that  when $y_{33}=0$ the texture of $\mathbf{Y}^\nu$ given
in Eq.~\eqref{Ytri4} corresponds to one of the textures considered
in~\cite{Frampton:2002qc}. In this case, the heavy Majorana neutrino $N_3$ completely
decouples, rendering this situation phenomenologically equivalent to the two right-handed
neutrino cases discussed in Sec.~\ref{sec4.2.3}.

\subsubsection{Orthogonal parametrization}
\label{sec4.2.2}

A particularly useful parametrization in the context of type I seesaw leptogenesis was
proposed by~\cite{Casas:2001sr}. Using a complex orthogonal matrix $\mathbf{R}$, the
Yukawa coupling matrix $\mathbf{Y}^\nu$ can be rewritten in the more convenient form for
leptogenesis calculations,
\begin{align} \label{YnuRmatrix}
\mathbf{Y}^\nu= v^{-1} \mathbf{U}^*\, \mathbf{d}_m^{1/2}\, \mathbf{R}\, \mathbf{d}_M^{1/2},
\end{align}
where $\mathbf{d}_M$ and $\mathbf{d}_m$ are the diagonal mass matrices defined in
Eqs.~\eqref{mRdiag} and \eqref{mnutypeIdiag}, respectively. In this parametrization,
\begin{align}\label{HnuRmatrix}
       \mathbf{H}^\nu_{ij} =(\mathbf{Y}^{\nu\dagger}\mathbf{Y}^\nu)_{ij} = \frac{M_i^{1/2} M_j^{1/2}}{v^2} \sum_k m_k \mathbf{R}^\ast_{ki} \mathbf{R}_{kj},
\end{align}
so that the flavored leptogenesis asymmetry given in Eq.~\eqref{cpasymIb} can be written
in the form
\begin{align}\label{cpasymIbR}
       \epsilon_1^\alpha \simeq  \frac{3\,M_1}{16\pi v^2}\frac{\sum_{j,k} m_j^{1/2}\,m_k^{3/2}\,\text{Im}\left[ \mathbf{U}_{\alpha j}^* \,\mathbf{U}_{\alpha k}\,\mathbf{R}_{j1}\,\mathbf{R}_{k1}\right] }{\sum_{k} m_k\,|\mathbf{R}_{k1}|^2}\,,
\end{align}
while the unflavored asymmetry \eqref{cpasymIc} reads
\begin{align}\label{cpasymIcR}
       \epsilon_1 \simeq  \frac{3\,M_1}{16\pi v^2}\frac{\sum_{j \neq 1} m_j^2\,\text{Im}\left[\mathbf{R}_{j1}^2\right]}{\sum_k m_k\,|\mathbf{R}_{k1}|^2}.
\end{align}

It becomes evident that the unflavored asymmetry \eqref{cpasymIcR} or, more generally,
the unflavored asymmetry defined in Eq.~\eqref{cpasymIa} does not depend on the low-energy
CP-violating phases of $\mathbf{U}$, since the matrix $\mathbf{U}$ cancels out in the
matrix $\mathbf{H}^\nu$, as can be seen from Eq.~\eqref{HnuRmatrix}. It should be noted,
however, that the above conclusion holds provided that the matrices $\mathbf{U}$ and
$\mathbf{R}$ are independent from each other, i.e., if no constraints or specific
\emph{ans\"{a}tze} are imposed on the matrix $\mathbf{Y}^\nu$. In particular, imposing
some flavor symmetries or texture zeros on the matrix $\mathbf{Y}^\nu$ may lead to
relations between the CP-violating phases in $\mathbf{U}$ and the CP-violating parameters
in $\mathbf{R}$. In such cases, the parametrization in Eq.~\eqref{YnuRmatrix} may not be
the most convenient for disentangling the CP violation responsible for leptogenesis from
CP violation at low energies.

If the matrix $\mathbf{R}$ is real, i.e., if the only source of high-energy CP violation
comes from the left-handed lepton sector, then the unflavored leptogenesis
CP-asymmetries $\epsilon_i$ vanish~\cite{Nardi:2006fx,Abada:2006ea}. The fact that the
matrix $\mathbf{R}$ is real when CP is an exact symmetry of the right-handed neutrino
sector is easily understood once the matrix $\mathbf{Y}^\nu$ is written in its singular
value decomposition, $\mathbf{Y}^\nu = \mathbf{V}_L^\dagger \mathbf{d}_\lambda
\mathbf{V}_R$, where $\mathbf{V}_{L,R}$ are unitary matrices and $\mathbf{d}_\lambda =
\text{diag}\,(\lambda_1,\lambda_2,\lambda_3)$ with $\lambda_i$ the corresponding
eigenvalues. The CP violation in the right-handed neutrino sector is thus encoded in the
phases of $\mathbf{V}_R$. On the other hand, using the
parametrization~\eqref{YnuRmatrix}, one can also write $\mathbf{H}^\nu=\mathbf{d}_M^{1/2}
\mathbf{R}^\dagger \mathbf{d}_m \mathbf{R} \mathbf{d}_M^{1/2}/v^2 = \mathbf{V}_R^\dagger
\mathbf{d}_\lambda^2 \mathbf{V}_R$, which clearly shows that the orthogonal matrix
$\mathbf{R}$ is real if and only if $\mathbf{V}_R$ is real.

The situation is, however, quite different when flavor effects are accounted for. We
consider, for definiteness, the $N_1$-dominated scenario with $M_1 \ll M_{2,3}$ at
temperatures $T \lesssim 10^{12}$~GeV. In this case, the flavored asymmetries are given
by Eq.~\eqref{cpasymIbR} and the relevant quantities are the combinations
$\text{Im}\left[ \mathbf{U}_{\alpha j}^* \,\mathbf{U}_{\alpha
k}\,\mathbf{R}_{j1}\,\mathbf{R}_{k1}\right]$, which explicitly depend on the PMNS matrix
elements. Therefore, provided that $\mathbf{R} \neq \openone$, the leptogenesis
asymmetries $\epsilon_1^\alpha$ do not vanish even if the matrix $\mathbf{R}$ is real.
Furthermore, in the latter case the CP-violating effects responsible for leptogenesis are
directly connected to the low-energy CP-violating phases in
$\mathbf{U}$~\cite{Pascoli:2006ie,Branco:2006ce}. This becomes evident from the
expression of the leptogenesis asymmetries,
\begin{align}\label{cpasymIbRreal}
       \epsilon_1^\alpha = \frac{3\,M_1}{16\pi v^2}\frac{\sum_{j} \sum_{k>j} \sqrt{m_j m_k}\, (m_k-m_j) \mathbf{R}_{j1}\,\mathbf{R}_{k1}\,\mathcal{I}_{jk}^\alpha}{\sum_{k} m_k\,|\mathbf{R}_{k1}|^2}\,,
\end{align}
where
\begin{align} \label{Ialphajk}
\mathcal{I}_{jk}^\alpha=\text{Im}\left[\mathbf{U}_{\alpha j}^* \mathbf{U}_{\alpha k}\right]
\end{align}
are rephasing invariant quantities.

At this point, one may wonder whether a real matrix $\mathbf{R}$ can be naturally realized in some model. In general, once CP violation is allowed through the introduction of complex Yukawa couplings, it will arise in both the left-handed and right-handed sectors, leading to a complex PMNS matrix $\mathbf{U}$ as well as a complex orthogonal matrix $\mathbf{R}$. The simplest way of restricting the number of CP-violating phases is through the assumption that CP is a good symmetry of the Lagrangian, only broken by the vacuum. A model with a complex leptonic mixing matrix $\mathbf{U}$ and real $\mathbf{R}$ can actually be constructed in a natural way. We consider the type I seesaw framework and impose CP invariance at the Lagrangian level. We also introduce three Higgs doublets, together with a $Z_3$ symmetry under which the left-handed fermion doublets $\psi_{Lj}$ transform as $\psi_{Lj} \rightarrow e^{-i 2\pi j/3} \psi_{Lj}$ and the Higgs doublets as $\phi_j \rightarrow e^{i 2\pi j/3} \phi_j$, while all other fields transform trivially. One can show that there is a region of the parameter space where the vacuum violates CP through complex vacuum expectation values. Yet, due to the $Z_3$ restrictions on Yukawa couplings, the combination  $\mathbf{Y}^{\nu\dagger}\mathbf{Y}^\nu$ turns out to be real, thus implying a real $\mathbf{R}$, while a complex $\mathbf{U}$ is generated. The drawback of such a scheme is
that leptogenesis must occur not far from the electroweak scale. However, one can envisage an alternative scenario where effective Yukawa couplings are generated by higher-order operators that involve singlet fields that acquire complex VEV at very high energies. From a different viewpoint, the case of a real matrix $\mathbf{R}$ can also be realized within a class of models based on the so-called sequential dominance~\cite{King:2006hn}.

To illustrate the possibility of a direct link between leptogenesis and low-energy CP
violation when the matrix $\mathbf{R}$ is real, we consider the following example. We
assume a normal hierarchical light neutrino mass spectrum with $m_1\simeq 0 \ll m_2\simeq
m_\text{sol} \ll m_3\simeq m_\text{atm}$. In this case, Eq.~\eqref{cpasymIbRreal} yields
\begin{align}\label{cpasymIbRreal1}
       \epsilon_1^\alpha \simeq \frac{3\,M_1}{16\pi v^2}\frac{m_\text{atm}\sqrt{m_\text{sol}\, m_\text{atm}}\, \mathbf{R}_{21}\,\mathbf{R}_{31}\,\mathcal{I}_{23}^\alpha}{m_\text{sol}\,|\mathbf{R}_{21}|^2+m_\text{atm}\,|\mathbf{R}_{31}|^2}\,.
\end{align}
We further assume that the CP-violating effects due to the low-energy Dirac-type
phase $\delta$ are subdominant and can be neglected ($\delta \simeq 0$). Then, using the
parametrization \eqref{Uparam}-\eqref{VPDG} of the mixing matrix $\mathbf{U}$, one can
show that
\begin{align}\label{imUUexample}
    &\mathcal{I}_{23}^e \simeq -c_{13}s_{12}s_{13}\sin(\alpha_{12}/2),\nonumber\\
    &\mathcal{I}_{23}^\mu \simeq c_{13}s_{23} (-c_{12}c_{23}+s_{12}s_{13}s_{23}) \sin(\alpha_{12}/2),\nonumber\\
    &\mathcal{I}_{23}^\tau \simeq c_{13}c_{23} (c_{23}s_{12}s_{13}+c_{12}s_{23}) \sin(\alpha_{12}/2),
\end{align}
with $\alpha_{12}=\alpha_1-\alpha_2$. Therefore, in this simple example, the flavored leptogenesis asymmetries depend on the
same Majorana-phase difference $\alpha_{12}$ that controls the effective Majorana
mass parameter $m_{ee}$ in $0\nu\beta\beta$ decay [cf. Eq.~\eqref{mee}]. We note, however,
that the sign of $\epsilon_1^\alpha$ cannot be uniquely predicted by the sign of
$\sin(\alpha_{12}/2)$ since the product $\mathbf{R}_{21}\,\mathbf{R}_{31}$ can be
positive or negative.

Before concluding this section, we briefly comment on the possibility of establishing
a connection between leptogenesis and low-energy CP violation taking into account other
effects (besides the leptonic CP asymmetries) that can affect the efficiency of
leptogenesis. Assuming a particular prior on the parameter space (e.g., by restricting the
orthogonal matrix $\mathbf{R}$ and the heavy and/or light neutrino mass spectra), it has
been shown that flavored leptogenesis can work for any value of the PMNS phases and,
therefore no direct connection can be established~\cite{Davidson:2007va}. On the other
hand, for an inverted-hierarchical light neutrino mass spectrum, one can show that there
exist regions in the leptogenesis parameter space where the purely high-energy
contribution to the baryon asymmetry is highly suppressed and a successful leptogenesis
can be achieved only if the necessary amount of CP violation is provided by the PMNS
Majorana phases~\cite{Molinaro:2008rg,Molinaro:2008cw}.

\subsubsection{Two right-handed neutrino case}
\label{sec4.2.3}

Neutrino oscillation data do not demand the presence of three right-handed neutrinos in
a type I seesaw framework. The solar and atmospheric neutrino mass scales could be
associated to just two heavy Majorana neutrino masses. Such a two right-handed neutrino
(2RHN) scenario has also the advantage of reducing the total number of free parameters so
that the analysis of neutrino phenomenology and leptogenesis becomes much
simpler~\cite{Frampton:2002qc,Raidal:2002xf,GonzalezFelipe:2003fi,Barger:2003gt,Ibarra:2003up,Guo:2006qa}.
To understand this, we recall that in the SM extended with three right-handed
neutrinos the Lagrangian of the neutrino sector contains 18 parameters at high energies:
3 heavy Majorana masses plus 15 real parameters (9 moduli and 6 phases) needed to specify the
Yukawa coupling matrix $\mathbf{Y}^\nu$. Of these, only 15 parameters are independent in
what concerns the light neutrino mass matrix $\mathbf{m}_\nu$ obtained through the seesaw
mechanism (the three Majorana masses $M_i$ can be absorbed into $\mathbf{Y}^\nu$ by an
appropriate rescaling of its elements). On the other hand, in the 2RHN case, there are
altogether 11 parameters: 2 heavy Majorana masses together with 9 real parameters (6 moduli and
3 phases) that specify the $3 \times 2$ matrix $\mathbf{Y}^\nu$. Once again, performing
the rescaling of the two heavy Majorana masses, the effective number is reduced to 9
parameters.

In the three right-handed neutrino case, the measurable quantities associated to the
light neutrino mass matrix are 3 masses, 3 mixing angles, and 3 phases, while for two
right-handed neutrinos this number is reduced by 2, since the lightest neutrino is
massless and its associated Majorana phase vanishes. Thus, in the latter case there is no
possibility of three quasidegenerate light neutrinos, and only two mass spectra are
allowed: a normal hierarchy with $m_1=0$, $m_2=m_\text{sol}$ and $m_3=m_\text{atm}$ or an
inverted hierarchy with $m_3=0$, $m_{1} = m_\text{atm}$ and $m_2 \approx
m_\text{atm}+m_\text{sol}^2/(2 m_\text{atm})$.

The parameters in $\mathbf{Y}^\nu$ which are associated with the seesaw but are not
determined by low-energy measurable quantities are most easily disentangled if this
matrix is written in terms of the orthogonal parametrization of Eq.~\eqref{YnuRmatrix}.
The six (two) undetermined parameters of the 3RHN (2RHN) model would correspond precisely
to those parameters that specify the complex orthogonal matrix $\mathbf{R}$. The 2RHN
model can then be thought of as the limiting case of the 3RHN model in which the heaviest
right-handed neutrino $N_3$ decouples from the theory because it is very heavy or its
Yukawa couplings are very small. From Eq.~\eqref{YnuRmatrix} one finds for the third
column of the matrix $\mathbf{R}$
\begin{align}
    \mathbf{R}_{i3}=\frac{v}{\sqrt{m_i M_3}}\,(\mathbf{U}^T\mathbf{Y}^\nu)_{i3}.
\end{align}
Thus, as $M_3 \rightarrow \infty$, $\mathbf{R}_{23}, \mathbf{R}_{33} \rightarrow 0$,
while $\mathbf{R}_{13} \rightarrow 1$ due to orthogonality. Consequently, in the 2RHN
model the orthogonal matrix $\mathbf{R}$ takes the simple $3\times2$ structure
\begin{align} \label{2RHNorth}
    \mathbf{R}=\begin{pmatrix}
    0&0\\
    \cos z & - \sin z\\
    \pm \sin z & \pm \cos z
    \end{pmatrix},
\end{align}
where $z$ is a complex angle and the $\pm$ signs account for a discrete indeterminacy in
$\mathbf{R}$. Using this form, the elements of the Dirac-neutrino Yukawa coupling matrix
read
\begin{align}\label{Ynu2RHN}
\begin{split}
\mathbf{Y}^{\nu}_{\alpha 1} &=\sqrt{M_{1}}(\sqrt{m_{2}}\cos z~U_{\alpha 2}^{\ast}\pm
\sqrt{m_{3}}\sin z~U_{\alpha 3}^{\ast })/v, \\
\mathbf{Y}^{\nu}_{\alpha 2} &=\sqrt{M_{2}}(-\sqrt{m_{2}}\sin z~U_{\alpha 2}^{\ast }\pm
\sqrt{m_{3}}\cos z~U_{\alpha 3}^{\ast })/v.
\end{split}
\end{align}
For an inverted hierarchy, the corresponding matrix $\mathbf{R}$ reads
\begin{align} \label{2RHNorthinv}
    \mathbf{R}=\begin{pmatrix}
    \cos z & - \sin z\\
    \pm \sin z & \pm \cos z\\
    0 & 0
    \end{pmatrix},
\end{align}
and Eqs.~\eqref{Ynu2RHN} become
\begin{align}\label{Ynu2RHNinv}
\begin{split}
\mathbf{Y}^{\nu}_{\alpha 1} &=\sqrt{M_{1}}(\sqrt{m_{1}}\cos z~U_{\alpha 1}^{\ast}\pm
\sqrt{m_{2}}\sin z~U_{\alpha 2}^{\ast })/v,\\
\mathbf{Y}^{\nu}_{\alpha 2} &=\sqrt{M_{2}}(-\sqrt{m_{1}}\sin z~U_{\alpha 1}^{\ast }\pm
\sqrt{m_{2}}\cos z~U_{\alpha 2}^{\ast })/v.
\end{split}
\end{align}

It is clear that without any assumption about the complex parameter $z$ there is no
direct link between the leptogenesis asymmetries and leptonic CP violation at low
energies. Nevertheless, the fact that the number of unknown parameters at high energies
is reduced with respect to the 3RHN case makes it possible to establish a connection
between thermal leptogenesis and low-energy neutrino parameters with simple assumptions
about the physics at high energies. For instance, assuming $M_1 \ll M_2$ and $z$ real,
the flavored leptogenesis asymmetries given in Eq.~\eqref{cpasymIbRreal1} for a normal
hierarchical neutrino mass spectrum read
\begin{align}\label{cpasymIbRreal2}
       \epsilon_1^\alpha \simeq \pm \frac{3\,M_1}{16\pi v^2}\frac{m_\text{atm}\sqrt{m_\text{sol}\, m_\text{atm}}\, \sin z \cos z\,\mathcal{I}_{23}^\alpha}{m_\text{sol}\,\cos^2 z +m_\text{atm}\,\sin^2 z}\,,
\end{align}
with the rephasing invariant quantities $\mathcal{I}_{23}^\alpha$ given by
Eqs.~\eqref{imUUexample} with the Majorana phase $\gamma_2=0$. On the other hand, the
total (unflavored) asymmetry $\epsilon_1$ would vanish in this case since $\sum_\alpha
\mathcal{I}_{23}^\alpha =0$.

We note that the asymmetry~\eqref{cpasymIbRreal2} is maximal when
\begin{align}
\sin z = \sqrt{\frac{m_\text{sol}}{m_\text{sol}+m_\text{atm}}} \approx \sqrt{\frac{m_\text{sol}}{m_\text{atm}}},
\end{align}
which implies the upper bound
\begin{align}\label{cpasymIbRreal2max}
       |\epsilon_1^\alpha| \leq \frac{3\,M_1 m_\text{atm}}{32\pi v^2}\,|\mathcal{I}_{23}^\alpha|\,.
\end{align}
Nevertheless, we remark that a maximal CP asymmetry does not necessarily correspond to a
maximal baryon asymmetry since leptogenesis also crucially depends on the subsequent
washout effects.

Yukawa coupling structures with texture zeros provide a well-motivated framework in which
the number of high-energy parameters is reduced and relations among low-energy neutrino
observable quantities may be implied. In the presence of a family symmetry, the charge
assignment under the symmetry to particles may lead to one or several Yukawa couplings
which are negligibly small compared to the others. It is clear that texture zeros are in general not WB invariant. This means that a given texture zero, which arises in a certain WB, may not be present or may appear in a different matrix entry in another WB. It is, however, important to distinguish among various types of texture zeros. Some of them have no physical meaning because they can be obtained through a WB transformation starting from arbitrary flavor matrices~\cite{Branco:2007nn}. On the other hand, there are texture zeros that do have physical implications. Among the latter, one should distinguish between zeros that result from a flavor symmetry from those that just reflect an ad-hoc assumption on the flavor structure. It should be emphasized that even when texture zeros result from a family symmetry imposed on the Lagrangian, they are manifest only in a particular basis, namely, the basis where the symmetry is transparent. Furthermore, it has been shown~\cite{Branco:2005jr} that a large class of sets of leptonic texture zeros imply the vanishing of certain CP-odd WB invariants. These invariants allow, for instance, one to recognize a flavor model, which is characterized by certain texture zeros in the matrix $\mathbf{Y}^{\nu}$ in the basis where the charged-lepton and right-handed neutrino mass matrices are diagonal, when the same model is written in an arbitrary WB where the zeros are not manifest.

The possibility of a texture zero in the (1,1) position is quite interesting from the phenomenological point
of view, since in the quark sector such a postulate, if applied to the up and down quark
matrices, leads to the remarkably successful prediction for the Cabibbo angle $\theta_C =
\theta_{12}=\sqrt{m_d/m_s}$~\cite{Gatto:1968ss}. Applying this rationale to the neutrino
sector of the 2RHN model, i.e., imposing $\mathbf{Y}^{\nu}_{11}=0$, would fix the value of
the unknown parameter $z$ in terms of low-energy neutrino data. From Eqs.~\eqref{Ynu2RHN}
and \eqref{Ynu2RHNinv}, one finds
\begin{align}\label{tanz}
    \tan z = \mp \sqrt{\frac{m_2}{m_3}}\, \frac{\mathbf{U}_{e2}^\ast}{\mathbf{U}_{e3}^\ast}, \quad
    \tan z = \mp \sqrt{\frac{m_1}{m_2}}\, \frac{\mathbf{U}_{e1}^\ast}{\mathbf{U}_{e2}^\ast},
\end{align}
for normal and inverted-hierarchical neutrino mass spectrum, respectively. Note also
that imposing additional texture zeros in the neutrino Yukawa coupling matrix would yield
relations among the mixing angles and neutrino masses. To see the implications for leptogenesis of a texture zero in the (1,1) position, we
consider the unflavored asymmetry given in Eq.~\eqref{cpasymIcR}, rewritten as
\begin{align}\label{cpasymIcRex}
       \epsilon_1 \simeq  \frac{3\,M_1}{16\pi v^2}\frac{(m_3^2-m_2^2)\,\text{Im}(\sin^2 z)}{m_2 |\cos^2 z| + m_3 |\sin^2 z|}.
\end{align}
Using the first relation in Eq.~\eqref{tanz}, $\epsilon_1$ can then be expressed in terms
of low-energy quantities as
\begin{align}\label{cpasymIcRex1}
       \epsilon_1 & \simeq  \frac{3\,M_1(m_3^2-m_2^2)}{16\pi v^2 m_{ee}}\frac{\text{Im}(\mathbf{U}_{e2}^{\ast 2}
       \mathbf{U}_{e3}^{2})}{|\mathbf{U}_{e2}|^{2} + |\mathbf{U}_{e3}|^{2}}\nonumber\\
        &\approx - \frac{3\,M_1m_\text{atm}^2}{16\pi v^2 m_{ee}}\, \sin^2 \theta_{13} \sin(2\delta+\alpha_1).
\end{align}
Thus, in this simple example, there is a correlation between the sign of the baryon asymmetry and low-energy leptonic
CP violation. Clearly, one texture zero is sufficient to establish such a link because the
sign of $\epsilon_1$ is determined by $\text{Im}(\tan^2 z)$, which in turn is fixed by
Eq.~\eqref{tanz}. If we consider the flavored asymmetries $\epsilon_1^\alpha$ given in
Eq.~\eqref{cpasymIbR}, it would still be possible to write them in terms of low-energy
observables. However, the direct connection between the sign of the baryon asymmetry and
CP violation at low energies would be lost since the phase contributions to the
individual asymmetries are more involved.

\begin{figure}[t]
\begin{center}
\includegraphics[width=9cm]{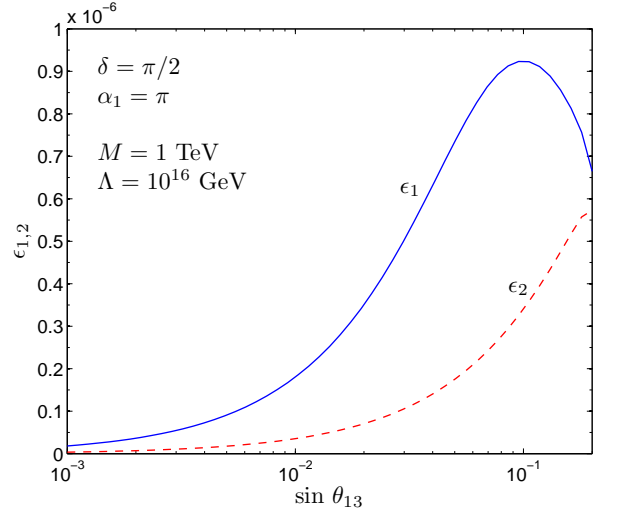}
\caption{The CP
asymmetries $\epsilon_1$ and $\epsilon_2$ as functions of $\sin\theta_{13}$ generated in a minimal radiative leptogenesis scenario. The curves correspond to the approximate analytic expressions given in Eqs.~(\ref{CPapp}).}
\label{fig:radasym}
\end{center}
\end{figure}

In the examples presented above, the heavy Majorana neutrinos have been assumed hierarchical in mass so that leptogenesis is dominated by the decays of $N_1$, the lightest of the heavy states. One can also envisage a situation when the heavy Majorana neutrino mass spectrum is exactly degenerate at energies above the leptogenesis scale. In this case, a small mass splitting among the heavy Majorana neutrino states can be generated, in a natural way, via the renormalization group running from the degeneracy scale down to the leptogenesis scale~\cite{GonzalezFelipe:2003fi}. To illustrate this, we consider again the minimal scenario with only two right-handed neutrinos, and assume that $M_1=M_2 \equiv M$, at a scale $\Lambda>M$. The evolution of the right-handed neutrino mass matrix $\mathbf{m}_R$ as a function of the energy scale $\mu$ is governed by the renormalization group equation
\begin{align}
\label{RGEMR} \frac{d \mathbf{m}_R}{dt}=\mathbf{H}^{\nu\,T} \mathbf{m}_R+\mathbf{m}_R\, \mathbf{H}^\nu\,,\quad
t=\dfrac{1}{16\,\pi^2}\ln\left(\mu/\Lambda\right)\,.
\end{align}
Then, defining $\delta_N \equiv M_2/M_1-1$, which quantifies the degree of degeneracy between $M_1$ and $M_2$,
the radiatively induced mass splitting at the decoupling scale $M$ will be approximatively given by
\begin{align}
\label{deltaNM}
\delta_N \simeq \frac{M m_\text{atm}}{8\pi^2 v^2}\,\left(1-\rho\right)\ln\left(\Lambda/M\right),
\end{align}
where $\rho \equiv m_\text{sol}/m_\text{atm}$.

To analyze the implications for leptogenesis, we impose the particular texture zero $\mathbf{Y}^\nu_{12}=0$~\cite{GonzalezFelipe:2003fi}. Then, in terms of the low-energy neutrino parameters, the  unflavored CP asymmetries $\epsilon_i$ generated by the $N_i$ decays read as
\begin{align}
\label{CPapp} \epsilon_1 &\simeq-\frac{3y_\tau^2}{64\pi} \frac{1+\rho}{(1-\rho)(\rho+x^2-\Delta)}\tan\theta_{13} \sin(\alpha_1/2) \nonumber\\
&\times\left[\cot\theta_{12} \cos(\delta-\alpha_1/2) + \tan\theta_{13} \cos(\alpha_1/2)\right]\,, \nonumber\\
\nonumber\\
\epsilon_2 &\simeq \frac{\rho+x^2-\Delta}{1+\rho\,x^2+\Delta}\times \epsilon_1\,,
\end{align}
where $x = \tan\theta_{13}/(\sqrt{\rho}\,\sin\theta_{12})$, $y_\tau$ is the $\tau$ Yukawa coupling, and
\begin{align}
\Delta = \frac{1}{2}(1-\rho)\left[-1+x^2+\sqrt{1+2x^2\cos \alpha_1+x^4} \right].
\end{align}

Taking, for instance, $\alpha_1= \pi$ and $\delta = \pi/2$, the CP
asymmetry $\epsilon_1$ reaches its maximum value for $x = \sqrt{\rho}\,$. This corresponds to $\tan\theta_{13} = \rho
\sin\theta_{12} \simeq 0.1$ and
\begin{align}
    |\epsilon_1^{\text{max}}| \simeq \frac{3y_\tau^2\,\cos\theta_{12}}{128\pi} \frac{
1+\rho}{1-\rho} \simeq 10^{-6}\,.
\end{align}
In Fig.~\ref{fig:radasym}, the CP asymmetries $\epsilon_i$ are plotted as functions of $\sin\theta_{13}$
taking $\Lambda=10^{16}$~GeV, $M=1$~TeV, $\delta=\pi/2$, $\alpha=\pi$, and
assuming $y_\tau =0.01$ in the analytical estimates. The curves correspond
to the approximate expressions given in Eqs.~\eqref{CPapp}. It is interesting to note that, in this case, the maximum of the leptogenesis asymmetry $\epsilon_1$ is reached for $s_{13}\simeq 0.1$, which is the sensitivity range of future reactor and superbeam neutrino oscillation experiments.

\subsubsection{Leptogenesis and flavor symmetries}
\label{sec4.2.4}

Present neutrino data (see Table~\ref{Tabnudata}) are in good
agreement with the so-called tribimaximal (TB) leptonic mixing~\cite{Harrison:2002er},
\begin{align} \label{UTB}
\mathbf{U}_\text{TB}=\begin{pmatrix}
\sqrt{\frac{2}{3}}&\sqrt{\frac{1}{3}}&0\\
-\sqrt{\frac{1}{6}}&\sqrt{\frac{1}{3}}&-\sqrt{\frac{1}{2}}\\
-\sqrt{\frac{1}{6}}&\sqrt{\frac{1}{3}}&\sqrt{\frac{1}{2}}
\end{pmatrix},
\end{align}
corresponding to the mixing angles $\theta_{12} = \arcsin (1/\sqrt{3})$, $\theta_{23}=
-\pi/4$ and $\theta_{13}=0$ in the standard PDG parametrization~\cite{Nakamura:2010zzi}
given in Eq.~\eqref{VPDG}. Since the above mixing matrix does not depend on any mass
parameter, it is usually referred to as a mass-independent mixing scheme. If one assumes
that the leptonic mixing is described at leading order by $\mathbf{U}_\text{TB}$, it is
natural to consider that this special structure arises due to a family symmetry. In
particular, discrete symmetries are quite attractive, and the tetrahedral (alternating)
group $A_4$, corresponding to even permutations of four objects, has been especially
popular and featured in a large number of models of leptonic
mixing~\cite{Altarelli:2010gt}.

From the phenomenological viewpoint, one of the attractive features of the
mass-independent mixing schemes is that they lead to a predictive neutrino mass matrix
structure which contains just a few parameters. The latter can then be directly related
to neutrino observables such as the neutrino mass-squared differences, the absolute
neutrino mass scale, and the effective mass parameter in $0\nu\beta\beta$ decays.

Besides restricting the number of relevant parameters, the imposition of certain flavor
symmetries in the lepton sector of the theory may lead to constraints on the CP
asymmetries in the framework of seesaw leptogenesis. In particular, it has been recently
shown that type I and type III seesaw flavor models that lead to an exact
mass-independent leptonic mixing have a vanishing leptogenesis CP asymmetry in leading
order~\cite{Jenkins:2008rb,Bertuzzo:2009im,AristizabalSierra:2009ex,Felipe:2009rr}. To
illustrate this fact, we consider the standard type I seesaw framework with three
right-handed neutrinos $\nu_R$. In this case, the relevant Lagrangian terms are given by
Eq.~\eqref{LtypeIseesaw}, and the effective neutrino mass matrix $\mathbf{m}_\nu$ is
obtained through the standard seesaw formula of Eq.~\eqref{mnutypeIseesaw}.

We assume that the type I seesaw Lagrangian is invariant under the transformations of
a given flavor symmetry group $\mathcal{G}$, so that left-handed and right-handed lepton
fields transform as $\nu_L \rightarrow \mathcal{G}_L \nu_L$ and $\nu_R \rightarrow
\mathcal{G}_R \nu_R$, respectively. Clearly, the generators $\mathcal{G}_L$ and
$\mathcal{G}_R$ are unitary matrices built from the columns of the unitary matrices
$\mathbf{U}$ and $\mathbf{U}_R$ that diagonalize the matrices $\mathbf{m}_\nu$ and
$\mathbf{m}_R$, respectively. The Lagrangian invariance then implies that the
Dirac-neutrino Yukawa coupling matrix $\mathbf{Y}^\nu$ should satisfy the symmetry
relation $\mathcal{G}_L^T \mathbf{Y}^\nu \,\mathcal{G}_R^\ast=\mathbf{Y}^\nu$. To analyze
the consequences of this relation for leptogenesis, we rewrite the symmetry equations in
the basis in which the right-handed neutrino mass matrix is diagonal,
\begin{align}\label{GH}
\mathcal{G}_R^{^\prime \dagger} \mathbf{d}_M\,\mathcal{G}_R^{\prime\ast}=\mathbf{d}_M,\quad \mathcal{G}_R^{\prime T}
\mathbf{H}^\nu \mathcal{G}_R^{\prime\ast}=\mathbf{H}^\nu,
\end{align}
with $\mathcal{G}_R^\prime = \mathbf{U}_R^\dagger\mathcal{G}_R \mathbf{U}_R$. Assuming a
nondegenerate heavy neutrino mass spectrum, the first relation in Eq.~(\ref{GH}) requires
the symmetry generators $\mathbf{G}_{R,i}^\prime\, (i=1,2,3)$ to be diagonal. Their
explicit forms are thus given by $\mathbf{G}_{R,1}^\prime = \text{diag}(1,1,-1)$,
$\mathbf{G}_{R,2}^\prime = \text{diag}(1,-1,1)$ and $\mathbf{G}_{R,3}^\prime =
\text{diag}(-1,1,1)$. The action of any two of these matrices in the second relation of
Eq.~(\ref{GH}) would then enforce $\mathbf{H}^\nu$ to be diagonal, which in turn implies
that the leptogenesis asymmetries~\eqref{cpasymI} and \eqref{cpasymIa} are equal to zero.
The case of a degenerate heavy neutrino mass spectrum can be analyzed in a similar way.
In the latter case, no leptogenesis CP asymmetry can be generated in leading order
either~\cite{Felipe:2009rr}. Note also that, due to the specific form of the matrix
combination $\mathbf{H}^\nu$ that appears in the leptogenesis CP asymmetries, only the
symmetry generators $\mathcal{G}_R$ are really needed in the above proof of vanishing
leptogenesis.

Clearly, if the complete mass matrix symmetry is not imposed as the residual symmetry of
the type I seesaw Lagrangian, the above conclusions do not necessarily remain valid. For
instance, requiring the right-handed sector of the Lagrangian to be invariant just under
the transformation $\nu_R \rightarrow \mathbf{G}_{R,1}\, \nu_R$ would lead to vanishing
$\mathbf{H}^\nu_{13}$ and $\mathbf{H}^\nu_{23}$ off-diagonal elements. Yet a leptogenesis
asymmetry could in principle be generated with a nonzero $\mathbf{H}^\nu_{12}$ matrix
element.

In a type II seesaw framework, the interplay between flavor symmetries and the
leptogenesis CP asymmetries is actually different~\cite{Varzielas:2011tp}. In the latter
case, we can see from Eq.~\eqref{cpasymIIa} that the flavored leptonic CP asymmetries
$\epsilon_i^{\alpha\beta}$  are proportional to the combination
$\text{Im}\bigl[\mu_i^\ast\mu_j \mathbf{Y}^{\Delta_i}_{\alpha\beta}
\mathbf{Y}^{\Delta_j\ast}_{\alpha\beta} \bigr]$, while the unflavored asymmetry
$\epsilon_i$ depends on $\text{Im}\bigl[\mu_i^\ast\mu_j \text{Tr}( \mathbf{Y}^{\Delta_i}
\mathbf{Y}^{\Delta_j\ast}) \bigr]$. To analyze the implications of discrete flavor
symmetries for type II seesaw leptogenesis, it is convenient to rewrite the light
neutrino mass matrix $\mathbf{m}_{\nu}=\mathbf{U}_\text{TB} \, \mathbf{d}_m  \,
\mathbf{U}_\text{TB}^T$ in terms of three contributions~\cite{Varzielas:2011tp},
\begin{align}\label{mnuCPD}
\mathbf{m}_{\nu} = x\, \mathbf{C} + y\, \mathbf{P} + z\, \mathbf{D},
\end{align}
where $x,y$, and $z$ are complex numbers;
\begin{align}\label{CPstruct}
\mathbf{C}=\frac{1}{3}\begin{pmatrix}
2&-1&-1\\
-1&2&-1\\
-1&-1&2
\end{pmatrix}\,,\quad \mathbf{P}=\begin{pmatrix}
1&0&0\\
0&0&1\\
0&1&0
\end{pmatrix},
\end{align}
denote the well-known magic and $\mu$-$\tau$ symmetric matrices, and $\mathbf{D}$
is the democratic matrix with all entries equal to $1/3$.

As it turns out, the type II seesaw leptonic asymmetry is in general nonvanishing. For
leptogenesis to be viable at least two scalar $SU(2)$ triplets are needed. Suppose, for
instance, that both triplets are singlets under the family symmetry. Then, one of them
can be associated to the $\mathbf{P}$ contribution and the other one to the $\mathbf{C}$
contribution in Eq.~\eqref{mnuCPD}. If a third scalar triplet is available, it may be
associated to the democratic component $\mathbf{D}$. In this minimal setup, unless a
democratic contribution is present, the unflavored asymmetry $\epsilon_i$  is
zero\footnote{Note that, if each scalar triplet is simultaneously associated to the
magic and $\mu$-$\tau$ symmetric contributions, the unflavored asymmetry is, in general,
nonvanishing.} because the product of the matrices $\mathbf{C}$ and $\mathbf{P}$ is
traceless, which then implies $\text{Tr}( \mathbf{Y}^{\Delta_i}
\mathbf{Y}^{\Delta_j\ast})=0$. On the other hand, the flavored leptogenesis asymmetries
do not necessarily vanish even when the democratic component is absent.

In addition to TB mixing, there are other mass-independent structures that can reproduce
the observed leptonic mixing angles. Below we give some examples of such mass-independent schemes.

The transposed TB mixing has the mixing matrix~\cite{Fritzsch:1995dj}
\begin{equation}\label{UtTB}
\mathbf{U}_\text{tTB}=
\begin{pmatrix}
\sqrt{\frac{1}{2}}&-\sqrt{\frac{1}{2}}&0\\
-\sqrt{\frac{1}{6}}&-\sqrt{\frac{1}{6}}&\sqrt{\frac{2}{3}}\\
\sqrt{\frac{1}{3}}&\sqrt{\frac{1}{3}}&\sqrt{\frac{1}{3}}
\end{pmatrix},
\end{equation}
where the solar and atmospheric mixing angles are given by $\theta_{12} = \pi/4$ and
$\theta_{23}= \arctan \sqrt{2}$, respectively. The well-known bimaximal structure has the
mixing matrix~\cite{Barger:1998ta}
\begin{equation} \label{UB}
\mathbf{U}_\text{B}=
\begin{pmatrix}
\sqrt{\frac{1}{2}}&-\sqrt{\frac{1}{2}}&0\\
\frac{1}{2}&\frac{1}{2}&-\sqrt{\frac{1}{2}}\\
\frac{1}{2}&\frac{1}{2}&\sqrt{\frac{1}{2}}
\end{pmatrix},
\end{equation}
and the corresponding mixing angles are in this case $\theta_{12} = \theta_{23}= \pi/4$.
There are also two golden ratio proposals related to the
quantity $\Phi=(1+\sqrt{5})/2$. The first matrix is~\cite{Kajiyama:2007gx}
\begin{equation}\label{UGR1}
\mathbf{U}_\text{GR}=
\begin{pmatrix}
\sqrt{\frac{1}{2}+\frac{1}{2\sqrt{5}}}&\sqrt{\frac{2}{5+\sqrt{5}}}
&0\\
-\sqrt{\frac{1}{5+\sqrt{5}}}&\sqrt{\frac{1}{5-\sqrt{5}}}&\sqrt{\frac{1}{2}}\\
\sqrt{\frac{1}{5+\sqrt{5}}}&-\sqrt{\frac{1}{4}+\frac{1}{4\sqrt{5}}}&
\sqrt{\frac{1}{2}}
\end{pmatrix},
\end{equation}
with the associated angles $\theta_{12} = \arctan (1/\Phi)$ and $\theta_{23}=\pi/4$,
while the second matrix reads~\cite{Rodejohann:2008ir}
\begin{equation}\label{UGR2}
\mathbf{U}_\text{GR}=
\begin{pmatrix}
\frac{1+\sqrt{5}}{4}& \frac{\sqrt{5-\sqrt{5}}}{2\sqrt{2}} & 0\\
-\frac{\sqrt{5-\sqrt{5}}}{4} & \frac{1+\sqrt{5}}{4 \sqrt{2}}
& -\sqrt{\frac{1}{2}} \\
-\frac{\sqrt{5-\sqrt{5}}}{4} & \frac{1+\sqrt{5}}{4 \sqrt{2}} &
\sqrt{\frac{1}{2}}
\end{pmatrix},
\end{equation}
with $\theta_{12} = \arccos (\Phi/2)$ and $\theta_{23}=-\pi/4$. Finally, the so-called
hexagonal mixing~\cite{Giunti:2002sr,Xing:2002az,Albright:2010ap} is described by the
matrix
\begin{equation}\label{UH}
\mathbf{U}_\text{H}=
\begin{pmatrix}
\sqrt{\frac{3}{4}}&\frac{1}{2}&0\\
-\sqrt{\frac{1}{8}}&\sqrt{\frac{3}{8}}&-\sqrt{\frac{1}{2}}\\
-\sqrt{\frac{1}{8}}&\sqrt{\frac{3}{8}}&\sqrt{\frac{1}{2}}
\end{pmatrix},
\end{equation}
which corresponds to the mixing angles $\theta_{12} = \pi/6$ and $\theta_{23}=-\pi/4$.

As in the TB case, the above mixing schemes predict the mixing angle $\theta_{13}=0$ and therefore no Dirac-type CP violation. The conclusions for the leptogenesis asymmetries previously drawn are equally valid in all these cases.

\subsection{CP-odd invariants for leptogenesis}
\label{sec4.3}

Based on the most general CP transformations in the lepton sector, that leave invariant the gauge interactions, we constructed WB invariants that need to vanish in order for CP invariance to hold at low energies (cf. Sec.~\ref{sec2.4}). CP-odd conditions derived from WB invariants are a powerful tool for model building, since they can be applied to any model without the need to go to a special basis. In this section, we are particularly interested in the construction of WB invariants which are sensitive to the CP-violating phases of leptogenesis.

In the case of unflavored leptogenesis, the CP asymmetry is only sensitive to phases
appearing in the matrix $\mathbf{H}^\nu$ so that the relevant WB invariant conditions can
be readily derived~\cite{Branco:2001pq}:
\begin{align}\label{wbinv}
\begin{split}\textcelsius
I_1 &\equiv \text{Im Tr}[\mathbf{H}^\nu (\mathbf{m}_R^{\dagger}\mathbf{m}_R)\, \mathbf{m}_R^{*}\, \mathbf{H}^{\nu*} \mathbf{m}_R]=0, \\
I_2 &\equiv \text{Im Tr}[\mathbf{H}^\nu (\mathbf{m}_R^{\dagger}\mathbf{m}_R)^2 \mathbf{m}_R^{*}\, \mathbf{H}^{\nu*} \mathbf{m}_R] = 0,\\
I_3 &\equiv \text{Im Tr}[\mathbf{H}^\nu (\mathbf{m}_R^{\dagger}\mathbf{m}_R)^2 \mathbf{m}_R^{*}\, \mathbf{H}^{\nu*} \mathbf{m}_R\, (\mathbf{m}_R^{\dagger}\mathbf{m}_R)] = 0.
\end{split}
\end{align}
The choice of these invariant conditions is not unique. For instance, by replacing $\mathbf{m}_R$ by $\mathbf{m}_R^{\ast-1}$ in the invariants $I_n$, one can construct another set of invariants which, for hierarchical right-handed neutrinos, are more suitable for leptogenesis~\cite{Davidson:2003yk}.

The quantities given in Eq.~\eqref{wbinv} can be evaluated in any convenient weak basis. In the WB in which the right-handed neutrino mass matrix $\mathbf{m}_R$ is diagonal and real, one obtains
\begin{align}\label{wbinvdiag}
\begin{split}
I_1&= \sum_{i=1}^3 \sum_{j>i}^3 M_i M_j (M_j^2-M_i^2)\, \text{Im}\left[(\mathbf{H}_{ij}^{\nu})^2\right]=0,\\
I_2&= \sum_{i=1}^3 \sum_{j>i}^3 M_i M_j (M_j^4-M_i^4)\, \text{Im}\left[(\mathbf{H}_{ij}^{\nu})^2\right]=0,\\
I_3&= \sum_{i=1}^3 \sum_{j>i}^3 M_i^3 M_j^3 (M_j^2-M_i^2)\, \text{Im}\left[(\mathbf{H}_{ij}^{\nu})^2\right]=0.\\
\end{split}
\end{align}

The appearance of the quadratic combination $(\mathbf{H}_{ij}^{\nu})^2$ in the above
expressions simply reflects the well-known fact that phases of $\pi/2$ in
$\mathbf{H}_{ij}^{\nu}$ do not imply CP violation. Note that Eqs.~\eqref{wbinvdiag}
constitute a set of linear equations in terms of the quantities
$\text{Im}\left[(\mathbf{H}_{ij}^{\nu})^2\right]$, where the coefficients are functions
of the right-handed neutrino masses $M_i$. The determinant of this system is equal to
${M_1}^2 {M_2}^2 {M_3}^2 (M_2^2-M_1^2)^2(M_3^2-M_1^2)^2(M_3^2-M_2^2)^2$. It then follows
that, if none of the $M_i$ vanish and there is no degeneracy in the masses $M_i$, the
simultaneous vanishing of $I_1, I_2$ and $I_3$ implies the vanishing of
$\text{Im}\left[(\mathbf{H}_{12}^{\nu})^2\right]$,
$\text{Im}\left[(\mathbf{H}_{13}^{\nu})^2\right]$ and
$\text{Im}\left[(\mathbf{H}_{23}^{\nu})^2\right]$. This implies, in turn, that the
unflavored type I leptogenesis asymmetries given in Eq.~\eqref{cpasymIa} are all equal to
zero.

We note that the WB invariants $I_i$ defined in Eq.~\eqref{wbinv} vanish if the heavy Majorana neutrinos are degenerate in mass. It is nevertheless possible to construct WB invariants which control the strength of CP violation in the latter case. For instance, the weak-basis invariant
\begin{equation}\label{highCPinv}
\mathcal{J}^{CP}_\text{deg}=M^{-6}\operatorname{Tr}\left[ \mathbf{Y}^{\nu} \mathbf{Y}^{\nu\, T}
\mathbf{Y}^{\ell} \mathbf{Y}^{\ell \dagger} \mathbf{Y}^{\nu \ast}
\mathbf{Y}^{\nu \dagger},\mathbf{Y}^{\ell \ast} \mathbf{Y}^{\ell\, T}\right]^{3},
\end{equation}
where $M$ is the common heavy Majorana neutrino mass, does not vanish in the case of an exactly degenerate heavy Majorana neutrino mass spectrum. Thus, $\mathcal{J}^{CP}_\text{deg} \neq 0$ would signal the violation of CP in this case.

For flavored leptogenesis, the phases appearing in $\mathbf{H}^\nu$ are still relevant. There is, however, the possibility of generating the required CP asymmetry even for $\mathbf{H}^\nu$ real. In this case, additional CP-odd WB invariant conditions are required, since $I_i$ cease to be necessary and sufficient. A possible choice are the CP-odd WB invariant conditions obtained from $I_i$ through the substitution of $\mathbf{H}^\nu$ by $\mathbf{\widehat{H}}^\nu={\mathbf{Y}^\nu}^{\dagger} \mathbf{Y}^\ell {\mathbf{Y}^\ell}^{\dagger} \mathbf{Y}^\nu$, and which are sensitive to the additional phases appearing in flavored leptogenesis~\cite{Branco:2009by}.

\section{CONCLUSIONS AND OUTLOOK}
\label{sec:conclusion}

After almost 50 years since its discovery, CP violation is still at the core of particle physics and cosmology. In the quark sector, CPV has been established in both the kaon and $B$ meson sectors, and the results obtained so far are compatible with the standard complex CKM mixing picture. With the discovery of neutrino masses, the natural expectation is that CP is also broken in the lepton sector. Indeed, in a unified description of fundamental particle physics, it is hard to imagine a scenario with CPV in the quark sector and not in the leptonic one.

The prospects for discovering LCPV in neutrino oscillation experiments mainly depend on the value of the reactor neutrino mixing angle $\theta_{13}$. The smaller this angle is, the longer we will have to wait until experiments become sensitive to CP violating effects. In the best-case scenario, CPV could be discovered in the near future by combining the data of reactor neutrino and superbeam experiments (see Sec.~\ref{sec3.2.4}), if $\theta_{13}$ is not too small.  If this is not the case, then we will probably have to wait for upgraded superbeams, $\beta$ or electron-capture beams, or neutrino factories. The recent data from the T2K experiment in Japan indicate the appearance of $\nu_{e}$ from the original $\nu_\mu$ neutrino beam with a number of observed $e$-like events, which exceed the expected ones. The probability of explaining the results with $\theta_{13}=0$ is less than $1\%$ and the obtained $90\%$~C.L. interval for $\sin^2(\theta_{13})$ is $[0.03(0.04),0.28(0.34)]$, with the numbers in parenthesis referring to the case in which $\dmatm < 0$. Such an indication of a nonzero (and not very small) value of $\theta_{13}$ is a good omen for the prospects of discovering CP violation in the near future. With some luck, a hint for CPV could be provided by combining the data of superbeam (NO$\nu$A and T2K )and reactor neutrino experiments (Double Chooz, Daya Bay and RENO). In any case, upgraded superbeams, $\beta$ beams, or neutrino factories will be for sure necessary to confirm such a hint and measure the CP-violating phase $\delta$.

It has been advocated that $\ndbd$ decays could, in principle, provide some information about Majorana-type CP violation in the lepton sector. Although this is true in theory, the task of extracting information about the Majorana phases using $\ndbd$ results is nontrivial. This holds even in the simplest scenario in which $\ndbd$ is induced by the exchange of light Majorana neutrinos. As discussed in Sec.~\ref{sec3.3}, the main difficulty in the Majorana phase determination from $\ndbd$ measurements resides on the uncertainties inherent to the nuclear matrix element determination. In particular, the precision required to make conclusive statements about Majorana CP violation seems to be far from what can be achieved. The observation of $\ndbd$ would establish the Majorana nature of neutrinos, and therefore would favor some neutrino mass generation mechanisms over others. In the near future, the experimental sensitivity of $\ndbd$ experiments will cover the region where $\dmatm < 0$, covering the IH and QD neutrino spectrum cases. The combined study of $\ndbd$ and $\beta$ decay, neutrino oscillations, and also cosmological data, will be crucial to improve the knowledge of neutrino fundamental parameters and test the minimal $\ndbd$ mechanism.

If neutrino masses are generated at an energy scale not far from the electroweak scale, there is a hope to test the neutrino mass mechanism at high-energy colliders like the LHC. In such a case, it is straightforward to infer that the presence of CP violation in the neutrino sector would have an impact on the physical processes involving the neutrino mass mediators. In Sec.~\ref{sec3.4} and \ref{sec3.5}, we illustrated how the leptonic CP phases affect the rates of several lepton decays in the context of the type II seesaw mechanism, in which neutrino masses are generated by the tree-level exchange of scalar triplets. The fact that the effective neutrino mass matrix is linear in the triplet-lepton-lepton couplings, allows one to write in a model-independent way the decay rates in terms of the low-energy neutrino parameters. In particular, we have seen that some decays are only sensitive to a particular set of CP phases. Therefore the detection of such decays complemented with neutrino data could provide extra information on leptonic CPV.

Another important question to be answered by future experiments is whether CP violation in the lepton sector follows the traditional CKM-like form with a unitary lepton mixing matrix. As discussed in Sec.~\ref{sec3.6}, deviations from unitarity in leptonic mixing appear in several extensions of the SM. Therefore, the detection of such effects would definitely point towards nonstandard physics. Nonunitarity effects are, in some cases, severely constrained by electroweak processes like radiative and three-body charged-lepton decays or leptonic $W$ and $Z$ decays. In Sec.~\ref{sec3.6}, we reviewed the present constraints on the unitarity of the leptonic mixing matrix in the context of the simple MUV hypothesis. In this framework, deviations from the standard CP violation scenario can be observed in future neutrino oscillation experiments like neutrino factories.

CP violation also plays a crucial role in cosmology, since the dynamical generation of the observed baryon asymmetry of the Universe requires that CP is violated. Once the SM is augmented with heavy states which can explain the smallness of neutrino masses, leptogenesis arises as the most natural and appealing mechanism to generate the excess of matter over antimatter. The CP violation present in the decays of the heavy Majorana neutrinos, not only gives rise to a leptonic asymmetry but it is also present in the effective neutrino mass matrix determined by the seesaw mechanism. Thus, one would expect that a connection between CP violation at low energies and the one relevant for leptogenesis could be established. Unfortunately, establishing this connection in a model-independent way is not possible. In general, assumptions about the flavor structure of the neutrino couplings and/or masses have to be considered in order to make predictions. In Sec.~\ref{sec4}, we showed a few examples in which a bridge between LCPV at low and high energies can be established. Obviously, the ultimate goal would be to test the leptogenesis mechanism at low energies, but this would be only possible if the lepton asymmetry is generated in the decays of particles that could be produced in accelerators. For certain, this will not be the case in a conventional leptogenesis framework in which the decaying seesaw mediators have masses much larger than the electroweak scale. However, if the origin of lepton-number violation is related to physics within our reach, then there may be a hope to test the leptogenesis mechanism or, at least, get a hint for it.

The answers to many of the open questions discussed in this review depend on the capability of future experiments to explore the unknown. In the neutrino sector, the milestones achieved in recent years have already excluded many theoretical ideas. Still, there are important questions like the ones concerning leptonic CP violation which are waiting for answers. We hope to find them just around the corner.

\section*{Acknowledgments}

We thank J. A. Aguilar-Saavedra, E. Fernandez-Martinez, S. Palomares-Ruiz, T. Schwetz, and M. T\'{o}rtola for the reading of parts of the manuscript and for the numerous comments and suggestions. We are also grateful to the CERN Theoretical Physics Division for hospitality during our visits to CERN where part of this work was accomplished. This work was partially supported by \textit{Funda\c c\~ao para a Ci\^encia e a Tecnologia} (FCT, Portugal) under Projects No. CERN/FP/116328/2010, PTDC/FIS/098188/2008, No. PTDC/FIS/111362/2009, and No. CFTP-FCT Unit 777, which are partially funded through POCTI (FEDER) and by Marie Curie Initial Training Network UNILHC PITN-GA-2009-237920.

\emph{Note added in proof.-} Recently, through the observation of electron-antineutrino disappearance, the Daya Bay Reactor Neutrino Experiment has measured the nonzero value~\cite{An:2012eh} $\sin^2(2\theta_{13})=0.092\pm0.016({\rm stat})\pm 0.005({\rm syst})$ with a significance of $5.2\sigma$. This confirms the T2K and MINOS experimental data presented in Sec.~\ref{sec3.1} and reinforces the prospects of a possible discovery of leptonic CP violation in the near future.

\end{document}